\documentclass[10pt,journal,cspaper,compsoc,onecolumn]{IEEEtran}

\usepackage{setspace}
\usepackage{negbinom}
\catcode`\@=11%

\def\ind{\protect\mathpalette{\protect\ind}{\perp}}
\def\ind#1#2{\mathrel{\setbox0\hbox{$#1#2$}%
\copy0\kern-\wd0\mkern4mu\box0}}

\allowdisplaybreaks

\def\notanonymous{1}
\def\figsintext{1}

\begin{document}
\abovedisplayskip=4pt plus3pt minus6pt
\belowdisplayskip=4pt plus3pt minus6pt

\title{Combinatorial Clustering and the Beta Negative Binomial Process}
\ifdefined\notanonymous
\author{Tamara Broderick,
Lester Mackey,
John Paisley, 
Michael I. Jordan
\IEEEcompsocitemizethanks{\IEEEcompsocthanksitem T. Broderick, L. Mackey, J. Paisley, and M. Jordan are with the Department
of Electrical Engineering and Computer Sciences and the Department of Statistics, University of California, Berkeley, CA 94705.
}
\thanks{}}
\else
\author{Anonymous}
\fi

\IEEEcompsoctitleabstractindextext{%
\begin{abstract}
We develop a Bayesian nonparametric approach to a general family
of latent class problems in which 
individuals can belong simultaneously to multiple 
classes and where each class can be exhibited multiple times by an 
individual.  We introduce a combinatorial stochastic process known 
as the \emph{negative binomial process} ($\nbp$) as an infinite-dimensional
prior appropriate for such problems.  We show that the $\nbp$ is conjugate 
to the beta process, and we characterize the posterior distribution under 
the beta-negative binomial process ($\bnbp$) and hierarchical models based 
on the $\bnbp$ (the $\hbnbp$).  We study the asymptotic properties of the 
$\bnbp$ and develop a three-parameter extension of the $\bnbp$ that
exhibits power-law behavior.  We derive MCMC algorithms for posterior 
inference under the $\hbnbp$, and we present experiments using these
algorithms in the domains of image segmentation, object recognition, and document analysis.
\end{abstract}}

\maketitle

%
%




\section{Introduction} \label{sec:introduction}

In traditional clustering problems the goal is to induce a set of latent 
classes and to assign each data point to one and only one class.  
This problem has been approached within a model-based framework via the 
use of finite mixture models, where the mixture components characterize
the distributions associated with the classes, and the mixing proportions capture 
the mutual exclusivity of the classes~\citep{FraleyRaftery02,MclachlanBasford88}.  
In many domains in which the notion of latent classes is natural, however, it 
is unrealistic to assign each individual to a single class.  For example, in 
genetics, while it may be reasonable to assume the existence of underlying ancestral 
populations that define distributions on observed alleles, each individual in an existing
population is likely to be a blend of the patterns associated with the ancestral 
populations.  Such a genetic blend is known as an \emph{admixture}~\citep{PritchardStDo00}.
A significant literature on model-based approaches to admixture has arisen
in recent years~\citep{blei:2003:latent,erosheva:2005:bayesian,PritchardStDo00}, 
with applications to a wide variety of domains in genetics and beyond, 
including document modeling and image analysis.\footnote{While we 
refer to such models generically as ``admixture models,'' we note that they 
are also often referred to as \emph{topic models} or \emph{mixed membership 
models}.}

Model-based approaches to admixture are generally built on the foundation 
of mixture modeling.  The basic idea is to treat each individual as a 
collection of data, with an exchangeability assumption imposed for the 
data within an individual but not between individuals.  For example, in 
the genetics domain the intra-individual data might be a set of genetic 
markers, with marker probabilities varying across ancestral populations.  
In the document domain the intra-individual data might be the set of words 
in a given document, with each document (the individual) obtained as a blend 
across a set of underlying ``topics'' that encode probabilities for the words.  
In the image domain, the intra-individual data might be visual characteristics like edges, hue,  
and location extracted from image patches. 
Each image is then a blend of object classes (e.g., grass, sky, or car),
each defining a distinct distribution over visual characteristics.
In general, this blending is achieved by making use of the probabilistic structure 
of a finite mixture but using a different sampling pattern.  In particular, 
mixing proportions are treated as random effects that are drawn once per 
individual, and the data associated with that individual are obtained by 
repeated draws from a mixture model having that fixed set of mixing proportions.  
The overall model is a hierarchical model, in which mixture components are 
shared among individuals and mixing proportions are treated as random 
effects.  

Although the literature has focused on using finite mixture models in this 
context, there has also been a growing literature on Bayesian nonparametric 
approaches to admixture models, notably the \emph{hierarchical Dirichlet 
process} ($\hdp$)~\citep{teh:2006:hierarchical_dirichlet}, where the number 
of shared mixture components is infinite.  Our focus in the current paper is 
also on nonparametric methods, given the open-ended nature of the inferential
objects with which real-world admixture modeling is generally concerned.

Although viewing an admixture as a set of repeated draws from a mixture
model is natural in many situations, it is also natural to take a different
perspective, akin to latent trait modeling, in which the individual 
(e.g., a document or a genotype) is characterized by the set of ``traits'' 
or ``features'' that it possesses, and where there is no assumption of mutual
exclusivity.  Here the focus is on the individual and not on the ``data''
associated with an individual.  Indeed, under the exchangeability assumption 
alluded to above it is natural to reduce the repeated draws from a mixture 
model to the counts of the numbers of times that each mixture component is 
selected, and we may wish to model these counts directly.  We may further 
wish to consider hierarchical models in which there is a linkage among the 
counts for different individuals.

This idea has been made explicit in a recent line of work based on the
\emph{beta process}.  Originally developed for survival analysis, where 
an integrated form of the beta process was used as a model for random 
hazard functions~\citep{hjort:1990:nonparametric}, more recently it has 
been observed that the beta process also provides a natural framework
for latent feature modeling~\citep{ThibauxJo07}.  In particular, as we discuss in 
detail in \mysec{bnp_priors}, a draw from the beta process yields an 
infinite collection of coin-tossing probabilities.  Tossing these 
coins---a draw from a \emph{Bernoulli process}---one obtains a set 
of binary features that can be viewed as a description of an admixed 
individual.
A key advantage of this approach is the conjugacy between the beta and Bernoulli processes:
this property allows for tractable inference, despite the countable infinitude of coin-tossing probabilities.
A limitation of this approach, however, is its restriction 
to binary features; indeed, one of the virtues of the mixture-model-based 
approach is that a given mixture component can be selected more than once, 
with the total number of selections being random.

To develop a more generally useful tool for modeling admixture within
a feature-based approach, we note that in the setting of classical random 
variables, beta-Bernoulli conjugacy is not the only form of conjugacy 
involving the beta distribution---the negative binomial distribution is 
also conjugate to the beta.  Anticipating the value of conjugacy in 
the setting of nonparametric models, we are motivated to develop a
stochastic process analogue of the negative binomial distribution,
a stochastic process that is conjugate to the beta process.  
It is one of the contributions of the current paper to define this process, 
which we refer to as the \emph{negative binomial process} ($\nbp$),\footnote{
\citet{zhou:2012:beta} have independently investigated negative binomial processes in the context of integer matrix factorization.
We discuss their concurrent contributions in more detail in \mysec{mix_hier}.}
and to provide a rigorous proof of its conjugacy to the beta process.
We then derive a new nonparametric mixture model based on the beta process and the $\nbp$ and a new model
of admixture based on the $\nbp$ and the \emph{hierarchical beta process}~\citep{ThibauxJo07}.
Unlike admixture models based on the HDP, our models allow for a random total number of features (e.g., words or traits)
per individual (e.g., a document or genotype).
We justify these modeling choices theoretically, by characterizing the prior behavior of the beta-negative binomial process hierarchy, 
and empirically on learning tasks from document analysis and computer vision.

The beta process and the $\nbp$ are not the only
way to generate infinite vectors of counts, and indeed there has been 
previous work on nonparametric count models based on the gamma process and 
the Poisson likelihood process~\citep{Thibaux08,Titsias07}.  A second contribution 
of the current paper is to explore the connections between these 
stochastic processes and the beta process and $\nbp$.  
Indeed, although some of the connections among the stochastic processes 
used in Bayesian nonparametrics are well known (e.g., that the Dirichlet 
process can be obtained from the gamma process by normalization), in
general there is a far less clear view of the linkages between these processes 
than there is of the linkages between the corresponding classical random 
variables.  We are able to establish several novel connections, including 
a new connection between the beta process and the gamma process. 

The remainder of the paper is organized as follows.  In \mysec{bnp_priors}
we present the framework of completely random measures that provides the
formal underpinnings for our work.  We discuss the Bernoulli process,
the $\nbp$, and their conjugacy to the beta process  
in \mysec{conjugacy}.  \mysec{mix_hier} focuses on the problem of modeling 
admixture and on general hierarchical modeling based on the negative binomial 
process.  \mysec{asymptotics} 
and \mysec{simulation} are devoted to a study of the asymptotic behavior 
of the $\nbp$ with a beta process prior, which we call the beta-negative binomial process ($\bnbp$).  
We describe algorithms for posterior inference in \mysec{posterior}.
Finally, we present experimental results. 
First, we use the $\bnbp$ to define a generative model for summaries of terrorist incidents with the goal of identifying the perpetrator of a given terrorist attack in \mysec{exp_doc}.
Second,
we demonstrate the utility of a finite approximation to the $\bnbp$ in the domain of automatic image segmentation
in \mysec{exp_image}. \mysec{conclusions} presents our conclusions.

\section{Completely Random Measures} \label{sec:bnp_priors}

In this section we review the notion of a completely random measure (CRM),
a general construction that yields random measures that are closely tied
to classical constructions involving sets of independent random variables.
We present CRM-based constructions of several of the stochastic processes
used in Bayesian nonparametrics, including the beta process, gamma process,
and Dirichlet process.  In the following section we build on the foundations
presented here to consider additional stochastic processes.

Consider a probability space $(\atomsp, \atomsig, \mathbb{P})$.
A \emph{random measure} is a random element $\mu$ such that $\mu(A)$ 
is a non-negative random variable for any $A$ in the sigma algebra $\atomsig$.
A \emph{completely random measure} (CRM) $\mu$ is a random measure such that,
for any disjoint, measurable sets $A, A' \in \atomsig$, we have that $\mu(A)$ and
$\mu(A')$ are independent random variables~\citep{Kingman67}. 
Completely random measures can be shown to be composed of at most three
components:
\begin{enumerate*}
	\item A \emph{deterministic measure}. For deterministic $\mu_{det}$, it is trivially the case that
		$\mu_{det}(A)$ and $\mu_{det}(A')$ are independent for disjoint $A,A'$.
	\item A \emph{set of fixed atoms}. Let $(\fatom_{1},\ldots,\fatom_{L}) \in \atomsp^{L}$ be a collection of deterministic locations, and let 
	$(\fweight_{1},\ldots,\fweight_{L}) \in \mathbb{R}_{+}^{L}$
	be a collection of independent random weights for the atoms. 
	The collection may be countably infinite, in which case we say $L = \infty$. Then
	let $\mu_{fix} = \sum_{l=1}^{L} \fweight_{l} \delta_{\fatom_{l}}$. The independence of the $\fweight_{l}$
	implies the complete randomness of the measure.
	\item An {\em ordinary component}. Let $\nu_{\pp}$ be a Poisson process intensity on the space
	$\atomsp \times \mathbb{R}_{+}$. Let $\{(\oatom_{1},\oweight_{1}), (\oatom_{2}, \oweight_{2}),\ldots\}$ be a draw from
	the Poisson process with intensity $\nu_{\pp}$.
	Then the ordinary component is the measure
	$\mu_{ord} = \sum_{j=1}^{\infty} \oweight_{j} \delta_{\oatom_{j}}$. 
	Here, the complete randomness follows from properties of the Poisson process.
\end{enumerate*}

One observation from this componentwise breakdown of CRMs
is that we can obtain a countably infinite collection of random variables, the $\oweight_{j}$, from the
Poisson process component if $\nu_{\pp}$ has infinite total mass (but is still sigma-finite).
Consider again the criterion that a CRM $\mu$ yield independent
random variables when applied to  
disjoint sets. In light of the observation about the collection $\{\oweight_{j}\}$,
this criterion may now be seen
as an extension of an independence assumption in the case of a finite set
of random variables. We cover specific examples next.

\subsection{Beta process} \label{sec:bp}

We have seen that CRMs have three components. Therefore, in order
to describe any CRM, it is enough to specify the deterministic measure,
fixed atoms, and ordinary component. The
{\em beta process}~\citep{hjort:1990:nonparametric, Kim99, ThibauxJo07}
is an example of a CRM. It has the following parameters: a \emph{mass 
parameter} $\bpmass > 0$, a \emph{concentration parameter} $\bpconc > 0$,
a purely atomic measure $\base_{fix} = \sum_{l} \bpfcpw_{l} \delta_{\fatom_{l}}$ 
with $\bpmass \bpfcpw_{l} \in (0,1)$ for all $l$ a.s., and a purely continuous
probability measure 
$\base_{ord}$ on $\atomsp$. Note that we
have explicitly separated out the mass parameter $\bpmass$ so 
that, e.g., $\base_{ord}$ is a probability measure; in \citet{ThibauxJo07}, these two
parameters are expressed as a single measure with total mass
equal to $\bpmass$. Typically, though, the normalized measure $\base_{ord}$ is used
separately from the mass parameter $\bpmass$ (as we will see below), so the 
notational separation is convenient.
Often the final two measure parameters are abbreviated as their sum: 
$\base = \base_{fix} + \base_{ord}$.

Given these parameters, the beta process has the following description as a CRM:
\begin{enumerate*}
	\item The deterministic measure is uniformly zero.
	\item The fixed atoms have locations $(\fatom_{1},\ldots,\fatom_{L}) \in \atomsp^{L}$,
	where $L$ is potentially infinite though typically finite. Atom weight $\fweight_{l}$ has distribution
	\begin{equation}
		\label{eq:bp_atom_weights}
		\fweight_{l} \indep \tb\left(\bpconc \bpmass \bpfcpw_{l},
			\bpconc (1 - \bpmass \bpfcpw_{l}) \right),
	\end{equation}
	where the $\bpfcpw_{l}$ parameters are the weights in the purely atomic measure $\base_{fix}$.
	\item The ordinary component has Poisson process intensity $\base_{ord} \times \nu$, where $\nu$ is the 
	measure
	\begin{equation}
		\label{eq:bp-intensity}
		\nu(d\bpweight) = \bpmass \bpconc \bpweight^{-1} (1 - \bpweight)^{\bpconc - 1} \; d\bpweight,
	\end{equation}
	which is sigma-finite with finite mean. It follows that the number of atoms in this component
	will be countably infinite with finite sum.
\end{enumerate*}
As in the original specification of \citet{hjort:1990:nonparametric} and
\citet{Kim99}, \eq{bp-intensity} can be generalized by allowing $\bpconc$ to depend on the $\atomsp$ coordinate. The homogeneous intensity in \eq{bp-intensity} seems to be used predominantly in practice~\citep{ThibauxJo07, fox:2009:sharing} though, and we focus on it here for ease of exposition. Nonetheless, we note that our results below extend easily to the non-homogeneous case.

The CRM is the sum of its components. Therefore, we may write a draw
from the beta process as
\begin{equation}
	\label{eq:bp_draw}
	\bpdraw = \sum_{k=1}^{\infty} \bpweight_{k} \delta_{\atom_{k}}
		\defeq \sum_{l=1}^{L} \fweight_{l} \delta_{\fatom_{l}}
			+ \sum_{j=1}^{\infty} \oweight_{j} \delta_{\oatom_{j}},
\end{equation}
with atom locations equal to the union of the fixed atom and ordinary component 
atom locations $\{\atom_k\}_{k} = \{\fatom_{l}\}_{l=1}^{L} \cup \{\oatom_{j}\}_{j=1}^{\infty}$. Notably, $\bpdraw$ is a.s.\ discrete. We denote a draw from the beta process 
as $\bpdraw \sim \bp(\bpconc, \bpmass, \base)$.
The provenance of the name ``beta process'' is now clear; each atom weight in the 
fixed atomic component is beta-distributed, and the Poisson process intensity generating 
the ordinary component is that of an improper beta distribution.

From the above description, the beta process provides a prior on a potentially infinite
vector of weights, each in $(0,1)$ and each associated with a corresponding parameter
$\atom \in \atomsp$. The potential countable infinity comes from the Poisson process
component. The weights in $(0,1)$ may be interpreted as probabilities, though
not as a distribution across the indices as we note that they need not sum to one.
We will see in \mysec{mix_hier} that the beta process is appropriate for
feature modeling~\citep{ThibauxJo07,griffiths:2006:infinite}.
In this context, each atom, indexed by $k$, of $\bpdraw$ corresponds to a feature. 
The atom weights $\{\bpweight_{k}\}$, which are each in $[0,1]$ a.s., can be
viewed as representing the frequency with which each feature occurs in the data set.
The atom locations $\{\atom_{k}\}$ represent parameters associated with the features
that can be used in forming a likelihood.

In \mysec{asymptotics}, we will show that an extension to the beta process
called the \emph{three-parameter beta process} has certain desirable properties 
beyond the classic beta process, in particular its ability to generate
power-law behavior~\citep{teh:2009:indian,broderick:2012:beta},
which roughly says that the number of features grows as 
a power of the number of data points.  In the three-parameter case, 
we introduce a {\em discount parameter} $\bpdisc \in (0,1)$
with $\bpconc > -\bpdisc$ and $\bpmass > 0$ such that:
\begin{enumerate*}
	\item There is again no deterministic component.
	\item The fixed atoms have locations $(\fatom_{1},\ldots,\fatom_{L}) \in \atomsp^{L}$, with $L$ potentially infinite but typically finite.
	Atom weight $\fweight_{l}$ has distribution $\fweight_{l} \indep \tb\left(\bpconc \bpmass \bpfcpw_{l} - \bpdisc,
			\bpconc (1 - \bpmass \bpfcpw_{l}) + \bpdisc \right)$,
	where the $\rho_{l}$ parameters are the weights in the purely atomic measure $\base_{fix}$
	and we now have the constraints $\bpconc \bpmass \bpfcpw_{l} - \bpdisc, \bpconc (1 - \bpmass \bpfcpw_{l}) + \bpdisc \ge 0$.
	\item The ordinary component has Poisson process intensity $H_{ord} \times \nu$, where $\nu$ is the 
	measure:
	$$
		\nu(d\bpweight) = \bpmass \frac{\Gamma(1+\bpconc)}{
			\Gamma(1-\bpdisc) \Gamma(\bpconc + \bpdisc)}
			\bpweight^{-1-\bpdisc} (1 - \bpweight)^{\bpconc + \bpdisc - 1} \; d\bpweight.
	$$
\end{enumerate*}
Again, we focus on the homogeneous intensity $\nu$ as in the beta process case though it is straightforward to allow $\bpconc$ to depend on coordinates in $\atomsp$.

In this case, we again have the full process draw $\bpdraw$ as in \eq{bp_draw}, and we say $\bpdraw \sim \tbp(\bpdisc, \bpconc, \bpmass, \base)$.

\subsection{Reparameterized beta process} \label{sec:rbp}

The specification that the atom parameters in the beta process be of the 
form $\bpconc \bpmass \rho_{l}$ and $\bpconc (1 - \bpmass \rho_{l})$ can 
be unnecessarily constraining.  Indeed, the classical beta distribution has 
two free parameters. Yet, in the beta process as described above, $\bpconc$ 
and $\bpmass$ are determined as part of the Poisson process intensity, 
so there is essentially one free parameter for each of the beta-distributed 
weights associated with the atoms (\eq{bp_atom_weights}).  A related 
problematic issue is that the beta process forces the two parameters in 
the beta distribution associated with each atom to sum to $\bpconc$, 
which is constant across all of the atoms.  

One way to remove these restrictions is to allow $\bpconc = \bpconc(\atom)$,
a function of the position $\atom \in \atomsp$ as mentioned above.
However, we demonstrate in \app{text_connections} that there are reasons to
prefer a fixed concentration parameter $\bpconc$ for the ordinary component; there 
is a fundamental relation between this parameter and similar parameters
in other common CRMs (e.g., the Dirichlet process, which
we describe in \mysec{dp}).  Moreover, the concern here is entirely centered 
on the behavior of the fixed atoms of the process, and letting $\bpconc$ 
depend on $\atom$ retains the unusual---from a classical parametric perspective---form 
of the beta distribution in \eq{bp_atom_weights}.  As an alternative, we
provide a specification of the beta process that more closely aligns
with the classical perspective in which we allow two general beta parameters 
for each atom.  As we will see, this reparameterization is natural,
and indeed necessary, in considering conjugacy.

We thus define the \emph{reparameterized beta process} ($\rbp$) as having the 
following parameterization: a {\em mass parameter} $\bpmass > 0$,
a {\em concentration parameter} $\bpconc > 0$, a number of fixed atoms 
$L \in \{0, 1,2,\ldots\} \cup \{\infty\}$ with locations $(\fatom_{1},\ldots,\fatom_{L}) \in \atomsp^{L}$,
two sets of strictly positive atom weight parameters $\{\rho_{l}\}_{l=1}^{L}$ 
and $\{\sigma_{l}\}_{l=1}^{L}$, and a purely continuous measure $\base_{ord}$ on $\atomsp$.
In this case, the atom weight parameters satisfy the simple condition $\rho_{l}, \sigma_{l} > 0$ for all $l \in \{1,\ldots,L\}$.
This specification is the same as the beta process specification introduced 
above with the sole exception of a more general parameterization for the 
fixed atoms.  We obtain the following CRM:
\begin{enumerate*}
	\item There is no deterministic measure.
	\item There are $L$ fixed atoms with locations $(\fatom_{1},\ldots,\fatom_{L}) \in \atomsp^{L}$
		and corresponding weights 
		$
			\fweight_{l} \indep \tb\left(\rho_{l}, \sigma_{l} \right).
		$
	\item The ordinary component has Poisson process intensity $H_{ord} \times \nu$, where $\nu$ is the 
	measure
	$
		\nu(d\bpweight) = \bpmass \bpconc \bpweight^{-1} (1 - \bpweight)^{\bpconc - 1} \; d\bpweight.
	$
\end{enumerate*}
As discussed above, we favor the homogeneous intensity $\nu$ in exposition but note the straightforward extension to allow $\bpconc$ to depend on $\atomsp$ location.

We denote this CRM by $\bpdraw \sim \rbp(\bpconc, \bpmass, \tbf{\fatom}, \tbf{\rho}, \tbf{\sigma}, H_{ord})$.

\subsection{Gamma process} \label{sec:gap}

While the beta process provides a countably infinite vector of frequencies in 
$(0,1]$ with associated parameters $\atom_{k}$, it is sometimes useful to have 
a countably infinite vector of positive, real-valued quantities that can be
used as rates rather than frequencies for features.  We can obtain such a 
prior with the \emph{gamma process}~\citep{Ferguson73}, a CRM with the
following parameters: a \emph{concentration parameter} $\gpconc > 0$,
a \emph{scale parameter} $\gpscale > 0$, a purely atomic measure 
$\base_{fix} = \sum_{l} \rho_{l} \delta_{\fatom_{l}}$ with 
$\forall l, \rho_{l} > 0$, and a purely continuous 
measure $\base_{ord}$ with support on $\Psi$.
Its description as a CRM is as follows~\citep{Thibaux08}:
\begin{enumerate*}
	\item There is no deterministic measure.
	\item The fixed atoms have locations $(\fatom_{1},\ldots,\fatom_{L}) \in \atomsp^{L}$, where $L$ is potentially infinite but typically finite.
	Atom weight $\fweight_{l}$ has distribution
	$
		\fweight_{l} \indep \ga(\gpconc \rho_{l}, \gpscale),
	$
	where we use the shape-inverse-scale parameterization of the gamma distribution and where the $\rho_{l}$ parameters are the weights in the purely atomic measure $\base_{fix}$.
	\item The ordinary component has Poisson process intensity $H_{ord} \times \nu$, where $\nu$ is the 
	measure:
	\begin{equation}
		\label{eq:gap_intensity}
		\nu(d\gapweight) = \gpconc \gapweight^{-1} \exp\left( - \gpscale \gapweight \right) \; d\gapweight.
	\end{equation}
\end{enumerate*}

As in the case of the beta process, the gamma process can be expressed
as the sum of its components:
$
	\gapdraw
		= \sum_{k} \gapweight_{k} \delta_{\atom_{k}}
		\defeq \sum_{l=1}^{L} \fweight_{l} \delta_{\fatom_{l}}
			+ \sum_{j} \oweight_{j} \delta_{\oatom_{j}}.
$
We denote this CRM as $\gapdraw \sim \gap(\gpconc, \gpscale, \base)$,
for $\base = \base_{fix} + \base_{ord}$.

\subsection{Dirichlet process} \label{sec:dp}

While the beta process has been used as a prior in featural models, the 
Dirichlet process is the classic Bayesian nonparametric prior for clustering
models~\citep{Ferguson73,maceachern:1998:estimating,mccloskey:1965:model,neal:2000:markov,
west:1992:hyperparameter}.  The Dirichlet process itself is not a CRM; 
its atom weights, which represent cluster frequencies, must sum to
one and are therefore correlated. But it can be obtained by normalizing 
the gamma process~\citep{Ferguson73}.

In particular, using facts about the Poisson process~\citep{Kingman93},
one can check that, when there are finitely many fixed atoms, we have
$\gapdraw(\atomsp) < \infty$ a.s.; that is, the total 
mass of the gamma process is almost surely finite despite
having infinitely many atoms from the ordinary component. Therefore, normalizing 
the process by dividing its weights by its total mass is well-defined. 
We thus can define a \emph{Dirichlet process} as 
\[
\dpdraw = \sum_{k} \dpweight_{k} \delta_{\atom_{k}} \defeq \gapdraw / \gapdraw(\atomsp), 
\]
where $\gapdraw \sim \gap(\gpconc, 1, \base)$, and where there are two
parameters: a \emph{concentration parameter} $\dpconc$ and a \emph{base 
measure} $\base$ with finitely many fixed atoms. 
Note that while we have chosen the scale parameter 
$\gpscale = 1$ in this construction, the choice is in fact arbitrary 
for $\gpscale > 0$ and does not affect the $\dpdraw$
distribution (\eqw{4.15} and \pw{83} of \citet{pitman:2006:combinatorial}).

From this construction, we see immediately that the Dirichlet process is 
almost surely atomic, a property inherited from the gamma process. 
Moreover, not only are the weights of the Dirichlet process all contained
in $(0,1)$ but they further sum to one. Thus, the Dirichlet process may 
be seen as providing a probability distribution on a countable set. 
In particular, this countable set is often viewed as a countable number 
of clusters, with cluster parameters $\atom_{k}$. 

\section{Conjugacy and combinatorial clustering}
\label{sec:conjugacy}

In \mysec{bnp_priors}, we introduced CRMs and showed how a number 
of classical Bayesian nonparametric priors can be derived from CRMs.
These priors provide infinite-dimensional vectors of real values, 
which can be interpreted as feature frequencies, feature rates,
or cluster frequencies.  To flesh out such interpretations we need
to couple these real-valued processes with discrete-valued processes
that capture combinatorial structure.  In particular, viewing the 
weights of the beta process as feature frequencies, it is natural 
to consider binomial and negative binomial models that transform 
these frequencies into binary values or nonnegative integer counts.  
In this section we describe stochastic processes that achieve such 
transformations, again relying on the CRM framework.  

The use of a Bernoulli likelihood whose frequency parameter is obtained
from the weights of the beta process has been explored in the context
of survival models by~\citet{hjort:1990:nonparametric} and~\citet{Kim99}
and in the context of feature modeling by~\citet{ThibauxJo07}.  After
reviewing the latter construction, we discuss a similar construction based
on the negative binomial process.  Moreover, recalling that
\citet{ThibauxJo07}, building on work of
\citet{hjort:1990:nonparametric} and \citet{Kim99}, have shown that the Bernoulli
likelihood is conjugate to the beta process, we demonstrate an analogous
conjugacy result for the negative binomial process.

\subsection{Bernoulli process} \label{sec:bep}

One way to make use of the beta process is to couple it to a 
\emph{Bernoulli process} \citep{ThibauxJo07}.
The Bernoulli process, denoted $\bep(\bernbase)$, has a single parameter, a 
\emph{base measure} $\bernbase$; $\bernbase$ is any discrete measure
with atom weights in $(0,1]$.  Although our focus will be on models in which $\bernbase$ 
is a draw from a beta process, as a matter of the general definition of the
Bernoulli process the 
base measure $\bernbase$ need not be a CRM or even random---just 
as the Poisson distribution is defined relative to a parameter
that may or may not be random in general but which is sometimes given a gamma
distribution prior.  Since $\bernbase$ is discrete by assumption, we may write 
\begin{equation}
	\label{eq:base_bep_disc}
	\bernbase = \sum_{k=1}^{\infty} \bpweight_{k} \delta_{\atom_{k}}
\end{equation}
with $\bpweight_{k} \in (0,1]$.
We say that the random measure $\likedraw$ is drawn from a Bernoulli process, 
$\likedraw \sim \bep(\bernbase)$, if $\likedraw = \sum_{k=1}^{\infty} \likeweight_{k} 
\delta_{\atom_{k}}$ with $\likeweight_{k} \indep \bern(\bpweight_{k}) \textrm{ for } 
k = 1,2,\ldots$. That is, to form the Bernoulli process,
we simply make a Bernoulli random variable draw for every one of the
(potentially countable) atoms of the base measure.
This definition of the Bernoulli process was proposed by
\citet{ThibauxJo07}; it differs from a precursor introduced by
\citet{hjort:1990:nonparametric} in the context of survival analysis.

One interpretation for this construction is that the atoms of the base measure 
$\bernbase$ represent potential features of an individual, with feature frequencies equal 
to the atom weights and feature characteristics defined by the atom locations.  
The Bernoulli process draw can be viewed as characterizing the individual by 
the set of features that have weights equal to one.  Suppose $\bernbase$ is derived
from a Poisson process as the ordinary component of a completely random measure
and has finite mass;
then the number of features exhibited by the Bernoulli process, i.e.\ the total
mass of the Bernoulli
process draw, is a.s.\ finite.  Thus the Bernoulli process can be viewed as
providing a Bayesian nonparametric model of sparse binary feature vectors.

Now suppose that the base measure parameter is a draw from a beta process with 
parameters $\bpconc > 0$, $\bpmass > 0$, and base measure $\base$. That is,
$\bpdraw \sim \bp(\bpconc, \bpmass, \base)$ and $\likedraw \sim \bep(\bpdraw)$.
We refer to the overall process as the \emph{beta-Bernoulli process} 
($\bbep$). Suppose that the beta process $\bpdraw$ has a finite number of fixed atoms.
Then
we note that the finite mass of the ordinary component of $\bpdraw$ implies that
$\likedraw$ has support on a finite set. That is, even though $B$ has a countable
infinity of atoms, $I$ has only a finite number of atoms. This observation is
important since, in any practical model, we will want an individual to exhibit only
finitely many features.


\citet{hjort:1990:nonparametric} and \citet{Kim99} originally established that
the posterior distribution of $\bpdraw$ under a constrained form of the $\bbep$
was also a beta process with known parameters.  \citet{ThibauxJo07}
went on to extend this analysis to the full $\bbep$.
We cite the result by \cite{ThibauxJo07} here,
using the completely random measure notation established above.
\begin{theorem} \label{thm:bp_bep_conjugacy} \quad \\
	\textbf{Summary:} The beta process prior is conjugate to the Bernoulli process likelihood. \\
	\textbf{Detailed:} Let $\base$ be a measure with atomic component
	$
		\base_{fix} = \sum_{l=1}^{L} \bpfcpw_{l} \delta_{\fatom_{l}}
	$
	and continuous component $H_{ord}$. Let $\bpconc$ and $\bpmass$ be
	strictly positive scalars. Consider $N$ conditionally-independent
	draws from the Bernoulli process: $\likedraw_{n} = \sum_{l=1}^{L} 
	\likeweight_{fix,n,l} \delta_{\fatom_{l}}
			+ \sum_{j=1}^{J} \likeweight_{ord,n,j} \delta_{\oatom_{j}}
			\iid \bep(\bpdraw), \textrm{ for } n = 1,\ldots,N$
	with $\bpdraw \sim \bp(\bpconc, \bpmass, \base)$.
	That is, the Bernoulli process draws have $J$ atoms that are not located at the atoms of $\base_{fix}$.
	Then, $\bpdraw | \likedraw_{1},\ldots,\likedraw_{N} \sim \bp(\bpconc_{post}, \bpmass_{post}, \base_{post})$ with 
	$\bpconc_{post} = \bpconc + N$,
	$\bpmass_{post} = \bpmass \frac{\bpconc}{\bpconc + N}$, and
	$\base_{post,ord} = \base_{ord}$. Further,
			$
				\base_{post,fix} = \sum_{l=1}^{L} \bpfcpw_{post,l} \delta_{\fatom_{l}}
					+ \sum_{j=1}^{J} \oweight_{post,j} \delta_{\oatom_{j}},
			$
			where
			$
				\bpfcpw_{post,l} = \bpfcpw_{l} + (\bpconc_{post} \bpmass_{post})^{-1} \sum_{n=1}^{N} \likeweight_{fix,n,l}
			$
			and
			$
				\oweight_{post,j} = (\bpconc_{post} \bpmass_{post})^{-1} \sum_{n=1}^{N} \likeweight_{ord,n,j}.
			$
\end{theorem}
Note that the posterior beta-distributed fixed atoms are well-defined
since $\oweight_{post,j} > 0$ follows from $\sum_{n=1}^{N} \likeweight_{ord,n,j} > 0$,
which holds by
construction.
As shown by~\citet{ThibauxJo07}, if the underlying beta process is integrated out 
in the $\bbep$, we recover the \emph{Indian buffet process} of~\cite{griffiths:2006:infinite}.

An easy consequence of \thm{bp_bep_conjugacy} is the following.
\begin{corollary} \label{cor:rbp_bep_conjugacy} \quad \\
	\textbf{Summary:} The $\rbp$ prior is conjugate to the Bernoulli process likelihood. \\
	\textbf{Detailed:}
	Assume the conditions of \thm{bp_bep_conjugacy}, and 
	consider $N$ conditionally-independent Bernoulli process draws:
	$
		\likedraw_{n} = \sum_{l=1}^{L} \likeweight_{fix,n,l} \delta_{\fatom_{l}}
			+ \sum_{j=1}^{J} \likeweight_{ord,n,j} \delta_{\oatom_{j}}
			\iid \bep(\bpdraw), \textrm{ for } n = 1,\ldots,N
	$
	with
	$
		\bpdraw \sim \rbp(\bpconc, \bpmass, \tbf{\fatom}, \tbf{\rho}, \tbf{\sigma}, H_{ord}) 
	$
	and $\{\rho_{l}\}_{l=1}^{L}$ and $\{\sigma_{l}\}_{l=1}^{L}$ strictly positive scalars. 
    Then, $\bpdraw | \likedraw_{1},\ldots,\likedraw_{N} \sim \rbp(\bpconc_{post}, \bpmass_{post}, \tbf{\fatom}_{post}, \tbf{\rho}_{post}, \tbf{\sigma}_{post}, H_{post,ord})$, for 
		$\bpconc_{post} = \bpconc + N$,
		$\bpmass_{post} = \bpmass \frac{\bpconc}{\bpconc + N}$,
		$\base_{post,ord} = \base_{ord}$, and $L+J$ fixed atoms, $\{\fatom_{post,l'}\}= \{\fatom_{l}\}_{l=1}^{L} \cup \{\oatom_{j}\}_{j=1}^{J}.$
		The $\tbf{\rho}_{post}$ and $\tbf{\sigma}_{post}$ parameters satisfy 
			$
				\rho_{post, l} = \rho_{l} + \sum_{n=1}^{N} \likeweight_{fix,n,l}
			$
			and
			$
				\sigma_{post, l} = \sigma_{l} + N - \sum_{n=1}^{N} \likeweight_{fix,n,l}
			$
			for $l \in \{1,\dots,L\}$ and 
			$
				\rho_{post, L+j} = \sum_{n=1}^{N} \likeweight_{ord,n,j}
			$
			and
			$
				\sigma_{post, L+j} = \bpconc + N - \sum_{n=1}^{N} \likeweight_{ord,n,j}
			$
			for $j \in \{1,\dots,J\}$.
\end{corollary}
The usefulness of the $\rbp$ becomes apparent in the posterior parameterization;
the distributions associated with the fixed atoms
more closely mirror the classical parametric conjugacy between the Bernoulli distribution 
and the beta distribution.  This is an issue of convenience in the
case of the $\bbep$, but it is more significant in the case of the negative 
binomial process, as we show in the following section, where conjugacy is 
preserved only in the $\rbp$ case (and not for the traditional $\bp$).

\subsection{Negative binomial process} \label{sec:nbp}

The Bernoulli distribution is not the only distribution that yields 
conjugacy when coupled to the beta distribution in the classical parametric setting; 
conjugacy holds for the negative binomial distribution as well.  As we show
in this section, this result can be extended to stochastic processes via 
the CRM framework.  

We define the \emph{negative binomial process} as a CRM with two parameters: 
a shape parameter $r > 0$ and a discrete base measure
$\nbbase = \sum_{k} \bpweight_{k} \delta_{\atom_{k}}$ whose 
weights $\bpweight_{k}$ take values in $(0,1]$.  As in the case of the Bernoulli process, 
$\nbbase$ need not be random at this point. Since $\nbbase$ is discrete, 
we again have a representation for $\nbbase$ as in \eq{base_bep_disc}, and we say that the
random measure $\likedraw$ is drawn from a negative binomial process, 
$\likedraw \sim \nbp(r, \nbbase)$, if $\likedraw = \sum_{k=1}^{\infty} \likeweight_{k} 
\delta_{\atom_{k}}$ with $\likeweight_{k} \indep \negbin(r, \bpweight_{k})
\textrm{ for}\ k = 1,2,\ldots$. That is, the negative binomial process is formed
by simply making a single draw from a negative binomial distribution at each
of the (potentially countably infinite) atoms of
$\nbbase$. This construction generalizes the geometric process studied by
\citet{Thibaux08}.

As a Bernoulli process draw can be interpreted as assigning a set of features to a data point, so can
we interpret a draw from the negative binomial process as assigning a set of feature counts to a data point. In
particular, as for the Bernoulli process, we assume that each data point has its own 
draw from the negative binomial process. Every atom with strictly positive mass in the
this draw corresponds to a feature that is exhibited by this data point. Moreover,
the size of the atom, which is a positive integer by construction, dictates how many times the feature
is exhibited by the data point. For example, if the data point is a document, and each feature represents
a particular word, then the negative binomial process draw would tell us how many occurrences of each
word there are in the document.

If the base measure for a negative binomial process is a beta process, we say that the combined process
is a {\em beta-negative binomial process} ($\bnbp$). If the base measure is a three-parameter beta
process, we say that the combined process is a {\em three-parameter beta-negative binomial process} ($\tbnbp$). When either the $\bp$ or $\tbp$ has a finite number of fixed atoms, the 
ordinary component of the $\bp$ or $\tbp$ still has an infinite number of atoms, but the number of
atoms in the negative binomial process is a.s.\ finite. We prove this fact and more in \mysec{asymptotics}.

We now suppose that the base measure for the negative binomial process is a draw 
$\bpdraw$ from an $\rbp$ with parameters $\bpconc > 0$, $\bpmass > 0$, $\{\fatom_{l}\}_{l=1}^{L}$, $\{\rho_{l}\}_{l=1}^{L}$, $\{\sigma_{l}\}_{l=1}^{L}$, and $\base_{ord}$. 
The overall specification is $\bpdraw \sim \rbp(\bpconc, \bpmass, \tbf{\fatom}, \tbf{\rho}, \tbf{\sigma}, \base_{ord})$ and $\likedraw \sim \nbp(r, \bpdraw)$.  The following 
theorem characterizes the posterior distribution for this model.  The proof
is given in \app{conjugacy_proofs}.
\begin{theorem} \label{thm:rbp_nbp_conjugacy} \quad \\
	\textbf{Summary:} The $\rbp$ prior is conjugate to the negative binomial process likelihood. \\
	\textbf{Detailed:}
	Let $\bpconc$ and $\bpmass$ be
	strictly positive scalars. 
	Let $(\fatom_{1},\ldots,\fatom_{L}) \in \atomsp^{L}$. Let the members of
	$\{\rho_{l}\}_{l=1}^{L}$ and $\{\sigma_{l}\}_{l=1}^{L}$ be strictly positive scalars.
	Let $\base_{ord}$ be continuous measure on $\atomsp$.
	Consider the following model for $N$ draws from a negative binomial process:
	$
		\likedraw_{n} = \sum_{l=1}^{L} \likeweight_{fix,n,l} \delta_{\fatom_{l}}
			+ \sum_{j=1}^{J} \likeweight_{ord,n,j} \delta_{\oatom_{j}}
			\iid \nbp(\bpdraw), \textrm{ for } n = 1,\ldots,N
	$
	with
	$
		\bpdraw \sim \rbp(\bpconc, \bpmass, \tbf{\fatom}, \tbf{\rho}, \tbf{\sigma}, H_{ord}).
	$
	That is, the negative binomial process draws have $J$ atoms
	that are not located at the atoms of $\base_{fix}$.
	Then, $\bpdraw | \likedraw_{1},\ldots,\likedraw_{N} \sim \rbp(\bpconc_{post}, \bpmass_{post}, \tbf{\fatom}_{post}, \tbf{\rho}_{post}, \tbf{\sigma}_{post}, H_{post,ord})$ for 
		$\bpconc_{post} = \bpconc + N r$,
		$\bpmass_{post} = \bpmass \frac{\bpconc}{\bpconc + Nr}$,
		$\base_{post,ord} = \base_{ord}$, and $L + J$ fixed atoms, $\{\fatom_{post,l}\} = \{\fatom_{l}\}_{l=1}^{L} \cup \{\oatom_{j}\}_{j=1}^{J}$. 
		The $\tbf{\rho}_{post}$ and $\tbf{\sigma}_{post}$ parameters satisfy 
			$
				\rho_{post, l} = \rho_{l} + \sum_{n=1}^{N} \likeweight_{fix,n,l}
			$
			and
			$
				\sigma_{post, l} = \sigma_{l} + r N
			$
			for $l \in \{1,\dots,L\}$ and 
			$
				\rho_{post, L+j} = \sum_{n=1}^{N} \likeweight_{ord,n,j}
			$
			and
			$
				\sigma_{post, L+j} = \bpconc + r N
			$
			for $j \in \{1,\dots,J\}$.
\end{theorem}

\section{Mixtures and admixtures} \label{sec:mix_hier}

We now assemble the pieces that we have introduced and consider Bayesian
nonparametric models of admixture.  Recall that the basic idea of
an admixture is that an individual (e.g., an organism, a document,
or an image) can belong simultaneously to multiple classes.  This
can be represented by associating a binary-valued vector with each
individual; the vector has value one in components corresponding to classes to which
the individual belongs and zero in components corresponding to classes to which
the individual does not belong.  More generally, we wish to remove the restriction to
binary values and consider a general notion of admixture in which
an individual is represented by a nonnegative, integer-valued vector.
We refer to such vectors as \emph{feature vectors}, and view the
components of such vectors as counts representing the number of
times the corresponding feature is exhibited by a given individual.
For example, a document may exhibit a given word zero or more times.

As we discussed in \mysec{introduction}, the standard approach to modeling
an admixture is to assume that there is an exchangeable set of data associated with
each individual and to assume that these data are drawn from a finite
mixture model with individual-specific mixing proportions.  There is 
another way to view this process, however, that opens the door to a variety
of extensions.  Note that to draw a set of data from a mixture, we can 
first choose the number of data points to be associated with each mixture 
component (a vector of counts) and then draw the 
data point values independently from each selected mixture component.  That is, we 
randomly draw nonnegative integers $\likeweight_{k}$ for each mixture 
component (or \emph{cluster}) $k$.  Then, for each $k$ and each 
$n = 1, \ldots, \likeweight_{k}$, we draw a data point $x_{k,n} 
\sim F(\atom_{k})$, where $\atom_{k}$ is the parameter associated with 
mixture component $k$.  The overall collection of data
for this individual is $\{x_{k,n}\}_{k, n}$, with 
$N = \sum_k \likeweight_{k}$ total points.  One way to generate data
according to this decomposition is to make use of the $\nbp$.
We draw $\likedraw = \sum_{k} \likeweight_{k} \delta_{\atom_{k}} \sim 
\nbp(r, \bpdraw)$, where $\bpdraw$ is drawn from a beta process,
$\bpdraw \sim \bp(\bpconc, \bpmass, \base)$.  The overall model is 
a $\bnbp$ mixture model for the counts, coupled to a conditionally
independent set of draws for the individual's data points $\{x_{k,n}\}_{k, n}$.

An alternative approach in the same spirit is to make use of a gamma process 
(to obtain a set of rates) that is coupled to a Poisson likelihood process ($\plp$)\footnote{
We use the terminology ``Poisson likelihood process'' to distinguish
a particular process with Poisson distributions affixed to each atom
of some base distribution
from the more general Poisson point process of \citet{Kingman93}.}
to convert 
the rates into counts~\citep{Titsias07}.  In particular, given a base measure 
$\gapdraw = \sum_{k} \gapweight_{k} \delta_{\atom_{k}}$, let $\likedraw 
\sim \plp(\gapdraw)$ denote $\likedraw = \sum_{k} \likeweight_{k} 
\delta_{\atom_{k}}$, with $\likeweight_{k} \sim \pois(\gapweight_{k})$.
We then consider a \emph{gamma Poisson process} ($\gaplp$) as follows:
$\gapdraw \sim \gap(\gpconc, \gpscale, \base)$, $\likedraw = \sum_{k} 
\likeweight_{k} \delta_{\atom_{k}} \sim \plp(\gapdraw)$, and
$x_{k,n} \sim F(\atom_{k}), \  \textrm{for } n = 1,\ldots,\likeweight_{k}$ and each $k$.

Both the $\bnbp$ approach and the $\gaplp$ approach deliver a random measure,
$\likedraw = \sum_{k} \likeweight_{k} \delta_{\atom_{k}}$, as a representation
of an admixed individual.  While the atom locations, $(\atom_{k})$, are 
subsequently used to generate data points, the pattern of admixture inheres in
the vector of weights $(\likeweight_{k})$.  It is thus natural to view 
this vector as the representation of an admixed individual.  Indeed, in some 
problems such a weight vector might itself be the observed data.  In other 
problems, the weights may be used to generate data in some more complex 
way that does not simply involve conditionally i.i.d.\ draws.

This perspective on admixture---focusing on the vector of weights $(\likeweight_{k})$ 
rather than the data associated with an individual---is also natural when 
we consider multiple individuals.  The main issue becomes that of linking 
these vectors among multiple individuals, and this can readily be achieved 
in the Bayesian formalism via a hierarchical model.  In the remainder of
this section we consider examples of such hierarchies in the Bayesian 
nonparametric setting.

Let us first consider the standard approach to admixture in which an
individual is represented by a set of draws from a mixture model.
For each individual we need to draw a set of mixing proportions, and these
mixing proportions need to be coupled among the individuals.  This can
be achieved via a prior known as the \emph{hierarchical Dirichlet
process} ($\hdp$)~\citep{teh:2006:hierarchical_dirichlet}:
\begin{align*}
	\dpdraw_{0} &\sim \tdp(\dpconc, \base) \\
	\dpdraw_{d} &= \sum_{k} \dpweight_{d,k} \delta_{\atom_{k}} \indep \tdp(\dpconc_{d}, \dpdraw_{0}),
		\quad d = 1,2,\ldots,
\end{align*}
where the index $d$ ranges over the individuals.  Note that the global measure 
$G_0$ is a discrete random probability measure, given that it is drawn from 
a Dirichlet process.  In drawing the individual-specific random measure
$\dpdraw_d$ at the second level, we therefore resample from among the atoms 
of $G_0$ and do so according to the weights of these atoms in $G_0$.  
This shares atoms among the individuals and couples the individual-specific
mixing proportions $\dpweight_{d,k}$.  We complete the model specification
as follows:
\begin{align*}
	\topic_{d,n} &\iid (\dpweight_{d,k})_k \quad \textrm{for } n = 1,\ldots,N_{d} \\
	x_{d,n} &\indep F(\atom_{\topic_{d,n}}),
\end{align*}
which draws an index $\topic_{d,n}$ from the discrete distribution $(\dpweight_{d,k})_k$
and then draws a data point $x_{d,n}$ from a distribution indexed by $\topic_{d,n}$.
For instance, $(\dpweight_{d,k})$ might represent topic proportions in document $d$;
$\atom_{\topic_{d,n}}$ might represent a topic, i.e.\ a distribution over words; and
$x_{d,n}$ might represent the $n$th word in the $d$th document.

As before, an alternative view of this process is that we draw an individual-specific
set of counts from an appropriate stochastic process and then generate the appropriate 
number of data points for each individual.  We also need to couple the counts
across individuals.  This can be achieved by constructing hierarchical models
involving the $\nbp$.  One way to proceed is the following conditional independence
hierarchy:
\begin{align} \label{eqn:bnbp}
	\bpdraw_{0} &\sim \bp(\bpconc, \bpmass, \base) \\ \notag
	\likedraw_{d} &= \sum_{k} \likeweight_{d,k} \delta_{\atom_{k}} 
		\indep \nbp(r_{d}, \bpdraw_{0}),
\end{align}
where we first draw a random measure $\bpdraw_{0}$ from the beta process and then draw 
multiple times from an $\nbp$ with base measure given by $\bpdraw_{0}$.  
Although this 
conditional independence hierarchy does couple count vectors across multiple 
individuals, it does not have the flexibility of the $\hdp$, which draws 
individual-specific mixing proportions from an underlying set of population-wide 
mixing proportions and then converts these mixing proportions into counts.  
We can capture this flexibility within an $\nbp$-based framework by simply extending 
the hierarchy by one level:
\begin{align} \label{eqn:hbnbp}
	\bpdraw_{0} &\sim \bp(\bpconc, \bpmass, \base) \\ \notag
	\bpdraw_{d} &\indep \bp(\bpconc_{d}, \bpmass_{d}, \bpdraw_{0} / \bpdraw_{0}(\atomsp)) \\ \notag
	\likedraw_{d} &= \sum_{k} \likeweight_{d,k} \delta_{\atom_{k}} \indep \nbp(r_{d}, \bpdraw_d).
\end{align}
Since $\bpdraw_0$ is almost surely an atomic measure, the atoms of each $\bpdraw_{d}$ 
will coincide with those of $\bpdraw_{0}$ almost surely.  The weights associated with
these atoms can be viewed as individual-specific feature probability vectors.
We refer to this prior as the \emph{hierarchical beta-negative binomial 
process} ($\hbnbp$).

We also note that it is possible to consider additional levels of structure in 
which a population is decomposed into subpopulations and further decomposed into 
subsubpopulations and so on, bottoming out in a set of individuals.  This tree
structure can be captured by repeated draws from a set of beta processes at each 
level of the tree, conditioning on the beta process at the next highest level of
the tree.  Hierarchies of this form have previously been explored for beta-Bernoulli 
processes by~\citet{ThibauxJo07}.

\textbf{Comparison with \citet{zhou:2012:beta}.}
\citet{zhou:2012:beta} have independently proposed a (non-hierarchical) 
beta-negative binomial process prior
\begin{align*}
	\bpdraw_{0} &= \sum_{k} \bpweight_k\delta_{r_{k},\atom_{k}} \sim \bp(\bpconc, \bpmass, R\times\base) \\ \notag
	\likedraw_{d} &= \sum_{k} \likeweight_{d,k} \delta_{\atom_{k}}
	\quad\text{where}\quad
		\likeweight_{d,k} \indep \negbin(r_{k}, \bpweight_k),
\end{align*}
where $R$ is a continuous finite measure over $\R^+$ used to associate a distinct failure parameter $r_k$ with
each beta process atom.
Note that each individual is restricted to use the same failure parameters and the same beta process weights under this model.
In contrast, our $\bnbp$ formulation \eqref{eqn:bnbp} offers the flexibility of differentiating individuals by assigning each its own failure parameter $r_d$.
Our $\hbnbp$ formulation \eqref{eqn:hbnbp} further introduces heterogeneity in the individual-specific beta process weights 
by leveraging the hierarchical beta process.
We will see that these modeling choices are particularly well-suited for admixture modeling in the coming sections.

\citet{zhou:2012:beta} use their prior to develop a Poisson factor analysis model for integer matrix factorization, while
our primary motivation is mixture and admixture modeling.
Our differing models and motivating applications have led to different challenges and algorithms for posterior inference.
While \citet{zhou:2012:beta} develop an inexact inference scheme based on a finite approximation to the beta process,
we develop both an exact Markov chain Monte Carlo sampler and a finite approximation sampler for posterior inference under the 
$\hbnbp$ (see \mysec{posterior}).
Finally, unlike \citet{zhou:2012:beta}, we provide an extensive theoretical analysis of our priors including a proof of the conjugacy of the beta process and the $\nbp$ (given in \mysec{conjugacy}) and an asymptotic analysis
of the $\bnbp$ (see \mysec{asymptotics}).

\section{Asymptotics} \label{sec:asymptotics}

An important component of choosing a Bayesian prior is verifying that its behavior 
aligns with our beliefs about the behavior of the data-generating mechanism. 
In models of clustering, a particular measure of interest is the
\emph{diversity}---the dependence of the number of clusters on the 
number of data points.  In speaking of the diversity, we typically
assume a finite number of fixed atoms in a process derived from a CRM,
so that asymptotic
behavior is dominated by the ordinary component.

It has been observed in a variety of different contexts that the number
of clusters in a data set grows as a {\em power law} of the size of the data;
that is, the number of clusters is asymptotically proportional to the number of 
data points raised to some positive power \citep{gnedin:2007:notes}.
Real-world examples of such behavior are
provided by~\citet{newman:2005:power} and \citet{mitzenmacher:2004:brief}.

The diversity has been characterized for the 
Dirichlet process ($\tdp$) and a two-parameter extension to the Dirichlet 
process known as the \emph{Pitman-Yor process} ($\pyp$)~\citep{pitman:1997:two}, 
with extra parameter $\dpdisc \in (0,1)$ and concentration parameter 
$\dpconc > -\dpdisc$. We will see that while the number of clusters generated
according to a $\tdp$ grows as a logarithm of the size of the data, the number
of clusters generated according to a $\pyp$ grows as a power of the size of the data.
Indeed, the popularity of the Pitman-Yor process---as an alternative
prior to the Dirichlet process in the clustering domain---can be attributed to this power-law
growth~\citep{goldwater:2006:interpolating,
teh:2006:hierarchical_bayesian, wood:2009:stochastic}.
In this section, we derive analogous asymptotic results for 
the $\bnbp$ treated as a clustering model. 

We first highlight a subtle difference between our model and the Dirichlet 
process. For a Dirichlet process, the number of data points $\numdata$ is 
known a priori and fixed. An advantage of our model is that it models the 
number of data points $\numdata$ as a random variable and therefore has 
potentially more predictive power in modeling multiple populations. 
We note that a similar effect can be achieved for the Dirichlet process 
by using the gamma process for feature modeling as described in \mysec{mix_hier} 
rather than normalizing away the mass that determines the number of observations.
However, there is no such unnormalized completely random measure for the 
$\pyp$~\citep{pitman:1997:two}. We thus treat $\numdata$ as fixed for the 
$\tdp$ and $\pyp$, in which case the number of clusters $K(\numdata)$ is a 
function of $\numdata$. On the other hand, the number of data points 
$\numdata(r)$ depends on $r$ in the case of the $\bnbp$, and the number of clusters 
$K(r)$ does as well.  We also define $K_{j}(\numdata)$ to be the number of clusters 
with exactly $j$ elements in the case of the $\tdp$ and $\pyp$, and we define $K_{j}(r)$ 
to be the number of clusters with exactly $j$ elements in the $\bnbp$ case.

For the $\tdp$ and $\pyp$, $K(\numdata)$ and $K_{j}(\numdata)$ are random even though
$\numdata$ is fixed, so it will be useful to also define their expectations:
\begin{equation}
	\label{eq:exp_clusts_dp}
	\Phi(\numdata) \defeq \mbe[K(\numdata)], \quad \Phi_{j}(\numdata) \defeq \mbe[K_{j}(\numdata)].
\end{equation}
In the $\bnbp$ and $\tbnbp$ cases, all of $K(r)$, $K_{j}(r)$, and $\numdata(r)$ are random. So we further define
\begin{equation}
	\label{eq:exp_clusts_bp}
	\Phi(r) \defeq \mbe[K(r)], \quad \Phi_{j}(r) \defeq \mbe[K_{j}(r)], \quad \xi(r) \defeq \mbe[\numdata(r)].
\end{equation}

We summarize the results that we establish in this section in \tab{asymptotics},
where we also include comparisons to existing results for the $\tdp$ and $\pyp$.\footnote{
The reader interested in power laws may also note
that the generalized gamma process
is a completely random measure that, when normalized, provides a probability
measure for clusters that has asymptotic behavior similar to the $\pyp$;
in particular, the expected number of clusters grows
almost surely as a power of the size of the data \citep{lijoi:2007:controlling}.
} 
The full statements of our results, from which the table is derived, can be found 
in \app{asymptotics_state}, and proofs are given in \app{asymptotics_proof}.

The table shows, for example, that for the $\tdp$, 
$\Phi(\numdata) \sim \gpconc \log(\numdata)$ as $\numdata \rightarrow \infty$, 
and, for the $\bnbp$, $\Phi_{j}(r) \sim \bpmass \bpconc j^{-1}$ as $r \rightarrow \infty$ 
(i.e., constant in $r$).  The result for the expected number of clusters for
the $\tdp$ can be found in~\citet{korwar:1973:contributions}; results for
expected number of clusters for both the $\tdp$ and $\pyp$ can be found in 
\citet[][\eqw{3.24} on \pw{69} and \eqw{3.47} on \pw{73}]{pitman:2006:combinatorial}. 
Note that in all cases the expected 
counts of clusters of size $j$ are asymptotic expansions in terms of 
$r$ for fixed $j$ and should not be interpreted as asymptotic expansions in 
terms of $j$.
\ifdefined\figsintext
\begin{table}
\caption{ \label{tab:asymptotics} Let $\numdata$ be the number of data points when this number is fixed and $\xi(r)$ be the expected number of data points when $\numdata$ is random. Let $\Phi(\numdata)$, $\Phi_{j}(\numdata)$, $\Phi(r)$, and $\Phi_{j}(r)$ be the expected number of clusters under various scenarios and defined as in \eqs{exp_clusts_dp} and \eqss{exp_clusts_bp}.
The upper part of the table gives the asymptotic behavior of $\Phi$ up to a multiplicative constant, and the bottom part of the table gives the multiplicative constants. For the $\tdp$, $\dpconc > 0$. For the $\pyp$, $\dpdisc \in (0,1)$ and $\dpconc > -\dpdisc$. For the $\bnbp$, $\dpconc > 1$. For the $\tbnbp$, $\dpdisc \in (0,1)$ and $\dpconc > 1 - \dpdisc$.}
\centering
\fbox{%
\begin{tabular}{ r c c}
Process & Expected number of clusters & Expected number of clusters of size $j$ \\ \hline
	& \multicolumn{2}{c}{Function of $\numdata$ or $\xi(r)$} \\
	\cline{2-3}
$\tdp$ & $\log(\numdata)$
	& $1$ \\ 
$\pyp$ & $\numdata^{\dpdisc}$
	& $\numdata^{\dpdisc}$ \\
$\bnbp$ & $\log(\xi(r))$
	& $1$ \\
$\tbnbp$ & $(\xi(r))^{\bpdisc}$
	& $(\xi(r))^{\bpdisc}$ \\
	\cline{2-3}
	& \multicolumn{2}{c}{Constants} \\
	\cline{2-3}
$\tdp$ & $\gpconc$
	& $\dpconc j^{-1}$ \\ 
$\pyp$ & $\frac{\Gamma(\dpconc+1)}{\dpdisc \Gamma(\dpconc+\dpdisc)}$
	& $\frac{\Gamma(\bpconc+1)}{\Gamma(1-\bpdisc) \Gamma(\bpconc+\bpdisc)} \frac{\Gamma(j-\bpdisc)}{\Gamma(j+1)}$ \\
$\bnbp$ & $\bpmass \bpconc$
	& $\bpmass \bpconc j^{-1}$ \\
$\tbnbp$ & $\frac{\bpmass^{1-\bpdisc}}{\bpdisc} \frac{\Gamma(\bpconc+1)}{\Gamma(\bpconc+\bpdisc)} \left( \frac{\bpconc+\bpdisc-1}{\bpconc} \right)^{\bpdisc}$
	& $\bpmass^{1-\bpdisc} \frac{\Gamma(\bpconc+1)}{\Gamma(1-\bpdisc) \Gamma(\bpconc+\bpdisc)} \frac{\Gamma(j-\bpdisc)}{\Gamma(j+1)} \left( \frac{\bpconc+\bpdisc-1}{\bpconc} \right)^{\bpdisc}$ \\
\end{tabular}}
\end{table}
\fi

We conclude that, just as for the Dirichlet process, the $\bnbp$ can achieve both
logarithmic cluster number growth in the basic model and power law cluster number
growth in the expanded, three-parameter model.

\section{Simulation} \label{sec:simulation}

Our theoretical results in \mysec{asymptotics} are supported by simulation results, 
summarized in \fig{asymptotics}; in particular, our simulation
corroborated the existence of power laws in the three-parameter beta process
case examined in \mysec{asymptotics}.
The simulation was performed as follows. 
For values of the negative binomial parameter $r$ evenly spaced between 
1 and 1,001, we generated beta process weights according to a beta process
(or three-parameter beta process) 
using a stick-breaking representation~\citep{paisley:2010:stick,broderick:2012:beta}. For each of the 
resulting atoms, we simulated negative binomial draws to arrive at a sample 
from a $\bnbp$. For each such $\bnbp$, we can count the resulting total number of 
data points $\numdata$ and total number of clusters $K$. Thus, each $r$ gives 
us an $(r,\numdata,K)$ triple. 

In the simulation, we set the mass parameter $\bpmass = 3$. We set the concentration parameter $\bpconc=3$; in particular, we note that the analysis in \mysec{asymptotics} implies that we should always have $\bpconc > 1$. Finally, we ran the simulation for both the $\bpdisc = 0$ case, where we expect no power law behavior, and the $\bpdisc=0.5$ case, where we do expect power law behavior. The results are shown in \fig{asymptotics}. Is this figure, we scatter plot the $(r,K)$ tuples from the generated $(r,\numdata,K)$ triples on the left and plot the $(\numdata,K)$ tuples on the right.

In the left plot, the upper black points represent the simulation with $\bpdisc=0.5$, and the lower blue data points represent the $\bpdisc=0$ case. The lower red line illustrates the theoretical result corresponding to the $\bpdisc=0$ case (\lem{Phi_r_bnbp} in \app{asymptotics_state}), and we can see that the anticipated logarithmic growth behavior agrees with our simulation. The upper red line illustrates the theoretical result for the $\bpdisc=0.5$ case (\lem{Phi_r_3bnbp} in \app{asymptotics_state}). The agreement between simulation and theory here demonstrates that, in contrast to the $\bpdisc=0$ case, the $\bpdisc=0.5$ case exhibits power law growth in the number of clusters $K$ as a function of the negative binomial parameter $r$. 

Our simulations also bear out that the expectation of the random number of data points $\numdata$ increases linearly with $r$ (\lems{xi_r_bnbp} and \lemss{xi_r_3bnbp} in \app{asymptotics_state}). We see, then, on the right side of \fig{asymptotics} the behavior of the number of clusters $K$ now plotted as a function of $\numdata$. As expected given the asymptotics of the expected value of $\numdata$, the behavior in the right plot largely mirrors the behavior in the left plot. Just as in the left plot, the lower red line (\thm{Phi_xi_bnbp} in \app{asymptotics_state}) shows the anticipated logarithmic growth of $K$ and $\numdata$ when $\bpdisc=0$. And the upper red line (\thm{Phi_xi_3bnbp} in \app{asymptotics_state}) shows the anticipated power law growth of $K$ and $\numdata$ when $\bpdisc=0.5$.

We can see the parallels with the $\tdp$ and $\pyp$ here.
Clusters generated from the Dirichlet process (i.e., Pitman-Yor process with $\dpdisc=0$) exhibit logarithmic growth of the expected number of clusters $K$ as the (deterministic) number of data points $\numdata$ grows. And clusters generated from the Pitman-Yor process with $\dpdisc \in (0,1)$ exhibit power law behavior in the expectation of $K$ as a function of (fixed) $\numdata$. So too do we see that the $\bnbp$, when applied to clustering problems, yields asymptotic growth similar to the $\tdp$ and that the $\tbnbp$ yields asymptotic growth similar to the $\pyp$.

\ifdefined\figsintext
\begin{figure}
	\includegraphics[width=0.47\textwidth]{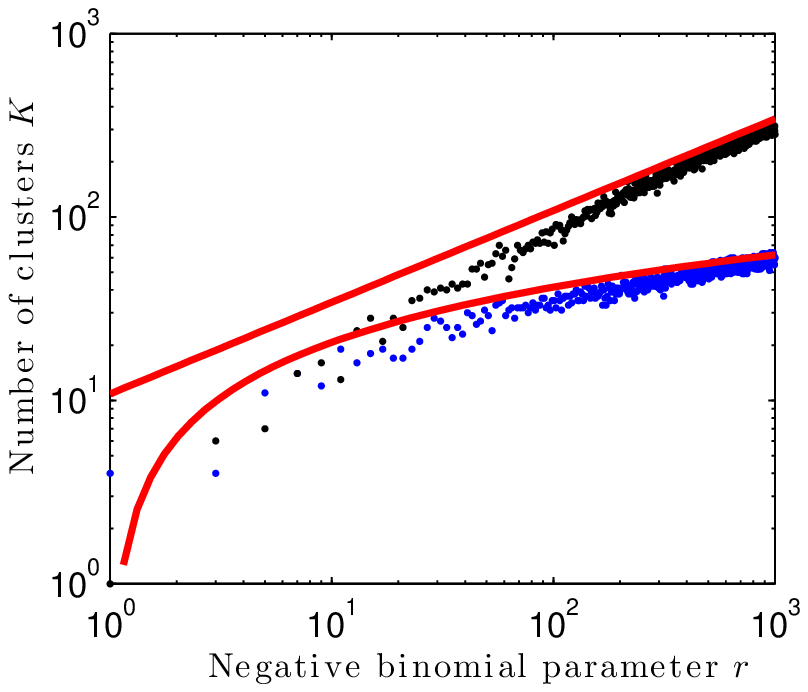}
	\includegraphics[width=0.47\textwidth]{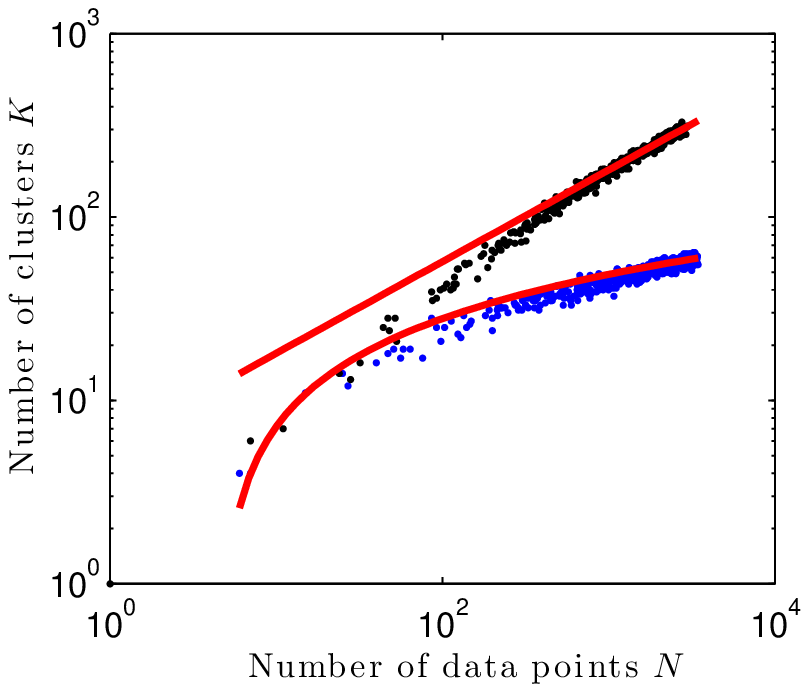}
\caption{\label{fig:asymptotics} For each $r$ evenly spaced between 1 and 1,001, we simulate (random) values of the number of data points $\numdata$ and number of clusters $K$ from the $\bnbp$ and $\tbnbp$. In both plots, we have mass parameter $\bpmass=3$ and concentration parameter $\bpconc=3$. On the {\em left}, we see the number of clusters $K$ as a function of the negative binomial parameter $r$ (see \lem{Phi_r_bnbp} and \lem{Phi_r_3bnbp} in \app{asymptotics_state}); on the {\em right}, we see the number of clusters $K$ as a function of the (random) number of data points $\numdata$ (see \thm{Phi_xi_bnbp} and \thm{Phi_xi_3bnbp} in \app{asymptotics_state}). In both plots, the upper black points show simulation results for the case $\bpdisc=0.5$, and the lower blue points show $\bpdisc=0$. Red lines indicate the theoretical asymptotic mean behavior we expect from \mysec{asymptotics}.}
\end{figure}
\fi

\section{Posterior inference} \label{sec:posterior}
In this section we present posterior inference algorithms for the $\hbnbp$.
We focus on the setting in which, for each individual $d$, there is an associated 
exchangeable sequence of observations $(\obs_{d,n})_{n=1}^{\numdata_{d}}$. 
We seek to infer both the admixture component responsible for each 
observation and the parameter $\atom_k$ associated with each component.
Hereafter, we let $\topic_{d,n}$ denote the unknown component index associated 
with $\obs_{d,n}$, so that $\obs_{d,n} \sim \obsdist(\atom_{\topic_{d,n}})$.

Under the $\hbnbp$ admixture model introduced in \mysec{mix_hier}, the 
posterior over component indices and parameters has the form
$$
	p(\tbf{\topic}_\doubcdot, \tbf{\atom}_\cdot\mid \tbf{\obs}_\doubcdot, \Theta) \propto 
	p(\tbf{\topic}_\doubcdot, \tbf{\atom}_\cdot, \tbf{\bpweight}_{0,\cdot}, \tbf{\bpweight}_\doubcdot \mid \tbf{\obs}_\doubcdot, \Theta),
$$
where $\Theta \defeq (\obsdist, \base, \bpmass_0,\bpconc_0,\tbf{\bpmass}_{\cdot},\tbf{\bpconc}_{\cdot},\tbf{r}_{\cdot})$ is the collection of all fixed hyperparameters.
As is the case with $\hdp$ admixtures~\citep{teh:2006:hierarchical_dirichlet}
and earlier hierarchical beta process featural models~\citep{ThibauxJo07},
the posterior of the $\hbnbp$ admixture cannot be obtained in analytical form
due to complex couplings in the marginal $p(\tbf{\obs}_\doubcdot\mid \Theta)$.
We therefore develop Gibbs sampling algorithms~\citep{GemanGe84} to draw samples 
of the relevant latent variables from their joint posterior.

A challenging aspect of inference in the nonparametric setting is the countable 
infinitude of component parameters and the countably infinite support of the 
component indices.  We develop two sampling algorithms that cope with this
issue in different ways.  In \mysec{infinite-sampler}, we use slice sampling 
to control the number of components that need be considered on a given round 
of sampling and thereby derive an exact Gibbs sampler for posterior inference 
under the $\hbnbp$ admixture model.  In \mysec{finite-sampler}, we describe an 
efficient alternative sampler that makes use of a finite approximation to the 
beta process.  Throughout we assume that the base measure $\base$ is continuous.
We note that neither procedure requires conjugacy between the base distribution 
$\base$ and the data-generating distribution $\obsdist$.

\subsection{Exact Gibbs slice sampler} \label{sec:infinite-sampler}

Slice sampling~\citep{Damien99,Neal03} has been successfully employed in 
several Bayesian nonparametric contexts, including Dirichlet process mixture 
modeling~\citep{Walker07,Papaspiliopoulos08,KalliGrWa11} and beta process
feature modeling~\citep{TehGoGh07}.  The key to its success lies in the 
introduction of one or more auxiliary variables that serve as adaptive 
truncation levels for an infinite sum representation of the stochastic 
process.

This adaptive truncation procedure proceeds as follows.  For each observation 
associated with individual $d$, we introduce an auxiliary variable $u_{d,n}$ 
with conditional distribution
$$
	u_{d,n} \sim \unif(0, \zeta_{d,\topic_{d,n}}),
$$
where $(\zeta_{d,k})_{k=1}^\infty$ is a fixed positive sequence with $\lim_{k\to\infty} \zeta_{d,k} = 0$.
To sample the component indices, we recall that a negative binomial draw 
$\likeweight_{d,k} \sim \negbin(r_{d}, \bpweight_{d,k})$ may be represented 
as a gamma-Poisson mixture:
\begin{align*}
	\lambda_{d,k} &\sim \ga\left(r_d,\frac{1-\bpweight_{d,k}}{\bpweight_{d,k}}\right)\\ 
	\likeweight_{d,k} &\sim \pois(\lambda_{d,k}).
\end{align*}
We first sample $\lambda_{d,k}$ from its full conditional.  By gamma-Poisson conjugacy, this has the simple form
\begin{align*}
	\lambda_{d,k} &\sim \ga\left(r_d + \likeweight_{d,k},1/\bpweight_{d,k}\right).
\end{align*}

We next note that, given $\tbf{\lambda}_{d,\cdot}$ and the total number of 
observations associated with individual $d$, the cluster sizes $\likeweight_{d,k}$ 
may be constructed by sampling each $\topic_{d,n}$ independently from  
$\tbf{\lambda}_{d,\cdot}/\sum_k\lambda_{d,k}$ and setting 
$\likeweight_{d,k} = \sum_n \mathbb{I}(\topic_{d,n} = k)$. 
Hence, conditioned on the number of data points $\numdata_d$,
the component parameters $\atom_k$, the auxiliary variables 
$\lambda_{d,k}$, and the slice-sampling variable $u_{d,n}$, 
we sample the index $\topic_{d,n}$ from a discrete distribution with
$$
	\mbp(\topic_{d,n}  =  k) \propto F(d\obs_{d,n}\mid \atom_k) \frac{\indic(u_{d,n} \leq \zeta_{d,k})}{\zeta_{d,k}} \lambda_{d,k}
$$
so that only the finite set of component indices $\{k : \zeta_{d,k} \geq u_{d,n}\}$ need be considered when sampling $\topic_{d,n}$.

Let $K_d \defeq \max \{k : \exists n \text{ s.t. } \zeta_{d,k} \geq u_{d,n}\}$ and $K \defeq \max_d K_d$.  
Then, on a given round of sampling, we need only explicitly represent $\lambda_{d,k}$ and $\bpweight_{d,k}$ for $k \leq K_d$ and $\atom_k$ and $\bpweight_{0,k}$ for $k \leq K$.
The simple Gibbs conditionals for $\bpweight_{d,k}$ and $\atom_k$ can be found in \app{infinite-sampler-details}.
To sample the shared beta process weights $\bpweight_{0,k}$, we leverage the size-biased construction of the beta process introduced by~\cite{ThibauxJo07}:
\begin{align*}
	\bpdraw_0 = \sum_{\round=0}^{\infty}\sum_{i=1}^{C_{\round}} \bpweight_{0,\round,i} \delta_{\atomloc{\round,i}},
\end{align*}
where 
\begin{equation*}
	C_{\round} \indep \pois\left(\frac{\bpconc_0\bpmass_0}{\bpconc_0+\round}\right),\quad  \bpweight_{0,\round,i} \indep \tb(1,\bpconc_0+\round),\quad  \text{ and }\quad \atomloc{\round,i} \iid \base,
\end{equation*}
and we develop a Gibbs slice sampler for generating samples from its posterior.
The details are deferred to \app{infinite-sampler-details}.

\subsection{Finite approximation Gibbs sampler}
\label{sec:finite-sampler}
\begin{figure}
\centering
\includegraphics[width=.45\textwidth]{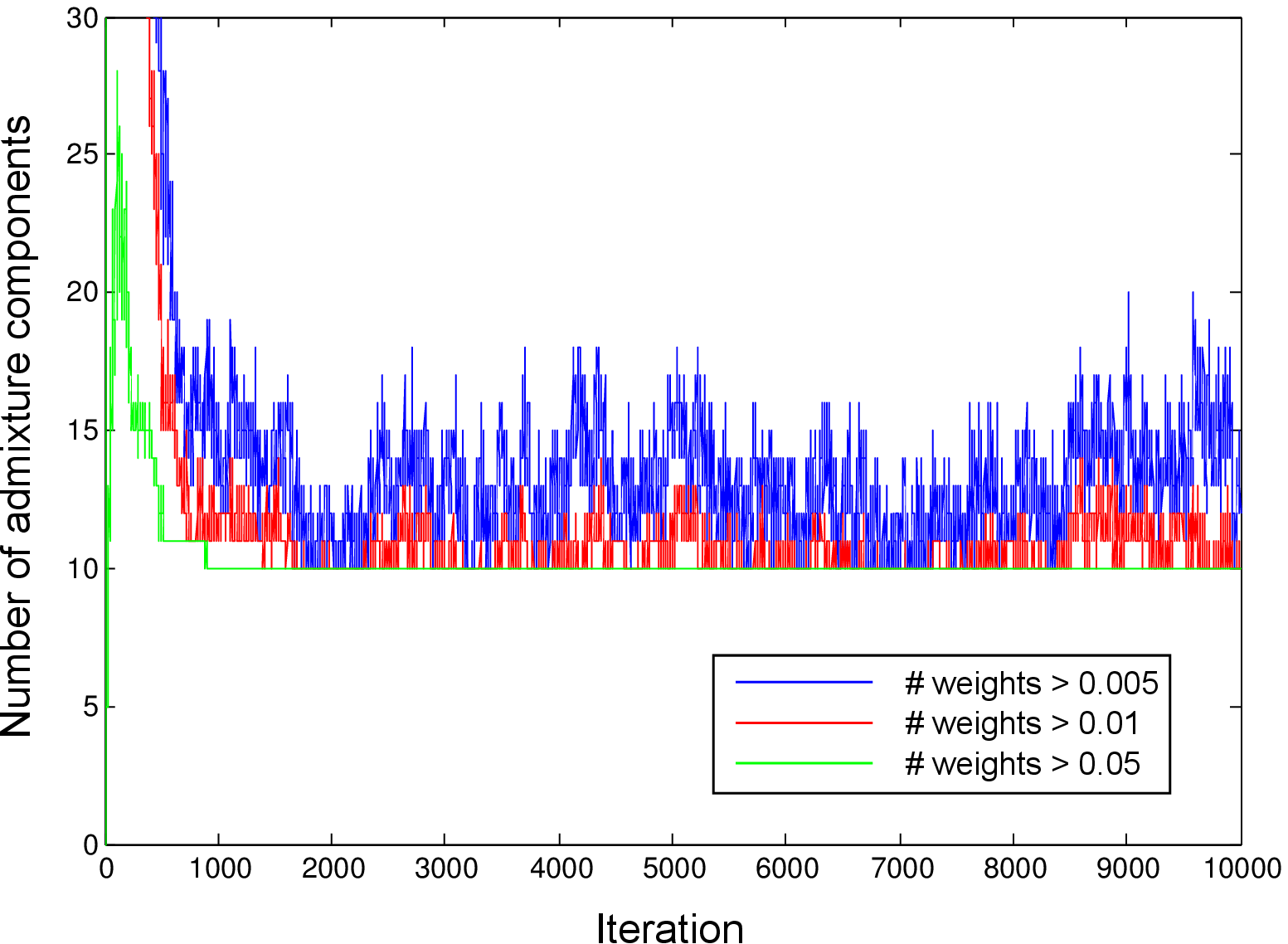}\quad\quad
\includegraphics[width=.45\textwidth]{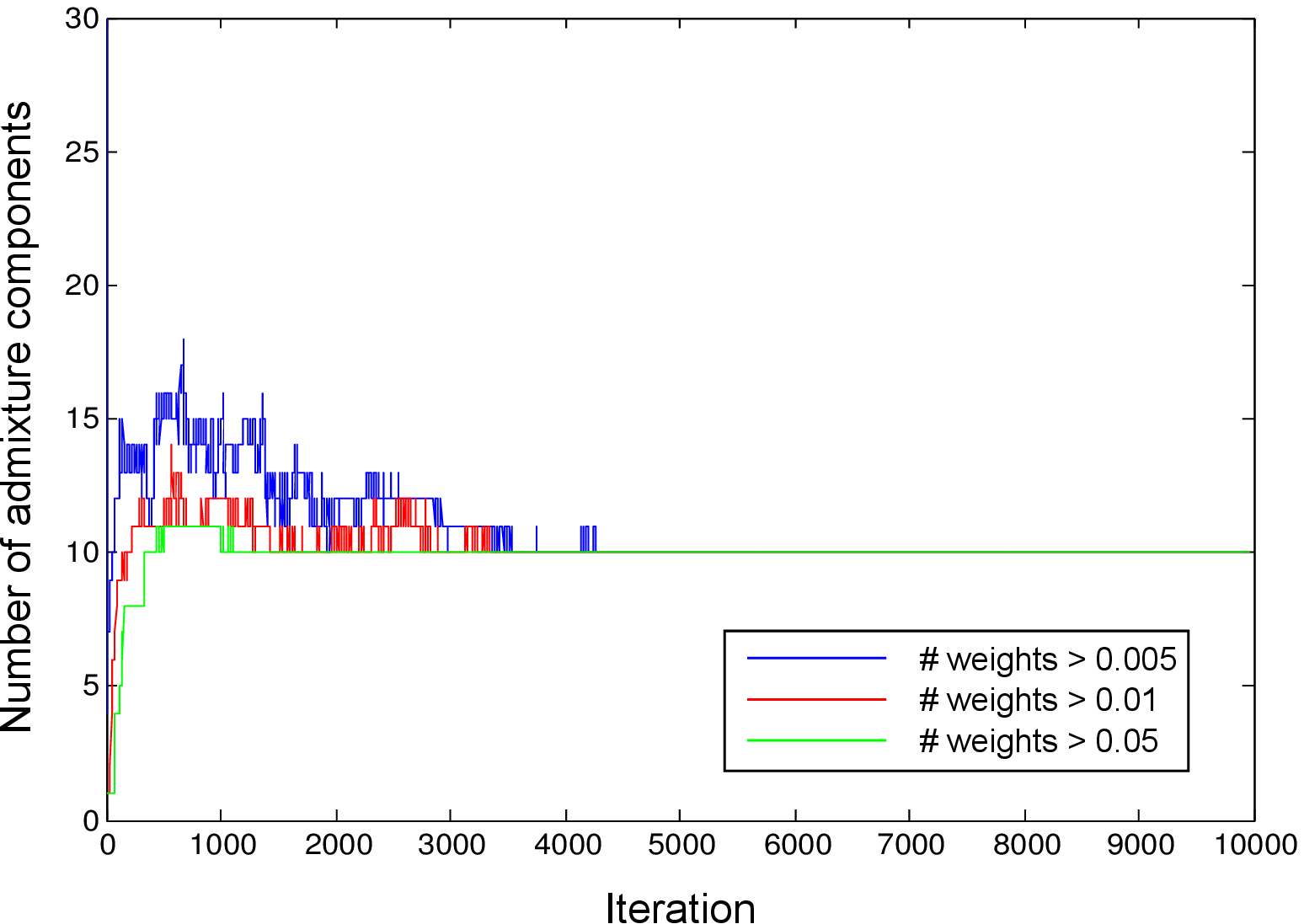}
\caption{\label{fig.toy_results}Number of admixture components used by the finite approximation sampler with $K=100$ (\emph{left}) and the exact Gibbs slice sampler (\emph{right}) on each iteration of $\hbnbp$ admixture model posterior inference. We use a standard ``toy bars'' data set with ten underlying admixture components (cf.\ \cite{GriffithsSt04}). 
We declare a component to be used by a sample if the sampled beta process weight, $b_{0,k}$, exceeds a small threshold. 
Both the exact and the finite approximation sampler find the correct underlying structure, while the finite sampler attempts to innovate more because of the larger number of proposal components available to the data in each iteration.}
\end{figure}

An alternative to the size-biased construction of $\bpdraw_0$ is a finite approximation to the beta process with a fixed number of components, $K$:
\begin{align}
\label{eq:bp-finite-approximation}
\bpweight_{0,k} \iid \tb(\bpconc_0\bpmass_0/K,\bpconc_0(1-\bpmass_0/K)),\quad \atom_k\iid H,\quad k\in\{1,\dots,K\}.
\end{align}
It is known that, when $H$ is continuous, the distribution of $\sum_{k=1}^K\bpweight_{0,k}\delta_{\atom_k}$ converges to $\bp(\bpconc_{0}, \bpmass_{0}, \base)$ as the number of components $K\to\infty$ (see the proof of \thmw{3.1} by \citet{hjort:1990:nonparametric} with the choice $A_{0}(t) = \bpmass$).
Hence, we may leverage the beta process approximation \eqref{eq:bp-finite-approximation} to develop an approximate posterior sampler for the $\hbnbp$ admixture model with an approximation level $K$ that trades off between computational efficiency and fidelity to the true posterior.
We defer the detailed conditionals of the resulting Gibbs sampler to \app{finite-sampler-details} and briefly compare the behavior of the finite and exact samplers on a toy data set in Figure \ref{fig.toy_results}. 
We note finally that the beta process approximation in \eq{bp-finite-approximation} also gives rise to a new finite admixture model that may be of interest in its own right; we explore the utility of this $\hbnbp$ approximation in \mysec{exp_image}.

\section{Document Topic Modeling} \label{sec:exp_doc}
In the next two sections, we will show how the $\hbnbp$ admixture model and its finite approximation can be used as practical building blocks for more complex supervised and unsupervised inferential tasks.

We first consider the unsupervised task of \emph{document topic modeling}, in which each individual $d$ is a 
document containing $\clustsize_d$ observations (words) and each word 
$\obs_{d,n}$ belongs to a vocabulary of size $V$.  The topic modeling
framework is an instance of admixture modeling in which we assume that 
each word of each document is generated from a latent admixture component 
or \emph{topic}, and our goal is to infer the topic underlying each word.

In our experiments, we let $\base_{ord}$, the $\atomsp$ dimension of the 
ordinary component intensity measure, be a Dirichlet distribution with 
parameter $\eta \onevec$ for $\eta = 0.1$ and $\onevec$ a $V$-dimensional 
vector of ones and let $\obsdist(\atom_k)$ be $\mult(1,\atom_k)$.
We use the setting $(\bpmass_0,\bpconc_0,\bpmass_{d},\bpconc_{d}) = (3,3,1,10)$ 
for the global and document-specific mass and concentration 
parameters and set the document-specific negative binomial shape 
parameter according to the heuristic 
$r_d = \clustsize_d (\bpconc_0-1)/(\bpconc_0\bpmass_0)$.
We arrive at this heuristic by matching $\clustsize_d$ to its expectation under a non-hierarchical $\bnbp$ model and solving for $r_d$:
$$
	\mbe[\clustsize_d] = r_d \mbe\left[ \sum\nolimits_{k=1}^{\infty} b_{d,k} / (1 - b_{d,k})\right] = \bpmass_0 \bpconc_0 / (\bpconc_0 - 1).
$$
When applying the exact Gibbs slice sampler, we let the slice sampling 
decay sequence follow the same pattern across all documents: 
$\zeta_{d,k} = 1.5^{-k}$.  

\subsection{Worldwide Incidents Tracking System}
\label{sec:wits}
We report results on the Worldwide Incidents Tracking System (WITS) data set.\footnote{\url{https://wits.nctc.gov}} This data set consists of reports on 79{,}754 terrorist attacks from the years 2004 through 2010. Each event contains a written summary of the incident, location information, victim statistics, and various binary fields such as ``assassination,'' ``IED,'' and ``suicide.'' We transformed each incident into a text document by concatenating the summary and location fields and then adding further words to account for other, categorical fields: e.g., an incident with seven hostages would have the word ``hostage'' added to the document seven times.  We used a vocabulary size of $V = 1{,}048$ words.

\textbf{Perpetrator Identification.}
Our experiment assesses the ability of the $\hbnbp$ admixture model to discriminate among incidents perpetrated by different organizations.
We first grouped documents according to the organization claiming responsibility for the reported incident.
We considered 5{,}390 claimed documents in total distributed across the ten organizations listed in \tab{perp-list}.
We removed all organization identifiers from all documents and randomly set aside $10\%$ of the documents in each group as test data.
Next, for each group, we trained an independent, organization-specific $\hbnbp$ model on the remaining documents in that group by drawing 10{,}000 MCMC samples.
We proceeded to classify each test document by measuring the likelihood of the document under each trained $\hbnbp$ model
and assigning the label associated with the largest likelihood.
The resulting confusion matrix across the ten candidate organizations is displayed in \tab{hbnbp-confusion}.
Results are reported for the exact Gibbs slice sampler; performance under the finite approximation sampler is nearly identical.

\ifdefined\figsintext
\begin{table}
\caption{\label{tab:perp-list} The number of incidents claimed by each organization in the WITS perpetrator identification experiment.}
\centering
\fbox{%
\begin{tabular}{c|l|c}
	Group ID & Perpetrator & $\#$ Claimed Incidents \\
	\hline
    1 & taliban & 2647 \\ 
    2 & al-aqsa &          417 \\
    3 & farc &           76 \\
    4 & izz al-din al-qassam &          478 \\
    5 & hizballah &           89 \\
    6 & al-shabaab al-islamiya &          426 \\
    7 & al-quds &          505 \\
    8 & abu ali mustafa & 249 \\
    9 & al-nasser salah al-din & 212 \\
    10 & communist party of nepal (maoist) & 291
\end{tabular}}
\end{table}
\fi

For comparison, we carried out the same experiment using the more standard $\hdp$ admixture model in place of the $\hbnbp$.
For posterior inference, we used the $\hdp$ block sampler code of Yee Whye Teh\footnote{\url{http://www.gatsby.ucl.ac.uk/~ywteh/research/npbayes/npbayes-r1.tgz}} and initialized the sampler with 100 topics and topic hyperparameter $\eta = 0.1$ (all remaining parameters were set to their default values).  
For each organization, we drew 250{,}000 MCMC samples and kept every twenty-fifth sample for evaluation.
The confusion matrix obtained through $\hdp$ modeling is displayed in \tab{hdp-confusion}.  
We see that, overall, $\hbnbp$ modeling leads to more accurate identification of perpetrators than its $\hdp$ counterpart.
Most notably, the $\hdp$ wrongly attributes more than half of all documents from group 1 (taliban) to group 3 (farc) or group 6 (al-shabaab al-islamiya). 
We hypothesize that the $\hbnbp$'s superior discriminative power stems from its ability to distinguish between documents both on the basis of word frequency and on the basis of document length. 

\ifdefined\figsintext
\begin{table}
\caption{Confusion matrices for WITS perpetrator identification.  See \tab{perp-list} for the organization names matching each group ID.}
\centering
\subfloat[$\hbnbp$ Confusion Matrix]
{\label{tab:hbnbp-confusion} \footnotesize%
\begin{tabular}{c|c|c|c|c|c|c|c|c|c|c|c|}
\multicolumn{12}{c}{Predicted Groups} \\ \cline{3-12}
\multicolumn{1}{c}{} & \multicolumn{1}{c}{} & \multicolumn{1}{|c}{1} & \multicolumn{1}{c}{2} & \multicolumn{1}{c}{3} & \multicolumn{1}{c}{4} & \multicolumn{1}{c}{5} & \multicolumn{1}{c}{6} & \multicolumn{1}{c}{7} & \multicolumn{1}{c}{8} & \multicolumn{1}{c}{9} & \multicolumn{1}{c|}{10} \\ \cline{2-12}
\multirow{10}{*}{\begin{sideways}Actual Groups\end{sideways}} &
1 &  \cellcolor[gray]{0.00} \textcolor{white}{1.00} & \cellcolor[gray]{1.00} 0.00 & \cellcolor[gray]{1.00} 0.00 & \cellcolor[gray]{1.00} 0.00 & \cellcolor[gray]{1.00} 0.00 & \cellcolor[gray]{1.00} 0.00 & \cellcolor[gray]{1.00} 0.00 & \cellcolor[gray]{1.00} 0.00 & \cellcolor[gray]{1.00} 0.00 & \cellcolor[gray]{1.00}  0.00 \\ \cline{3-12} &
2 &  \cellcolor[gray]{1.00} 0.00 & \cellcolor[gray]{0.62} 0.38 & \cellcolor[gray]{1.00} 0.00 & \cellcolor[gray]{0.98} 0.02 & \cellcolor[gray]{1.00} 0.00 & \cellcolor[gray]{1.00} 0.00 & \cellcolor[gray]{0.71} 0.29 & \cellcolor[gray]{0.71} 0.29 & \cellcolor[gray]{0.98} 0.02 & \cellcolor[gray]{1.00}  0.00 \\\cline{3-12} &
3 &  \cellcolor[gray]{1.00} 0.00 & \cellcolor[gray]{1.00} 0.00 & \cellcolor[gray]{0.00} \textcolor{white}{1.00} & \cellcolor[gray]{1.00} 0.00 & \cellcolor[gray]{1.00} 0.00 & \cellcolor[gray]{1.00} 0.00 & \cellcolor[gray]{1.00} 0.00 & \cellcolor[gray]{1.00} 0.00 & \cellcolor[gray]{1.00} 0.00 & \cellcolor[gray]{1.00}  0.00 \\\cline{3-12} &
4 &  \cellcolor[gray]{1.00} 0.00 & \cellcolor[gray]{1.00} 0.00 & \cellcolor[gray]{1.00} 0.00 & \cellcolor[gray]{0.46} \textcolor{white}{0.54} & \cellcolor[gray]{1.00} 0.00 & \cellcolor[gray]{1.00} 0.00 & \cellcolor[gray]{0.85} 0.15 & \cellcolor[gray]{0.73} 0.27 & \cellcolor[gray]{0.96} 0.04 & \cellcolor[gray]{1.00}  0.00 \\\cline{3-12} &
5 &  \cellcolor[gray]{0.89} 0.11 & \cellcolor[gray]{0.67} 0.33 & \cellcolor[gray]{1.00} 0.00 & \cellcolor[gray]{0.89} 0.11 & \cellcolor[gray]{0.56} 0.44 & \cellcolor[gray]{1.00} 0.00 & \cellcolor[gray]{1.00} 0.00 & \cellcolor[gray]{1.00} 0.00 & \cellcolor[gray]{1.00} 0.00 & \cellcolor[gray]{1.00}  0.00 \\\cline{3-12} &
6 &  \cellcolor[gray]{0.98} 0.02 & \cellcolor[gray]{1.00} 0.00 & \cellcolor[gray]{1.00} 0.00 & \cellcolor[gray]{1.00} 0.00 & \cellcolor[gray]{1.00} 0.00 & \cellcolor[gray]{0.02} \textcolor{white}{0.98} & \cellcolor[gray]{1.00} 0.00 & \cellcolor[gray]{1.00} 0.00 & \cellcolor[gray]{1.00} 0.00 & \cellcolor[gray]{1.00}  0.00 \\\cline{3-12} &
7 &  \cellcolor[gray]{1.00} 0.00 & \cellcolor[gray]{0.90} 0.10 & \cellcolor[gray]{1.00} 0.00 & \cellcolor[gray]{0.94} 0.06 & \cellcolor[gray]{0.98} 0.02 & \cellcolor[gray]{1.00} 0.00 & \cellcolor[gray]{0.52} 0.48 & \cellcolor[gray]{0.70} 0.30 & \cellcolor[gray]{0.96} 0.04 & \cellcolor[gray]{1.00}  0.00 \\\cline{3-12} &
8 &  \cellcolor[gray]{1.00} 0.00 & \cellcolor[gray]{0.96} 0.04 & \cellcolor[gray]{1.00} 0.00 & \cellcolor[gray]{1.00} 0.00 & \cellcolor[gray]{1.00} 0.00 & \cellcolor[gray]{1.00} 0.00 & \cellcolor[gray]{0.84} 0.16 & \cellcolor[gray]{0.24} \textcolor{white}{0.76} & \cellcolor[gray]{0.96} 0.04 & \cellcolor[gray]{1.00}  0.00 \\\cline{3-12} &
9 &  \cellcolor[gray]{1.00} 0.00 & \cellcolor[gray]{0.90} 0.10 & \cellcolor[gray]{1.00} 0.00 & \cellcolor[gray]{0.95} 0.05 & \cellcolor[gray]{0.90} 0.10 & \cellcolor[gray]{1.00} 0.00 & \cellcolor[gray]{0.71} 0.29 & \cellcolor[gray]{0.57} 0.43 & \cellcolor[gray]{0.95} 0.05 & \cellcolor[gray]{1.00}  0.00 \\\cline{3-12} &
10 &  \cellcolor[gray]{1.00} 0.00 & \cellcolor[gray]{1.00} 0.00 & \cellcolor[gray]{1.00} 0.00 & \cellcolor[gray]{1.00} 0.00 & \cellcolor[gray]{1.00} 0.00 & \cellcolor[gray]{1.00} 0.00 & \cellcolor[gray]{1.00} 0.00 & \cellcolor[gray]{1.00} 0.00 & \cellcolor[gray]{1.00} 0.00 & \cellcolor[gray]{0.00}  \textcolor{white}{1.00} \\\cline{2-12}
\end{tabular}
}
\\
\subfloat[$\hdp$ Confusion Matrix]
{\label{tab:hdp-confusion}\footnotesize%
\begin{tabular}{c|c|c|c|c|c|c|c|c|c|c|c|}
\multicolumn{12}{c}{Predicted Groups} \\ \cline{3-12}
\multicolumn{1}{c}{} & \multicolumn{1}{c}{} & \multicolumn{1}{|c}{1} & \multicolumn{1}{c}{2} & \multicolumn{1}{c}{3} & \multicolumn{1}{c}{4} & \multicolumn{1}{c}{5} & \multicolumn{1}{c}{6} & \multicolumn{1}{c}{7} & \multicolumn{1}{c}{8} & \multicolumn{1}{c}{9} & \multicolumn{1}{c|}{10} \\ \cline{2-12}
\multirow{10}{*}{\begin{sideways}Actual Groups\end{sideways}} &
1 &  \cellcolor[gray]{0.54} 0.46 & \cellcolor[gray]{1.00} 0.00 & \cellcolor[gray]{0.74} 0.26 & \cellcolor[gray]{1.00} 0.00 & \cellcolor[gray]{0.97} 0.03 & \cellcolor[gray]{0.77} 0.23 & \cellcolor[gray]{1.00} 0.00 & \cellcolor[gray]{1.00} 0.00 & \cellcolor[gray]{1.00} 0.00 & \cellcolor[gray]{0.99}  0.01 \\\cline{3-12} &
2 &  \cellcolor[gray]{1.00} 0.00 & \cellcolor[gray]{0.69} 0.31 & \cellcolor[gray]{0.98} 0.02 & \cellcolor[gray]{0.98} 0.02 & \cellcolor[gray]{1.00} 0.00 & \cellcolor[gray]{1.00} 0.00 & \cellcolor[gray]{0.71} 0.29 & \cellcolor[gray]{0.64} 0.36 & \cellcolor[gray]{1.00} 0.00 & \cellcolor[gray]{1.00}  0.00 \\\cline{3-12} &
3 &  \cellcolor[gray]{1.00} 0.00 & \cellcolor[gray]{1.00} 0.00 & \cellcolor[gray]{0.00} \textcolor{white}{1.00} & \cellcolor[gray]{1.00} 0.00 & \cellcolor[gray]{1.00} 0.00 & \cellcolor[gray]{1.00} 0.00 & \cellcolor[gray]{1.00} 0.00 & \cellcolor[gray]{1.00} 0.00 & \cellcolor[gray]{1.00} 0.00 & \cellcolor[gray]{1.00}  0.00 \\\cline{3-12} &
4 &  \cellcolor[gray]{1.00} 0.00 & \cellcolor[gray]{1.00} 0.00 & \cellcolor[gray]{1.00} 0.00 & \cellcolor[gray]{0.48} \textcolor{white}{0.52} & \cellcolor[gray]{0.96} 0.04 & \cellcolor[gray]{1.00} 0.00 & \cellcolor[gray]{0.94} 0.06 & \cellcolor[gray]{0.69} 0.31 & \cellcolor[gray]{0.94} 0.06 & \cellcolor[gray]{1.00}  0.00 \\\cline{3-12} &
5 &  \cellcolor[gray]{0.89} 0.11 & \cellcolor[gray]{1.00} 0.00 & \cellcolor[gray]{1.00} 0.00 & \cellcolor[gray]{1.00} 0.00 & \cellcolor[gray]{0.56} 0.44 & \cellcolor[gray]{1.00} 0.00 & \cellcolor[gray]{0.89} 0.11 & \cellcolor[gray]{0.89} 0.11 & \cellcolor[gray]{0.89} 0.11 & \cellcolor[gray]{0.89}  0.11 \\\cline{3-12} &
6 &  \cellcolor[gray]{1.00} 0.00 & \cellcolor[gray]{1.00} 0.00 & \cellcolor[gray]{1.00} 0.00 & \cellcolor[gray]{1.00} 0.00 & \cellcolor[gray]{1.00} 0.00 & \cellcolor[gray]{0.00} \textcolor{white}{1.00} & \cellcolor[gray]{1.00} 0.00 & \cellcolor[gray]{1.00} 0.00 & \cellcolor[gray]{1.00} 0.00 & \cellcolor[gray]{1.00}  0.00 \\\cline{3-12} &
7 &  \cellcolor[gray]{1.00} 0.00 & \cellcolor[gray]{0.90} 0.10 & \cellcolor[gray]{1.00} 0.00 & \cellcolor[gray]{0.96} 0.04 & \cellcolor[gray]{1.00} 0.00 & \cellcolor[gray]{1.00} 0.00 & \cellcolor[gray]{0.62} 0.38 & \cellcolor[gray]{0.58} 0.42 & \cellcolor[gray]{0.94} 0.06 & \cellcolor[gray]{1.00}  0.00 \\\cline{3-12} &
8 &  \cellcolor[gray]{1.00} 0.00 & \cellcolor[gray]{0.96} 0.04 & \cellcolor[gray]{1.00} 0.00 & \cellcolor[gray]{1.00} 0.00 & \cellcolor[gray]{1.00} 0.00 & \cellcolor[gray]{1.00} 0.00 & \cellcolor[gray]{0.92} 0.08 & \cellcolor[gray]{0.16} \textcolor{white}{0.84} & \cellcolor[gray]{0.96} 0.04 & \cellcolor[gray]{1.00}  0.00 \\\cline{3-12} &
9 &  \cellcolor[gray]{1.00} 0.00 & \cellcolor[gray]{0.95} 0.05 & \cellcolor[gray]{1.00} 0.00 & \cellcolor[gray]{0.90} 0.10 & \cellcolor[gray]{1.00} 0.00 & \cellcolor[gray]{1.00} 0.00 & \cellcolor[gray]{0.76} 0.24 & \cellcolor[gray]{0.38} \textcolor{white}{0.62} & \cellcolor[gray]{1.00} 0.00 & \cellcolor[gray]{1.00}  0.00 \\\cline{3-12} &
10 &  \cellcolor[gray]{1.00} 0.00 & \cellcolor[gray]{1.00} 0.00 & \cellcolor[gray]{1.00} 0.00 & \cellcolor[gray]{1.00} 0.00 & \cellcolor[gray]{1.00} 0.00 & \cellcolor[gray]{1.00} 0.00 & \cellcolor[gray]{1.00} 0.00 & \cellcolor[gray]{1.00} 0.00 & \cellcolor[gray]{1.00} 0.00 & \cellcolor[gray]{0.00}  \textcolor{white}{1.00} \\\cline{2-12}
\end{tabular}
}
\end{table}
\fi

We would expect the $\hbnbp$ to have greatest difficulty discriminating among perpetrators when both word usage frequencies and document length distributions are similar across groups.
To evaluate the extent to which this occurs in our perpetrator identification experiment, for each organization, we plotted the density histogram of document lengths in \fig{wits2-doc-len-hist} and the heat map displaying word usage frequency across all associated documents in \fig{wits2-word-freq}.
We  find that the word frequency patterns are nearly identical across groups 2, 7, 8, and 9 (al-aqsa, al-quds, abu ali mustafa, and al-nasser salah al-din, respectively) and that the document length distributions of these four groups are all well aligned. As expected, the majority of classification errors made by our $\hbnbp$ models result from misattribution among these same four groups.
The same group similarity structure is evidenced in a display of the ten most probable words from the most probable $\hbnbp$ topic for each group, \tab{topic1}.
There, we also find an intuitive summary of the salient regional and methodological vocabulary associated with each organization.

\ifdefined\figsintext
\begin{table}
\normalsize
\caption{The ten most probable words from the most probable topic in the final MCMC sample of each group in the WITS perpetrator identification experiment. 
The topic probability is given in parentheses. See \tab{perp-list} for the organization names matching each group ID.}\label{tab.topic1}
\label{tab:topic1}
\centering
\scriptsize
 \begin{tabular}{l l}\hline
\multicolumn{2}{c}{$\hbnbp$: Top topic per organization}\\\hline 
\vspace{-2mm}\\
group 1 (0.29) & afghanistan, assailants, claimed, responsibility, armedattack, fired, police, victims, armed, upon\\\vspace{-3mm}\\
group 2 (0.77) & israel, assailants, armedattack, responsibility, fired, claimed, district, causing, southern, damage\\\vspace{-3mm}\\
group 3 (0.95) & colombia, victims, facility, wounded, armed, claimed, forces, revolutionary, responsibility, assailants\\\vspace{-3mm}\\
group 4 (0.87) & israel, fired, responsibility, claimed, armedattack, causing, injuries, district, southern, assailants\\\vspace{-3mm}\\
group 5 (0.95) & victims, wounded, facility, israel, responsibility, claimed, armedattack, fired, rockets, katyusha\\\vspace{-3mm}\\
group 6 (0.54) & wounded, victims, somalia, civilians, wounding, facility, killing, mortars, armedattack, several\\\vspace{-3mm}\\
group 7 (0.83) & israel, district, southern, responsibility, claimed, fired, armedattack, assailants, causing, injuries\\\vspace{-3mm}\\
group 8 (0.94) & israel, district, southern, armedattack, claimed, fired, responsibility, assailants, causing, injuries\\\vspace{-3mm}\\
group 9 (0.88) & israel, district, southern, fired, responsibility, claimed, armedattack, assailants, causing, injuries\\\vspace{-3mm}\\
group 10 (0.80) & nepal, victims, hostage, assailants, party, communist, claimed, front, maoist/united, responsibility\\\vspace{-3mm}\\\hline
 \end{tabular}
\end{table}
\fi


\ifdefined\figsintext
\begin{figure}
	\centering
  	\subfloat[Density histograms of document lengths.]{\label{fig:wits2-doc-len-hist} \includegraphics[width=1\textwidth]{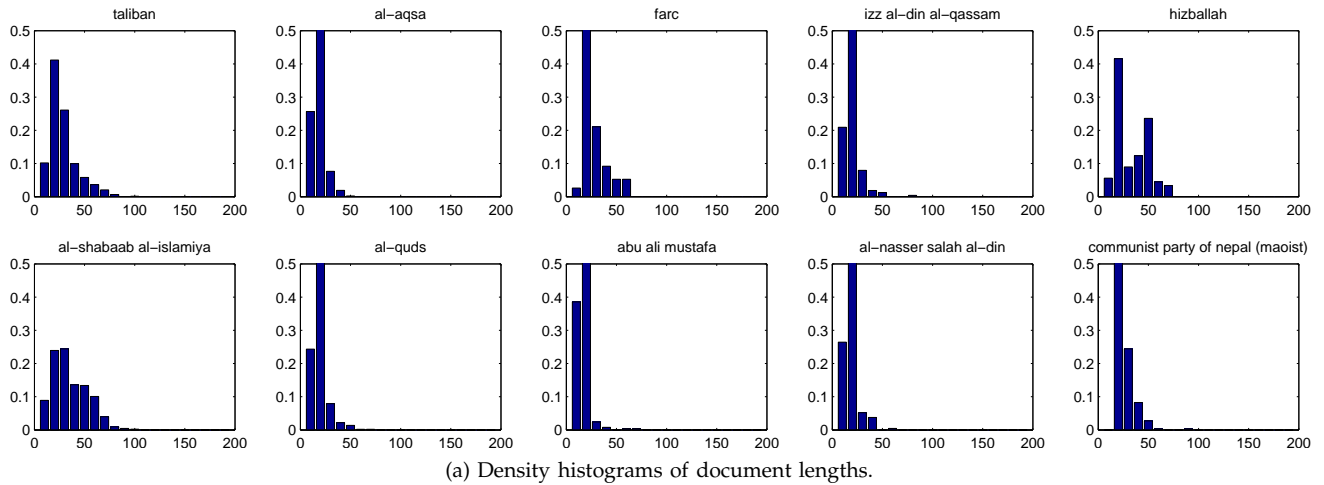}}\\
  	\subfloat[Heat map of word frequencies for the 200 most common words across all documents (best viewed in color).]{\label{fig:wits2-word-freq} \includegraphics[width=1\textwidth]{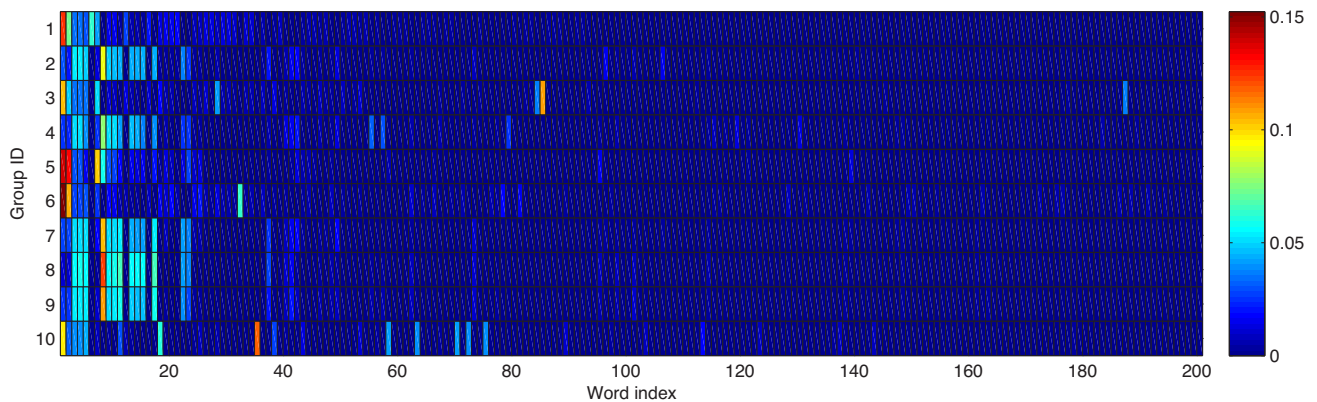}}
	\caption{Document length distributions and word frequencies for each organization in the WITS perpertrator identification experiment.}
	\label{fig:wits2}
\end{figure}
\fi

\section{Image Segmentation and Object Recognition} \label{sec:exp_image}
Two problems of enduring interest in the computer vision community are \emph{image segmentation}, dividing an image into its distinct, semantically meaningful regions, and \emph{object recognition}, labeling the regions of images according to their semantic object classes.  Solutions to these problems are at the core of applications such as content-based image retrieval, video surveying, and object tracking.  
Here we will take an admixture modeling approach to jointly recognizing and localizing objects within images \citep{CaoFe07,RussellFrEfSiZi06, Sivic05,VerbeekTr07}.
Each individual $d$ is an image comprised of $\clustsize_d$ image patches (observations), and each patch $\tbf{\obs}_{d,n}$ is assumed to be generated by an unknown object class (a latent component of the admixture).
Given a series of training images with image patches labeled, the problem of recognizing and localizing objects in a new image reduces to inferring the latent class associated with each new image patch.
Since the number of object classes is typically known \emph{a priori}, we will tackle this inferential task with the finite approximation to the $\hbnbp$ admixture model given in \mysec{finite-sampler} and  compare its performance with that of a more standard model of admixture, Latent Dirichlet Allocation (LDA)~\citep{blei:2003:latent}. 

\subsection{Representing an Image Patch}
We will represent each image patch as a vector of visual descriptors drawn from multiple modalities.  \citet{VerbeekTr07} suggest three complementary modalities: texture, hue, and location.  Here, we introduce a fourth: opponent angle.
To describe hue, we use the robust hue descriptor of \cite{VandeweijerSc06}, which grants invariance to illuminant variations, lighting geometry, and specularities.   For texture description we use ``dense SIFT'' features \citep{Lowe04,DalalTr05}, histograms of oriented gradients computed not at local keypoints but rather at a single scale over each patch.  To describe coarse location, we cover each image with a regular $c$ x $c$ grid of cells (for a total of $V^{\mathrm{loc}}=c^2$ cells) and assign each patch the index of the covering cell.  The opponent angle descriptor of \cite{VandeweijerSc06} captures a second characterization of image patch color.  These features are invariant to specularities, illuminant variations, and diffuse lighting conditions.

To build a discrete visual vocabulary from these raw descriptors, we vector quantize the dense SIFT, hue, and opponent angle descriptors using k-means, producing $V^{\mathrm{sift}}$, $V^{\mathrm{hue}}$, and $V^{\mathrm{opp}}$ clusters respectively.  Finally, we form the observation associated with a patch by concatenating the four modality components into a single vector, $\tbf{\obs}_{d,n} = (\obs_{d,n}^{\mathrm{sift}},\obs_{d,n}^{\mathrm{hue}},\obs_{d,n}^{\mathrm{loc}},\obs_{d,n}^{\mathrm{opp}})$.  As in \cite{VerbeekTr07}, we assume that the descriptors from disparate modalities are conditionally independent given the latent object class of the patch.  
Hence, we define our data generating distribution and our base distribution over parameters $\tbf{\psi}_k = (\psi_k^{\mathrm{sift}},\psi_k^{\mathrm{hue}},\psi_k^{\mathrm{loc}},\psi_k^{\mathrm{opp}})$ via
\begin{align*}
	\atom_{k}^m &\indep \dir(\eta \onevec_{V^m})\quad &\textrm{for } m \in \{\mathrm{sift}, \mathrm{hue}, \mathrm{loc}, \mathrm{opp}\} \\
	\obs_{d,n}^m \mid \topic_{d,n}, \tbf{\atom}_{\cdot} &\indep \mult(1,\atom_{\topic_{d,n}}^m)\quad &\textrm{for } m \in \{\mathrm{sift}, \mathrm{hue}, \mathrm{loc}, \mathrm{opp}\} 
\end{align*}
for a hyperparameter $\eta \in \R$ and $\onevec_{V^m}$ a $V^m$-dimensional vector of ones.

\subsection{Experimental Setup} \label{sec:setup-seg}
We use the Microsoft Research Cambridge pixel-wise labeled image database v1 in our experiments.\footnote{http://research.microsoft.com/vision/cambridge/recognition/}  The data set consists of 240 images, each of size 213 x 320 pixels.  Each image has an associated pixel-wise ground truth labeling, with each pixel labeled as belonging to one of 13 semantic classes or to the \emph{void} class.  Pixels have a ground truth label of \emph{void} when they do not belong to any semantic class or when they lie on the boundaries between classes in an image.  The dataset provider notes that there are insufficiently many instances of \emph{horse}, \emph{mountain}, \emph{sheep}, or \emph{water} to learn these classes, so, as in \cite{VerbeekTr07}, we treat these ground truth labels as \emph{void} as well.  Thus, our general task is to learn and segment the remaining nine semantic object classes.
	
From each image, we extract 20 x 20 pixel patches spaced at 10 pixel intervals across the image.  We choose the visual vocabulary sizes $(V^{\mathrm{sift}},V^{\mathrm{hue}}, V^{\mathrm{loc}},V^{\mathrm{opp}}) = (1000,100,100,100)$ and fix the hyperparameter $\eta = 0.1$.
As in \cite{VerbeekTr07}, we assign each patch a ground truth label $\topic_{d,n}$ representing the most frequent pixel label within the patch.
When performing posterior inference, we divide the dataset into training and test images.  We allow the inference algorithm to observe the labels of the training image patches, and we evaluate the algorithm's ability to correctly infer the label associated with each test image patch.  
	
Since the number of object classes is known \emph{a priori}, we employ the HBNBP finite approximation Gibbs sampler of \mysec{finite-sampler} to conduct posterior inference.
We again use the hyperparameters $(\bpmass_0,\bpconc_0,\bpmass_{d},\bpconc_{d}) = (3,3,1,10)$ for all documents $d$ and set $r_d$ according to the heuristic $r_d = \clustsize_d (\bpconc_0-1)/(\bpconc_0\bpmass_0)$. 
We draw 10,000 samples and, for each test patch, predict the label with the highest posterior probability across the samples.
We compare $\hbnbp$ performance with that of LDA using the standard variational inference algorithm of~\cite{blei:2003:latent} and maximum \emph{a posteriori} prediction of patch labels.
For each model, we set $K = 10$, allowing for the nine semantic classes plus \emph{void}, and, following \cite{VerbeekTr07}, we ensure that the \emph{void} class remains generic by fixing $\atom_{10}^m=(\frac{1}{V^m},\cdots,\frac{1}{V^m})$ for each modality $m$.
	
\subsection{Results}
Figure~\ref{fig:segs} displays sample test image segmentations obtained using the $\hbnbp$ admixture model.
Each pixel is given the predicted label of its closest patch center.
\ifdefined\figsintext
\begin{figure}
	\centering
	\includegraphics[width=.99\textwidth]{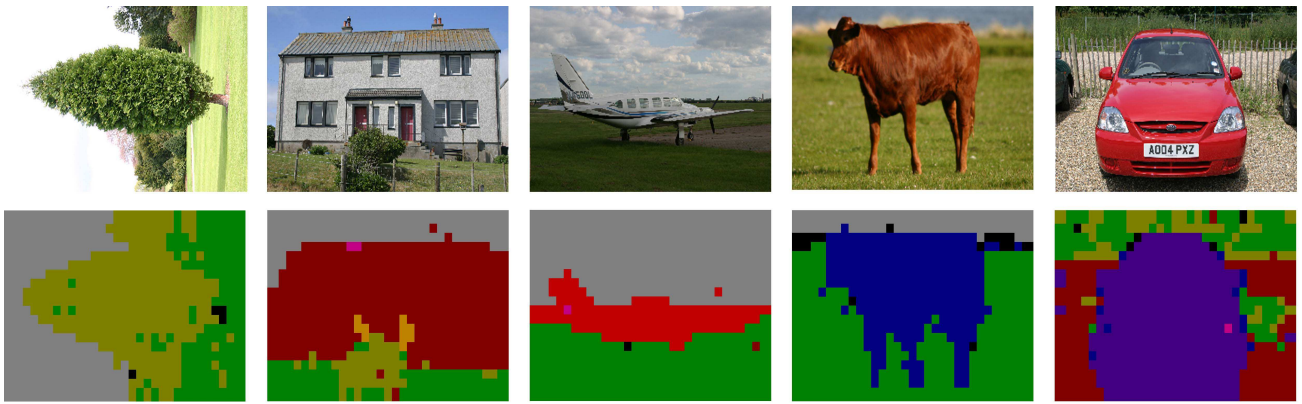}
  \caption{MSRC-v1 test image segmentations inferred by the HBNBP admixture model (best viewed in color).}
	\label{fig:segs}
\end{figure}
\fi
Test patch classification accuracies for the $\hbnbp$ admixture model and LDA are reported in \tabs{hbnbp-confusion-seg} and~\tabss{lda-confusion-seg} respectively.
All results are averaged over twenty randomly generated 90\% training / 10\% test divisions of the data set.  
The two methods perform comparably, with the $\hbnbp$ admixture model outperforming LDA in the prediction of every object class save \emph{building}.  Indeed, the mean object class accuracy is 0.79 for the $\hbnbp$ model versus 0.76 for LDA,
showing that the $\hbnbp$ provides a viable alternative to more classical approaches to admixture.
\begin{table}
\caption{Confusion matrices for patch-level image segmentation and object recognition on the MSRC-v1 database.
We report test image patch inference accuracy averaged over twenty randomly generated 90\% training / 10\% test divisions.}
\centering
\subfloat[$\hbnbp$ Confusion Matrix]
{\label{tab:hbnbp-confusion-seg} \footnotesize%
\begin{tabular}{c|c|c|c|c|c|c|c|c|c|c|}
\multicolumn{11}{c}{Predicted Class Label} \\ \cline{3-11}
\multicolumn{1}{c}{} & \multicolumn{1}{c}{} & \multicolumn{1}{|c}{building} & \multicolumn{1}{c}{grass} & \multicolumn{1}{c}{tree} & \multicolumn{1}{c}{cow} & \multicolumn{1}{c}{sky} & \multicolumn{1}{c}{aeroplane} & \multicolumn{1}{c}{face} & \multicolumn{1}{c}{car} & \multicolumn{1}{c|}{bicycle} \\ \cline{2-11}
\multirow{9}{*}{\begin{sideways}Actual Class Label\end{sideways}} &
building &  \cellcolor[gray]{0.34} \textcolor{white}{0.66} & \cellcolor[gray]{0.99} 0.01 & \cellcolor[gray]{0.95} 0.05 & \cellcolor[gray]{1.00} 0.00 & \cellcolor[gray]{0.97} 0.03 & \cellcolor[gray]{0.91} 0.09 & \cellcolor[gray]{0.99} 0.01 & \cellcolor[gray]{0.97} 0.03 & \cellcolor[gray]{0.91}  0.09 \\\cline{3-11} &
grass &  \cellcolor[gray]{1.00} 0.00 & \cellcolor[gray]{0.11} \textcolor{white}{0.89} & \cellcolor[gray]{0.94} 0.06 & \cellcolor[gray]{0.98} 0.02 & \cellcolor[gray]{1.00} 0.00 & \cellcolor[gray]{0.99} 0.01 & \cellcolor[gray]{1.00} 0.00 & \cellcolor[gray]{1.00} 0.00 & \cellcolor[gray]{1.00}  0.00 \\\cline{3-11} &
tree &  \cellcolor[gray]{0.99} 0.01 & \cellcolor[gray]{0.92} 0.08 & \cellcolor[gray]{0.25} \textcolor{white}{0.75} & \cellcolor[gray]{0.99} 0.01 & \cellcolor[gray]{0.96} 0.04 & \cellcolor[gray]{0.97} 0.03 & \cellcolor[gray]{1.00} 0.00 & \cellcolor[gray]{1.00} 0.00 & \cellcolor[gray]{0.93}  0.07 \\\cline{3-11} &
cow &  \cellcolor[gray]{1.00} 0.00 & \cellcolor[gray]{0.90} 0.10 & \cellcolor[gray]{0.96} 0.04 & \cellcolor[gray]{0.28} \textcolor{white}{0.72} & \cellcolor[gray]{1.00} 0.00 & \cellcolor[gray]{1.00} 0.00 & \cellcolor[gray]{0.95} 0.05 & \cellcolor[gray]{0.99} 0.01 & \cellcolor[gray]{0.99}  0.01 \\\cline{3-11} &
sky &  \cellcolor[gray]{0.96} 0.04 & \cellcolor[gray]{1.00} 0.00 & \cellcolor[gray]{0.99} 0.01 & \cellcolor[gray]{1.00} 0.00 & \cellcolor[gray]{0.07} \textcolor{white}{0.93} & \cellcolor[gray]{0.99} 0.01 & \cellcolor[gray]{1.00} 0.00 & \cellcolor[gray]{1.00} 0.00 & \cellcolor[gray]{1.00}  0.00 \\\cline{3-11} &
aeroplane &  \cellcolor[gray]{0.90} 0.10 & \cellcolor[gray]{0.96} 0.04 & \cellcolor[gray]{0.99} 0.01 & \cellcolor[gray]{1.00} 0.00 & \cellcolor[gray]{0.98} 0.02 & \cellcolor[gray]{0.19} \textcolor{white}{0.81} & \cellcolor[gray]{1.00} 0.00 & \cellcolor[gray]{0.98} 0.02 & \cellcolor[gray]{1.00}  0.00 \\\cline{3-11} &
face &  \cellcolor[gray]{0.96} 0.04 & \cellcolor[gray]{1.00} 0.00 & \cellcolor[gray]{0.99} 0.01 & \cellcolor[gray]{0.96} 0.04 & \cellcolor[gray]{1.00} 0.00 & \cellcolor[gray]{1.00} 0.00 & \cellcolor[gray]{0.16} \textcolor{white}{0.84} & \cellcolor[gray]{1.00} 0.00 & \cellcolor[gray]{1.00}  0.00 \\\cline{3-11} &
car &  \cellcolor[gray]{0.80} 0.20 & \cellcolor[gray]{1.00} 0.00 & \cellcolor[gray]{0.99} 0.01 & \cellcolor[gray]{1.00} 0.00 & \cellcolor[gray]{0.99} 0.01 & \cellcolor[gray]{0.99} 0.01 & \cellcolor[gray]{1.00} 0.00 & \cellcolor[gray]{0.27} \textcolor{white}{0.73} & \cellcolor[gray]{0.98}  0.02 \\\cline{3-11} &
bicycle &  \cellcolor[gray]{0.84} 0.16 & \cellcolor[gray]{1.00} 0.00 & \cellcolor[gray]{0.96} 0.04 & \cellcolor[gray]{1.00} 0.00 & \cellcolor[gray]{1.00} 0.00 & \cellcolor[gray]{1.00} 0.00 & \cellcolor[gray]{1.00} 0.00 & \cellcolor[gray]{0.98} 0.02 & \cellcolor[gray]{0.27}  \textcolor{white}{0.73} \\\cline{2-11} 
\end{tabular}
}
\\
\subfloat[LDA Confusion Matrix]
{\label{tab:lda-confusion-seg} \footnotesize%
\begin{tabular}{c|c|c|c|c|c|c|c|c|c|c|}
\multicolumn{11}{c}{Predicted Groups} \\ \cline{3-11}
\multicolumn{1}{c}{} & \multicolumn{1}{c}{} & \multicolumn{1}{|c}{building} & \multicolumn{1}{c}{grass} & \multicolumn{1}{c}{tree} & \multicolumn{1}{c}{cow} & \multicolumn{1}{c}{sky} & \multicolumn{1}{c}{aeroplane} & \multicolumn{1}{c}{face} & \multicolumn{1}{c}{car} & \multicolumn{1}{c|}{bicycle} \\ \cline{2-11}
\multirow{9}{*}{\begin{sideways}Actual Groups\end{sideways}} &
building &  \cellcolor[gray]{0.31} \textcolor{white}{0.69} & \cellcolor[gray]{0.99} 0.01 & \cellcolor[gray]{0.96} 0.04 & \cellcolor[gray]{0.99} 0.01 & \cellcolor[gray]{0.97} 0.03 & \cellcolor[gray]{0.93} 0.07 & \cellcolor[gray]{0.99} 0.01 & \cellcolor[gray]{0.97} 0.03 & \cellcolor[gray]{0.92}  0.08 \\\cline{3-11} &
grass &  \cellcolor[gray]{1.00} 0.00 & \cellcolor[gray]{0.12} \textcolor{white}{0.88} & \cellcolor[gray]{0.95} 0.05 & \cellcolor[gray]{0.98} 0.02 & \cellcolor[gray]{1.00} 0.00 & \cellcolor[gray]{0.99} 0.01 & \cellcolor[gray]{1.00} 0.00 & \cellcolor[gray]{1.00} 0.00 & \cellcolor[gray]{1.00}  0.00 \\\cline{3-11} &
tree &  \cellcolor[gray]{0.98} 0.02 & \cellcolor[gray]{0.92} 0.08 & \cellcolor[gray]{0.25} \textcolor{white}{0.75} & \cellcolor[gray]{0.99} 0.01 & \cellcolor[gray]{0.96} 0.04 & \cellcolor[gray]{0.98} 0.02 & \cellcolor[gray]{1.00} 0.00 & \cellcolor[gray]{1.00} 0.00 & \cellcolor[gray]{0.95}  0.05 \\\cline{3-11} &
cow &  \cellcolor[gray]{1.00} 0.00 & \cellcolor[gray]{0.90} 0.10 & \cellcolor[gray]{0.97} 0.03 & \cellcolor[gray]{0.30} \textcolor{white}{0.70} & \cellcolor[gray]{1.00} 0.00 & \cellcolor[gray]{1.00} 0.00 & \cellcolor[gray]{0.95} 0.05 & \cellcolor[gray]{0.99} 0.01 & \cellcolor[gray]{0.99}  0.01 \\\cline{3-11} &
sky &  \cellcolor[gray]{0.95} 0.05 & \cellcolor[gray]{1.00} 0.00 & \cellcolor[gray]{0.98} 0.02 & \cellcolor[gray]{1.00} 0.00 & \cellcolor[gray]{0.09} \textcolor{white}{0.91} & \cellcolor[gray]{0.99} 0.01 & \cellcolor[gray]{1.00} 0.00 & \cellcolor[gray]{1.00} 0.00 & \cellcolor[gray]{1.00}  0.00 \\\cline{3-11} &
aeroplane &  \cellcolor[gray]{0.88} 0.12 & \cellcolor[gray]{0.96} 0.04 & \cellcolor[gray]{0.99} 0.01 & \cellcolor[gray]{1.00} 0.00 & \cellcolor[gray]{0.98} 0.02 & \cellcolor[gray]{0.25} \textcolor{white}{0.75} & \cellcolor[gray]{1.00} 0.00 & \cellcolor[gray]{0.97} 0.03 & \cellcolor[gray]{1.00}  0.00 \\\cline{3-11} &
face &  \cellcolor[gray]{0.96} 0.04 & \cellcolor[gray]{1.00} 0.00 & \cellcolor[gray]{0.99} 0.01 & \cellcolor[gray]{0.95} 0.05 & \cellcolor[gray]{1.00} 0.00 & \cellcolor[gray]{1.00} 0.00 & \cellcolor[gray]{0.20} \textcolor{white}{0.80} & \cellcolor[gray]{1.00} 0.00 & \cellcolor[gray]{1.00}  0.00 \\\cline{3-11} &
car &  \cellcolor[gray]{0.81} 0.19 & \cellcolor[gray]{1.00} 0.00 & \cellcolor[gray]{0.99} 0.01 & \cellcolor[gray]{1.00} 0.00 & \cellcolor[gray]{0.99} 0.01 & \cellcolor[gray]{0.99} 0.01 & \cellcolor[gray]{1.00} 0.00 & \cellcolor[gray]{0.29} \textcolor{white}{0.71} & \cellcolor[gray]{0.97}  0.03 \\\cline{3-11} &
bicycle &  \cellcolor[gray]{0.81} 0.19 & \cellcolor[gray]{1.00} 0.00 & \cellcolor[gray]{0.96} 0.04 & \cellcolor[gray]{0.99} 0.01 & \cellcolor[gray]{1.00} 0.00 & \cellcolor[gray]{1.00} 0.00 & \cellcolor[gray]{1.00} 0.00 & \cellcolor[gray]{0.98} 0.02 & \cellcolor[gray]{0.32}  \textcolor{white}{0.68} \\\cline{2-11} 
\end{tabular}
}
\end{table}

\subsection{Parameter Sensitivity}
To test the sensitivity of the $\hbnbp$ admixture model to misspecification of the mass, concentration, and likelihood hyperparameters, we measure the fluctuation in test set performance as each hyperparameter deviates from its default value (with the remainder held fixed).
The results of this study are summarized in \tab{sensitivity}.
We find that the $\hbnbp$ model is rather robust to changes in the hyperparameters and maintains nearly constant predictive performance, even as the parameters vary over several orders of magnitude.


\begin{table}
\caption{\label{tab:sensitivity}Sensitivity of $\hbnbp$ admixture model to hyperparameter specification for joint image segmentation and object recognition on the MSRC-v1 database.
Each hyperparameter is varied across the specified range while the remaining parameters are held fixed to the default values reported in \mysec{setup-seg}.
We report test patch inference accuracy averaged across object classes and over twenty randomly generated 90\% training / 10\% test divisions.
For each test patch, we predict the label with the highest posterior probability across 2,000 samples.
}
\centering
\fbox{%
\begin{tabular}{c|c|c|c}
Hyperparameter & Parameter range & Minimum accuracy & Maximum accuracy \\\cline{1-4} 
$\bpmass_0$ & $[0.3,30]$ & 0.786 & 0.787 \\
$\bpconc_0$ & $[1.5,30]$ & 0.786 & 0.786 \\
$\eta$ & $[2\times10^{-16},1]$ & 0.778 & 0.788 
\end{tabular}
}
\end{table}

\section{Conclusions}
\label{sec:conclusions}

Motivated by problems of admixture, in which individuals are represented multiple times in multiple latent classes, 
we introduced the negative binomial process, an infinite-dimensional prior for vectors of counts.  
We developed new nonparametric admixture models based on the $\nbp$ and its conjugate prior, the beta process, 
and characterized the relationship between the $\bnbp$ and preexisting models for admixture.  
We also analyzed the asymptotics of our new priors, derived MCMC procedures for posterior inference, 
and demonstrated the effectiveness of our models in the domains of image segmentation and document analysis.

There are many other problem domains in which latent vectors of counts provide a natural modeling framework and where we believe that the $\hbnbp$
can prove useful.  
These include the computer vision task of \emph{multiple object recognition}, 
where one aims to discover which and how many objects are present in a given image~\citep{Titsias07}, 
and the problem of modeling \emph{copy number variation} in genomic regions, 
where one seeks to infer the underlying events responsible for large repetitions or deletions in segments of DNA~\citep{ChenXiZh11}.

\ifdefined\notanonymous
\section*{Acknowledgments} 
Support for this project was provided by IARPA under the ``Knowledge Discovery 
and Dissemination'' program (IARPA-BAA-09-10) and by ONR under the Multidisciplinary 
University Research Initiative (MURI) program (N00014-11-1-0688).  Tamara Broderick 
was supported by a National Science Foundation Graduate Research Fellowship.
Lester Mackey was supported by the National Defense Science and Engineering 
Graduate Fellowship.
\fi

\bibliographystyle{ECA_jasa}
{\small{\bibliography{refs}}}

\pagebreak
\appendices

\section{Connections}
\label{app:text_connections}

In \mysec{mix_hier} we noted that both the beta-negative binomial process
($\bnbp$) and the gamma Poisson process ($\gaplp$) provide nonparametric models for the
count vectors arising in admixture models.  In this section, we will elucidate
some of the deeper connections between these two stochastic processes.
We will see that understanding these connections can not only inspire
new stochastic process constructions but also lead to novel inference algorithms.

\ifdefined\figsintext
\begin{table}
\caption{\label{tab:connections} A comparison of two Bayesian nonparametric constructions of clusterings such that the clusters have conditionally independent, random sizes; hence the data set size itself is random.
$\pp$ indicates a Poisson point process draw with the given intensity.}
\centering
\fbox{%
\begin{tabular}{p{0.52\textwidth} | p{0.42\textwidth}}
Beta negative binomial process
& Gamma Poisson likelihood process \\ \hline
$\nu(d\bpweight, d\atom) =  \bpmass \bpconc \bpweight^{-1} (1-\bpweight)^{\bpconc-1} \; d\bpweight \; \base(d\atom)$
	& $\nu(d\gapweight, d\atom) = \gpconc \gapweight^{-1} e^{-\gpscale \gapweight} \; d\gapweight \; \base(d\atom)$ \\
$(\bpweight_{k}, \atom_{k}) \sim \pp( \nu(d\bpweight,d\atom) )$
	& $(\gapweight_{k}, \atom_{k}) \sim \pp( \nu(d\gapweight,d\atom) )$ \\
$\bpdraw = \sum_{k} \bpweight_{k} \delta_{\atom_{k}}$
	& $\gapdraw = \sum_{k} \gapweight_{k} \delta_{\atom_{k}}$ \\
$\lambda_{k} \indep \ga(r, \frac{1-\bpweight_{k}}{\bpweight_{k}})$
	& \\
$\likeweight_{k} \indep \pois(\lambda_{k})$
	& $\likeweight_{k} \indep \pois(\gapweight_{k})$ \\
\end{tabular}}
\end{table}
\fi
We are motivated by \tab{connections}, which indicates a strong parallel between the $\bnbp$ and $\gaplp$ 
constructions for clusterings where the
size of each cluster is independent and random
conditioned on some underlying process. The former requires an additional random stage
consisting of
a draw from a gamma distribution.  
Here, we use the representation of the 
negative binomial distribution, $\likeweight \sim \negbin(r, \bpweight)$, 
as a gamma mixture of Poisson distributions: 
$\bppweight \sim \ga(r, (1-\bpweight)/\bpweight)$ and $\likeweight \sim \pois(\bppweight)$.
However, this table mostly highlights the parallel on the level of the likelihood process
and therefore on the level of classic, one-dimensional distributions. The relations
between such distributions are well-studied.

Noting that many classic, one-dimensional distributions are easily obtained
from each other by a simple change of variables, we aim to find new, analogous 
transformations in the stochastic process setting. In particular, 
all of our results in this section, which apply to nonparametric Bayesian 
priors derived from Poisson point processes, have direct analogues in the setting 
of one-dimensional distributions. We start by reviewing these known
distributional relations.
First, consider a beta distributed random 
variable $x \sim \tb(a,b)$.
Then the variable $x/(1-x)$ has a 
\emph{beta prime distribution} with parameters $a$ and $b$; specifically, 
$\betaprime(a,b)$ denotes the beta prime distribution with density
$$
	\betaprime(z\mid a,b) = \frac{\Gamma(a+b)}{\Gamma(a)\Gamma(b)}z^{a-1}(1+z)^{-a-b}.
$$
The beta prime distribution can alternatively be derived from a gamma distribution. 
Namely, if $x \sim \ga(a,\gpscale)$ and $y \sim \ga(b,\gpscale)$ are independent, 
then $x/y \sim \betaprime(a,b)$. This connection is not the only one between the 
beta and gamma distributions though. Let
\begin{equation}
	\label{eq:gamma_dist}
	x \sim \ga(a,\gpscale), \quad y \sim \ga(b,\gpscale).
\end{equation}
Then
\begin{equation}
	\label{eq:beta_from_gamma}
	x/(x+y) \sim \tb(a,b).
\end{equation}
In the rest of this section, we present similar results but now for the process 
case---the beta process, gamma process and a new process we call the \emph{beta 
prime process}.  The proofs of these results appear in \app{connections}.

We start by defining a 
new completely random measure with nonnegative, real-valued feature weights.  
First, we note that, as for the processes defined in \mysec{bnp_priors},
there is no deterministic measure. Second,
we specify that the fixed atoms have distribution
$$
	\fweight_{l} \indep \betaprime(\bpconc \bpmass \rho_{l}, \bpconc(1-\bpmass \rho_{l}))
$$
at locations $(\fatom_{l})$. Here, $\bpconc > 0$, $\bpmass > 0$, $(\rho_{l})_{l=1}^{\infty}$,
and $(\fatom_{l})$ are
parameters. As usual, while the number of fixed atoms $L$ may be countably
infinite, it is typically finite. Finally, the ordinary component has Poisson process intensity
$\base_{ord} \times \nu$, where
\begin{align}
	\label{eq:bpp-intensity}
	\nu(d\bppweight) = \bpmass \bpconc \bppweight^{-1} (1+\bppweight)^{-\bpconc} \; d\bppweight,
\end{align}
which we note is sigma-finite with finite mean, guaranteeing that the number of atoms
generated from the ordinary component
will be countably infinite with finite sum.

We abbreviate by defining $\base = \sum_{l=1}^{L} \rho_{l} \delta_{\fatom_{l}} + \base_{ord}$ and say
that the resulting CRM
$
	\bppdraw \defeq \sum_{k} \bppweight_{k} \delta_{\atom_{k}}
$
is a draw from a \emph{beta prime process} 
($\bpp$) with base distribution $\base$: $\bppdraw \sim \bpp(\bpconc,\bpmass,\base)$.
The name ``beta prime process'' reflects the fact that the underlying intensity is an improper beta prime distribution as well as the beta prime distribution of the fixed atoms.

With this definition in hand, we can find the stochastic process analogues of the distributional results above (with proofs in \app{connections}). Just as a beta prime distribution can be derived from a beta random variable, we have the following result that a similar transformation of the atom weights of a beta process yields a beta prime process.
\begin{proposition}
	\label{prop:bpp_from_bp}
	Suppose $\bpdraw = \sum_k\bpweight_{k}\delta_{\atom_{k}} \sim \bp(\bpconc,\bpmass,\base)$.
	Then $\sum_k\frac{\bpweight_{k}}{1-\bpweight_{k}}\delta_{\atom_{k}} \sim \bpp(\bpconc,\bpmass,\base)$.
\end{proposition}

Just as a beta prime random variable can be derived as the ratio of gamma random variables, we find that the atoms of the beta prime process can be constructed as by taking ratios of gamma random variables and the atoms of a gamma process.
\begin{proposition}
	\label{prop:bpp_from_gp}
	Suppose $\gapdraw = \sum_k\gapweight_{k}\delta_{\atom_{k}} \sim \gap(\bpmass\bpconc,\gpscale,\base)$ and $\tau_{k} \sim \ga(\bpconc(1-\bpmass \base(\{\atom_{k}\})),\gpscale)$ independently for each $k$.
	Then $\sum_k\frac{\gapweight_{k}}{\tau_k}\delta_{\atom_{k}} \sim \bpp(\bpconc,\bpmass,\base)$.
\end{proposition}

And, finally, the analogue to constructing a beta random variable from two gamma random variables is the construction of a beta process from a gamma process and an infinite vector of independent gamma random variables.
\begin{proposition} \label{prop:bp_from_gp}
	Suppose $\gapdraw = \sum_k\gapweight_{k}\delta_{\atom_{k}} \sim \gap(\bpmass\bpconc,\gpscale,\base)$ and $\tau_{k} \sim \ga(\bpconc(1-\bpmass \base(\{\atom_{k}\})),\gpscale)$ independently for each $k$.
	Then $\sum_k\frac{\gapweight_{k}}{\tau_k+\gapweight_{k}}\delta_{\atom_{k}} \sim \bp(\bpconc,\bpmass,\base)$.	
\end{proposition}

The key to the manipulations above was the Poisson process framework of the ordinary component. In particular, we see that $\bp$ itself can be derived from $\gap$, and therefore the connection between the $\bnbp$ and $\gaplp$ is not restricted to just the negative binomial and Poisson likelihood. Moreover, besides introducing a further stochastic process in the form of the beta prime process, we emphasize that these relations potentially allow us to perform inference for a new stochastic process when inference for another, related stochastic process is already known---or to have available alternative, potentially faster or better mixing, inference algorithms.

\section{Proofs for Appendix~\ref{app:text_connections}} \label{app:connections}

\begin{proofof}{\prop{bpp_from_bp}}
	First, consider the ordinary component of a beta process. The Mapping Theorem of \citet{Kingman93} tells us that
if the collection of tuples $(\atom_{k}, \bpweight_{k})$ come from a Poisson process with intensity $H_{ord} \times \nu_{beta}$, where $\nu_{beta}$ is the beta process intensity of \eq{bp-intensity}, then the collection of tuples $(\atom_{k}, \bpweight_{k}/(1-\bpweight_{k}))$ are draws from a Poisson process with intensity $H_{ord} \times \nu$, where we apply a change of variables to find:
	\begin{align*}
		\nu(d\bppweight) &= \bpmass \bpconc \left(\frac{\bppweight}{1+\bppweight}\right)^{-1} \left(1-\frac{\bppweight}{1+\bppweight}\right)^{\bpconc - 1}\frac{1}{(1+\bppweight)^2} \; d\bppweight \\
		\nu(d\bppweight) &= \bpmass \bpconc \bppweight^{-1} (1+\bppweight)^{-\bpconc} \; d\bppweight,
	\end{align*}
	which matches \eq{bpp-intensity}.
	
	For any particular atom where $\bpweight_{k} \sim \tb(\bpconc \bpmass \rho_{k}, \bpconc(1-\bpmass \rho_{k}))$ and $\rho_k = \base(\{\atom_{k}\}) > 0$, we simply quote the well-known, one-dimensional change of variables $\bpweight_{k}/(1-\bpweight_{k}) \sim \betaprime(\bpconc \bpmass \rho_{k}, \bpconc(1-\bpmass \rho_{k}))$.
	
	Since there is no deterministic component, we have considered all components of the completely random measure.
\end{proofof}

\begin{proofof}{\prop{bpp_from_gp}}
	We again start with the ordinary component of a completely random measure. In particular, we assume the collection of tuples $(\atom_{k}, \gapweight_{k})$ is generated according to a Poisson process with intensity $H_{ord} \times \nu_{gamma}$, where $\nu_{gamma}$ is the gamma process intensity of \eq{gap_intensity}.
	
	Consider a random variable $\tau_{k} \sim \ga(\gpconc, \gpscale)$ associated with each such tuple. Then $1/\tau_{k} \sim \ig(\gpconc, \gpscale)$. We consider a marked Poisson process with mark $\bppweight_{k} \defeq \gapweight_{k} / \tau_{k}$ at tuple $(\atom_{k}, \gapweight_{k})$ of the original process. By the scaling property of the inverse gamma distribution, we note $\bppweight_{k} \sim \ig(\bpconc,\gpscale\gapweight_{k})$ given $\gapweight_{k}$. So the Marking Theorem~\citep{Kingman93} implies that the collection of tuples $(\atom_{k}, \bppweight_{k})$ is itself a draw from a Poisson point process with intensity $H_{ord} \times \nu$, where
	\begin{align*}
		\nu(d\bppweight) &= \int p(\bppweight\mid \bpconc,\gpscale,\gapweight) \; \nu(d\gapweight) \; d\gapweight \; d\bppweight \\
		&= d\bppweight\int \frac{1}{\Gamma(\bpconc)}(\gpscale\gapweight)^\bpconc \bppweight^{-\bpconc-1}\exp(-\gpscale\gapweight/\bppweight) 
			\cdot \bpmass \bpconc\gapweight^{-1} \exp(-\gpscale\gapweight) \; d\gapweight \\
		&= \bpmass \bpconc\gpscale^{\bpconc}\frac{1}{\Gamma(\bpconc)}\bppweight^{-\bpconc-1}d\bppweight\int \gapweight^{\bpconc-1} \exp(-\gapweight\gpscale(1+\bppweight)/\bppweight) \; d\gapweight \\
		&= \bpmass \bpconc\gpscale^{\bpconc}\frac{1}{\Gamma(\bpconc)}\bppweight^{-\bpconc-1}\Gamma(\bpconc)\left(\frac{\bppweight}{\gpscale(1+\bppweight)}\right)^\bpconc \; d\bppweight\\
		&= \bpmass \bpconc\bppweight^{-1}\left(1+\bppweight\right)^{-\bpconc} \; d\bppweight,
	\end{align*}
	which matches the beta prime process ordinary component intensity of \eq{bpp-intensity}.
	
	For any particular atom of the gamma process, $\gapweight_{k} \sim \ga(\bpconc \bpmass \rho_{k}, \gpscale)$ with $\rho_{k} = \base(\{\atom_{k}\}) > 0$, it is well known that $\gapweight_{k}/\tau_{k}$ has the $\betaprime(\bpconc \bpmass \rho_{k}, \bpconc(1-\bpmass \rho_{k}))$
	distribution, as desired.
	
	There is no deterministic component of the gamma process.
\end{proofof}

\begin{proofof}{\prop{bp_from_gp}}
	Before proceeding to prove \prop{bp_from_gp} in the manner of the proofs of \props{bpp_from_bp} and \propss{bpp_from_gp} above, we first note that \prop{bp_from_gp} can be derived from \prop{bpp_from_gp} and an inverse change of variables from that in \prop{bpp_from_bp}. 

	Taking the same direct route of proof as above, though, we begin with the ordinary component of the gamma process so that the collection of tuples $(\atom_{k}, \gapweight_{k})$ is generated according to a Poisson process with intensity $H_{ord} \times \nu_{gamma}$, where $\nu_{gamma}$ is the gamma process intensity of \eq{gap_intensity}. The Marking Theorem \citep{Kingman93} tells us that the marked Poisson process with points $(\atom_{k}, \gapweight_{k}, \tau_{k})$ has intensity $\base_{ord} \times \nu$, where
	\begin{align*}
		\nu( d\gapweight, d\tau)
			&= \bpmass \bpconc \gapweight^{-1} e^{-\gpscale \gapweight} \cdot (\Gamma(\bpconc))^{-1} \tau^{\bpconc-1} \exp(-\gpscale \tau) \; \gpscale^{\bpconc} \; d\gapweight \; d\tau.
	\end{align*}
Now consider the change of variables $u = \gapweight / (\gapweight + \tau), v = \gapweight + \tau$. 
The reverse transformation is $\gapweight = uv, \tau = (1-u)v$ with Jacobian $v$.
Then 
the Poisson point process with points
$(\atom_{k}, u_{k}, v_{k})$
has intensity $\base_{ord} \times \nu$, where
\begin{align*}
	\nu(d\atom, du, dv)
		&= (\Gamma(\bpconc))^{-1} \bpmass \bpconc \gpscale^{\bpconc} u^{-1} v^{-1} (1-u)^{\bpconc-1} v^{\bpconc - 1} e^{-\gpscale v} \cdot v \; du \; dv.
\end{align*}
So the Poisson point process with points
$(\atom_{k}, u_{k})$
has intensity $\base_{ord} \times \nu$, with
\begin{align*}
	\nu(d\atom, du)
		&= \int_{v} \mu( d\atom, du, dv) \\
		&= \int_{v} (\Gamma(\bpconc))^{-1} \bpmass \bpconc \gpscale^{\bpconc} u^{-1} (1-u)^{\bpconc-1} v^{\bpconc - 1} e^{-\gpscale v} \; du \; dv \\
		&= (\Gamma(\bpconc))^{-1} \bpmass \bpconc \gpscale^{\bpconc} u^{-1} (1-u)^{\bpconc-1} \Gamma(\bpconc) \gpscale^{-\bpconc} \; du \\
		&= \bpmass \bpconc u^{-1} (1-u)^{\bpconc - 1} \; du,
\end{align*}
which is the known beta process intensity.

In the discrete case with $\base(\{\atom_{k}\}) = \rho_{k} > 0$, we have by construction
$$
	\gapweight_{k} \sim \ga(\bpconc \bpmass \rho_{k}, \gpscale)
$$
and
$$
	\tau_{k} \sim \ga(\bpconc(1-\bpmass \rho_{k}), \gpscale).
$$
From classic finite distributional results, we have
$$
	\frac{\gapweight_{k}}{\tau_{k} + \gapweight_{k}} \sim \tb(\bpconc \bpmass \rho_{k}, \bpconc(1-\bpmass \rho_{k})),
$$
exactly as in the case of the beta process.

As the gamma process and beta process each have no deterministic components, this completes the proof.
\end{proofof}

\section{Full results for Section~\ref{sec:asymptotics}} \label{app:asymptotics_state}

In order to fill in \tab{asymptotics}, we start by briefly establishing the results for
expected number of clusters of size $j$ for the $\tdp$ and $\pyp$; the results
for the expected total number of clusters are cited in the main text. We then
move on to full results for the $\bnbp$ and $\tbnbp$. Proofs for all results in
this section appear in \app{asymptotics_proof}.

\begin{theorem} \label{thm:dp_cl_size_j}
Assume that the concentration parameter for the $\tdp$
satisfies $\dpconc > 0$. Then the expected number 
of data clusters of size $j$, $\Phi_{j}(\numdata)$,
has asymptotic growth
$$
	\Phi_{j}(\numdata)
		\sim \dpconc j^{-1}, \quad \numdata \rightarrow \infty.
$$
\end{theorem}

\begin{theorem} \label{thm:pyp_cl_size_j}
Assume that the discount parameter for the $\pyp$ satisfies 
$\dpdisc \in (0,1)$, and
the concentration parameter satisfies $\dpconc > 1 - \bpdisc$.
Then the expected number of data clusters of size $j$,
$\Phi_{j}(\numdata)$, has asymptotic growth
$$
	\Phi_{j}(\numdata)
		\sim \frac{\Gamma(\bpconc+1)}{\Gamma(1-\bpdisc) \Gamma(\bpconc+\bpdisc)} \frac{\Gamma(j-\bpdisc)}{\Gamma(j+1)}
			\numdata^{\dpdisc}, \quad \numdata \rightarrow \infty.
$$
\end{theorem}

Next we establish how the expected 
number of data points, $\xi(r)$, grows asymptotically with $r$ in the $\bnbp$ 
case (in \lem{xi_r_bnbp}) and the $\tbnbp$ case (in \lem{xi_r_3bnbp}). 
We begin by showing that the expected number of data points is infinite 
for the concentration parameter range $\bpconc \le 1 - \bpdisc$ in both 
the $\bnbp$  ($\bpdisc = 0$) and $\tbnbp$  models.

\begin{lemma} \label{lem:xi_r_conc_l1ma}
Assume that the discount parameter for three-parameter beta process satisfies 
$\bpdisc \in [0,1)$ (the beta process is the special case when $\bpdisc = 0$), 
the concentration parameter satisfies $\bpconc \le 1 - \bpdisc$, and the mass 
parameter satisfies $\bpmass > 0$.  Then the expected number of data points, 
$\xi(r) = \mbe[ \sum_{k} \likeweight_{k}]$, from a $\bnbp$  or $\tbnbp$ , as appropriate, 
is infinite.
\end{lemma}

\begin{lemma} \label{lem:xi_r_bnbp}
Assume that the concentration parameter for the beta process satisfies 
$\bpconc > 1$ and the mass parameter satisifies $\bpmass > 0$.  Then the 
expected number of data points $\xi(r) = \mbe[ \sum_{k} \likeweight_{k}]$ 
from a $\bnbp$ has asymptotic growth
$$
	\xi(r) \sim \bpmass \frac{\bpconc}{\bpconc - 1} r, \quad r \rightarrow \infty.
$$
\end{lemma}

\begin{lemma} \label{lem:xi_r_3bnbp}
Assume that a three-parameter beta process has discount parameter $\bpdisc \in (0,1)$ 
and concentration parameter $\bpconc > 1 - \bpdisc$.  Then the expected number of 
data points $\xi(r) = \mbe[ \sum_{k} \likeweight_{k}]$ from a $\tbnbp$ has asymptotic growth
$$
	\xi(r) \sim \bpmass \frac{\bpconc}{\bpconc+\bpdisc-1} r, \quad r \rightarrow \infty.
$$
\end{lemma}

Next, we establish how the expected number of clusters, $\Phi(r)$, grows 
asymptotically as $r \rightarrow \infty$ in the $\bnbp$ case (in \lem{Phi_r_bnbp}) 
and in the $\tbnbp$ case (in \lem{Phi_r_3bnbp}).

\begin{lemma} \label{lem:Phi_r_bnbp}
Let $\bpconc > 0$.
Then the expected number of clusters $\Phi(r) = \mbe[ \sum_{k} \mbo\{\likeweight_{k} > 0\}]$ from a $\bnbp$ has asymptotic growth
$$
	\Phi(r) \sim \bpmass \bpconc \log r, \quad r \rightarrow \infty.
$$
\end{lemma}

\begin{lemma} \label{lem:Phi_r_3bnbp}
Consider a three-parameter beta process.  Let the discount parameter 
satisfy $\alpha > 0$ and the concentration parameter satisfy $\bpconc > -\bpdisc$. 
Then the number of clusters $K(r) \sum_{k} \mbo\{\likeweight_{k} > 0\}$ from a 
$\tbnbp$ has almost sure asymptotic growth
$$
	K(r) \stackrel{a.s.}{\sim} \frac{\bpmass}{\bpdisc} \frac{\Gamma(\bpconc + 1)}{\Gamma(\bpconc + \bpdisc)} r^{\bpdisc}, \quad r \rightarrow \infty.
$$
\end{lemma}	

We are also interested in how the expected number of clusters of size $j$, $\Phi_{j}(r)$, grows as $r \rightarrow \infty$. To that end, we establish this asymptotic growth in the $\bnbp$ case in \lem{Phi_j_r_bnbp} and in the $\tbnbp$ case in \lem{Phi_j_r_3bnbp} below.

\begin{lemma} \label{lem:Phi_j_r_bnbp}
Let $\bpconc > 0$.
Then the expected number of clusters of size $j$, $\Phi_{j}(r)  = \mbe[ \sum_{k} \mbo\{\likeweight_{k} = j\}]$, from a $\bnbp$ has asymptotic growth
$$
	\Phi_{j}(r) \sim \bpmass \bpconc j^{-1}, \quad r \rightarrow \infty.
$$
That is, the number is asymptotically constant in $r$.
\end{lemma}

\begin{lemma} \label{lem:Phi_j_r_3bnbp}
Let $\bpconc > -\bpdisc$ and $\bpdisc \in (0,1)$.
Then the expected number of clusters of size $j$, $\Phi_{j}(r)  = \mbe[ \sum_{k} \mbo\{\likeweight_{k} = j\}]$, from a $\tbnbp$ has asymptotic growth
$$
	\Phi_{j}(r)  \sim \bpmass \frac{\Gamma(1+\bpconc)}{\Gamma(1-\bpdisc)\Gamma(\bpconc+\bpdisc)} \frac{\Gamma(j-\bpdisc)}{\Gamma(j+1)} r^{\bpdisc}, \quad r \rightarrow \infty.
$$
\end{lemma}

Finally, we wish to combine these results to establish asymptotic results for 
the diversity, i.e., the expected number of clusters (or clusters of size $j$) 
as the expected number of data points varies. We find the asymptotic growth 
in the number of clusters for the $\bnbp$ in \thm{Phi_xi_bnbp} and for the 
$\tbnbp$ in \thm{Phi_xi_3bnbp}. We find the asymptotic growth in the number of 
clusters of size $j$ for the $\bnbp$ (in fact, the result has already been shown 
in \lem{Phi_j_r_bnbp}) and for the $\tbnbp$ in \thm{Phi_j_xi_3bnbp}.

\begin{theorem} \label{thm:Phi_xi_bnbp}
	Let $\bpconc > 1$. Then the expected number of clusters $\Phi$ grows asymptotically as the log of the expected number of data points $\xi$:
	$$
		\Phi(r) \sim \bpmass \bpconc \log(\xi(r)), \quad r \rightarrow \infty.
	$$
\end{theorem}

\begin{theorem} \label{thm:Phi_xi_3bnbp}
	Let $\bpconc + \bpdisc > 1$ and $\bpdisc \in (0,1)$. Then the number of clusters $K$ grows asymptotically as a power of the expected number of data points $\xi$:
	$$
		K(r) \stackrel{a.s.}{\sim} \frac{\bpmass^{1-\bpdisc}}{\bpdisc} \frac{\Gamma(\bpconc+1)}{\Gamma(\bpconc+\bpdisc)} \left(\frac{\bpconc+\bpdisc-1}{\bpconc}\right)^{\alpha} (\xi(r))^{\bpdisc}, \quad r \rightarrow \infty.
	$$
\end{theorem}

\begin{theorem} \label{thm:Phi_j_xi_3bnbp}
	Let $\bpconc + \bpdisc > 1$ and $\bpdisc \in (0,1)$. Then the expected number of clusters of size $j$, $\Phi_{j}$, grows asymptotically as a power of the expected number of data points $\xi$:
	$$
		\Phi_{j}(r) \sim \bpmass^{1-\bpdisc} \frac{\Gamma(\bpconc+1)}{\Gamma(1-\bpdisc) \Gamma(\bpconc+\bpdisc)} \frac{\Gamma(j-\bpdisc)}{\Gamma(j+1)} \left( \frac{\bpconc+\bpdisc-1}{\bpconc} \right)^{\bpdisc} (\xi(r))^{\bpdisc}, \quad r \rightarrow \infty.
	$$
\end{theorem}

\section{Proofs for Appendix~\ref{app:asymptotics_state}} \label{app:asymptotics_proof}

\begin{proofof}{\thm{dp_cl_size_j}}
When cluster proportions are generated according to a Dirichlet process
and clustering belonging is generated according to draws from the resulting random measure,
the joint distribution of $(K_{1}(\numdata), \ldots, K_{\numdata}(\numdata))$
is described by the \emph{Ewens sampling formula},
which appears as \eqw{2.9} in \citep{watterson:1974:sampling}.
It follows that \eqw{2.22} in \citep{watterson:1974:sampling} gives $\Phi_{j}(\numdata) = \mbe[K_{j}(\numdata)]$:
$$
	\Phi_{j}(\numdata)
		= \frac{\dpconc}{j} \binom{\dpconc + \numdata - j - 1}{ \numdata - j }
			\cdot \binom{ \dpconc + \numdata - 1 }{ \numdata }^{-1}.
$$
Therefore,
\begin{align*}
	\Phi_{j}(\numdata)
		&= \frac{\dpconc}{j} \frac{
				\Gamma(\dpconc + \numdata - j )
				}{
				\Gamma( \numdata - j + 1) \Gamma(\dpconc )
			} \cdot \frac{
				\Gamma( \numdata + 1) \Gamma(\dpconc )
				}{
				\Gamma( \numdata + \dpconc )
			} \\
		&= \frac{\dpconc}{j} \cdot \frac{
				\Gamma(\numdata + \dpconc  - j)
			}{
				\Gamma(\numdata + \dpconc)
			} \cdot \frac{
				\Gamma( \numdata + 1)
			}{
				\Gamma( \numdata + 1 - j)
			} \\
		&\sim \frac{\dpconc}{j} \cdot (\numdata + \dpconc)^{-j} \cdot (\numdata + 1)^{j},
			\quad \numdata \rightarrow \infty \\
		&\sim \frac{\dpconc}{j}, \quad \numdata \rightarrow \infty,
\end{align*}
where the asymptotics for the ratios of gamma functions
follow from \citet{tricomi:1951:asymptotic}.
\end{proofof}

\begin{proofof}{\thm{pyp_cl_size_j}}
\citet{pitman:2006:combinatorial} establishes that, for the $\pyp$ with parameters
$\dpconc$ and $\dpdisc$ given in the result statement, we have
$\Phi(\numdata) \sim \frac{\Gamma(\gpconc+1)}{\dpdisc \Gamma(\gpconc+\dpdisc)} \numdata^{\dpdisc}$ as $\numdata \rightarrow \infty$.

Note that $\Phi(\numdata)$ is in the form of \eqw{48} on \pw{167} of \citep{gnedin:2007:notes}. The desired result follows by applying \eqw{51} on \pw{167} of \citep{gnedin:2007:notes}.
\end{proofof}

\begin{proofof}{\lem{xi_r_conc_l1ma}}
In this case, we have
\begin{align*}
	\mbe[ \sum_{k} \likeweight_{k} ] 
		&= \mbe\left[ \mbe[ \sum_{k} \likeweight_{k} | \tbf{\bpweight}_{\cdot} ] \right] \\
		& \textrm{by the tower property} \\
		&= \mbe\left[ \sum_{k} \mbe[ \likeweight_{k} | \tbf{\bpweight}_{\cdot} ] \right] \\
		& \textrm{by monotonicity} \\
		&= \mbe\left[ \sum_{k} \frac{\bpweight_{k} r}{(1-\bpweight_{k})} \right] \\
		& \textrm{using the mean of the negative binomial distribution} \\
		&= \int_{0}^{1} \frac{b r}{(1-b)} \; \nu(db) \\
		& \textrm{by Campbell's Theorem~\citep{Kingman93}} \\
		&= r \frac{\Gamma(1+\bpconc)}{\Gamma(1-\bpdisc)\Gamma(\bpconc+\bpdisc)} \int_{0}^{1} b^{-\bpdisc} (1-b)^{\bpconc+\bpdisc-2} \; db.
\end{align*}
The final line is finite iff
$$
	1 - \bpdisc > 0, \quad \textrm{and} \quad \bpconc + \bpdisc - 1 > 0.
$$
Equivalently, the final line is finite iff
$$
	\bpdisc < 1 \quad \textrm{and} \quad \bpconc > 1 - \bpdisc.
$$
\end{proofof}

\begin{proofof}{\lem{xi_r_bnbp}}
Let $\bpdraw = \sum_{k} \bpweight_{k} \atom_{k}$ be beta process distributed. Let $\likeweight_{k} \iid \negbin(r, \bpweight_{k})$.
By the Marking theorem~\citep{Kingman93}, the Poisson process $\{\bpweight_{k}, \atom_{k}, \likeweight_{k}\}$ has intensity
\begin{align}
	\label{eq:dens_pois_proc_and_neg_bin}
	\nu(d\bpweight, d\atom, \likeweight)
		&= \bpmass \bpconc \bpweight^{-1} (1-\bpweight)^{\bpconc - 1} \binom{\likeweight+r-1}{\likeweight} (1-\bpweight)^{r} \bpweight^{\likeweight} \; d\bpweight \; \base_{ord}(d\atom).
\end{align}
So the Poisson process $\{\likeweight_{k}\}$ has intensity
\begin{align*}
	\nu(\likeweight)
		&= \bpmass \bpconc \frac{\Gamma(\likeweight+r)}{\Gamma(\likeweight+1)\Gamma(r)} \frac{\Gamma(\likeweight) \Gamma(r+\bpconc)}{\Gamma(\likeweight+r+\bpconc)}.
\end{align*}
Thus, by Campbell's theorem~\citep{Kingman93}, 
\begin{align*}
	\mbe[\sum_{k} \likeweight_{k}]
		= \sum_{\likeweight=1}^{\infty} \likeweight \nu(\likeweight) 
		= \bpmass \bpconc \frac{\Gamma(r+\bpconc)}{\Gamma(r)} \sum_{\likeweight=1}^{\infty} \frac{\Gamma(\likeweight+r)}{\Gamma(\likeweight+r+\bpconc)}.
\end{align*}

To evaluate the sum $\sum_{\likeweight=1}^{\infty} \frac{\Gamma(\likeweight+r)}{\Gamma(\likeweight+r+\bpconc)}$, we appeal to a result from~\citet{tricomi:1951:asymptotic}:
\begin{align}
	\label{eq:gamma_fcn_ratio}
	\frac{\Gamma(x + a)}{\Gamma(x+b)} = x^{a - b} \left[ 1 + \frac{(a-b)(a+ b - 1)}{2x} + O(x^{-2}) \right], \quad x \rightarrow \infty.
\end{align}

In particular,
\begin{align*}
	\frac{\Gamma(\likeweight+r)}{\Gamma(\likeweight+r+\bpconc)}
		&\le (\likeweight+r)^{-\bpconc} \left[ 1 - \frac{\bpconc(\bpconc - 1)}{2(\likeweight+r)} + C (\likeweight+r)^{-2} \right]
		& \textrm{for some constant $C$}
\end{align*}
and
\begin{align*}
	\frac{\Gamma(\likeweight+r)}{\Gamma(\likeweight+r+\bpconc)}
		&\ge (\likeweight+r)^{-\bpconc} \left[ 1 - \frac{\bpconc(\bpconc - 1)}{2(\likeweight+r)} - C' (\likeweight+r)^{-2} \right]
		& \textrm{for some constant $C'$}.
\end{align*}

Before proceeding, we establish for $a > 1$,
\begin{align*}
	\sum_{\likeweight=1}^{\infty} (\likeweight+r)^{-a}
		\le \int_{x=0}^{\infty} (x+r)^{-a} \; dx 
		= (a-1)^{-1} r^{1-a}
\end{align*}
and
\begin{align*}
	\sum_{\likeweight=1}^{\infty} (\likeweight+r)^{-a}
		\ge \int_{x=1}^{\infty} (x+r)^{-a} \; dx 
		= (\alpha-1)^{-1} (r+1)^{1-a}.
\end{align*}
So
\begin{align*}
	\sum_{\likeweight=1}^{\infty} \frac{\Gamma(\likeweight+r)}{\Gamma(\likeweight+r+\bpconc)}
		&\le (\bpconc-1)^{-1} r^{1-\bpconc}  - \frac{\bpconc-1}{2} (r+1)^{-\bpconc} + C (\bpconc+1)^{-1} r^{-\bpconc-1}
\end{align*}
and
\begin{align*}
	\sum_{\likeweight=1}^{\infty} \frac{\Gamma(\likeweight+r)}{\Gamma(\likeweight+r+\bpconc)}
		&\ge (\bpconc-1)^{-1} (r+1)^{1-\bpconc} - \frac{\bpconc-1}{2} r^{-\bpconc} - C (\bpconc+1)^{-1} (r+1)^{-\bpconc-1}.
\end{align*}

Since, for $\bpconc > 1$, we have
\begin{equation}
	\frac{r^{1-\bpconc}}{(r+1)^{1-\bpconc}}
		\rightarrow 1, \quad r \rightarrow \infty,
\end{equation}
it follows that
\begin{equation}
	\sum_{\likeweight=1}^{\infty} \frac{\Gamma(\likeweight+r)}{\Gamma(\likeweight+r+\bpconc)}
		\sim (\bpconc-1)^{-1} r^{1-\bpconc}.
\end{equation}
From~\eq{gamma_fcn_ratio}, we also have $\frac{\Gamma(r+\bpconc)}{\Gamma(r)} \sim r^{\bpconc}$ as $r \rightarrow \infty$. So we conclude that
\begin{align*}
	\mbe[\sum_{k} \likeweight_{k}]
		&\sim \bpmass \frac{\bpconc}{\bpconc - 1} r, \quad r \rightarrow \infty,
\end{align*}
as desired.
\end{proofof}

\begin{proofof}{\lem{xi_r_3bnbp}}
The proof proceeds as above. In this case, we have that the Poisson process $\{\bpweight_{k}, \atom_{k}, \likeweight_{k}\}$ has intensity
\begin{align*}
	\nu(d\bpweight, d\atom, \likeweight)
		&= \bpmass \frac{\Gamma(1+\bpconc)}{\Gamma(1-\bpdisc) \Gamma(\bpconc+\bpdisc)} \bpweight^{-1-\bpdisc} (1-\bpweight)^{\bpconc+\bpdisc-1}
		\frac{\Gamma(\likeweight+r)}{\Gamma(\likeweight+1) \Gamma(r)} (1-\bpweight)^{r} \bpweight^{\likeweight} \; db \; \base(d\atom).
\end{align*}
So the Poisson process $\{\likeweight_{k}\}$ has intensity
\begin{align*}
	\nu(\likeweight) &= \bpmass \frac{\Gamma(1+\bpconc)}{\Gamma(1-\bpdisc) \Gamma(\bpconc+\bpdisc)} \frac{\Gamma(\likeweight+r)}{\Gamma(\likeweight+1) \Gamma(r)} \frac{\Gamma(\likeweight-\bpdisc)\Gamma(r+\bpconc+\bpdisc)}{\Gamma(\likeweight + r + \bpconc)}.
\end{align*}
By Campbell's theorem, 
\begin{align*}
	\mbe[\sum_{k} \likeweight_{k}]
		= \sum_{\likeweight=1}^{\infty} \likeweight \nu(\likeweight) 
		= \bpmass \frac{\Gamma(1+\bpconc)}{\Gamma(1-\bpdisc) \Gamma(\bpconc + \bpdisc)} \frac{ \Gamma(r+\bpconc+\bpdisc)}{\Gamma(r)} \sum_{\likeweight=1}^{\infty} \frac{\Gamma(\likeweight+r)}{\Gamma(\likeweight+r+\bpconc)} \frac{\Gamma(\likeweight-\bpdisc)}{\Gamma(\likeweight)}.
\end{align*}

We will find the following inequalities, with $\likeweight \ge 1$ and $\bpdisc \in (0,1)$, useful \citep[cf.\ \eqw{2.8} in][]{qi:2010:bounds}:
\begin{equation}
	\label{eq:ratio_gamma_ineq}
	(\likeweight-\bpdisc)^{-\bpdisc} \le \frac{\Gamma(\likeweight-\bpdisc)}{\Gamma(\likeweight)} \le (\likeweight-1)^{-\bpdisc}.
\end{equation}

We will also find the following integrals useful. Let $a > 1$.
\begin{align}
	\nonumber
	\sum_{\likeweight=2}^{\infty} (\likeweight+r)^{-a} (\likeweight-\bpdisc)^{-\bpdisc} 
		\nonumber
		&\le  \sum_{\likeweight=2}^{\infty} (\likeweight+r)^{-a} (\likeweight-1)^{-\bpdisc} \\
		\nonumber
		&\le \int_{x=0}^{\infty} (x+r)^{-a} x^{-\bpdisc} \; dx \\
		\nonumber
		&= r^{-a - \bpdisc + 1} \int_{y=0}^{\infty} (y+1)^{-a} y^{-\bpdisc} \; dy \\
		\label{eq:power_upper_int}
		&= r^{-a - \bpdisc + 1} \frac{\Gamma(1-\bpdisc)\Gamma(a+\bpdisc-1)}{\Gamma(a)}.
\end{align}
Similarly,
\begin{align}
	\sum_{\likeweight=2}^{\infty} (\likeweight+r)^{-a} (\likeweight-1)^{-\bpdisc} 
		\nonumber
		&\ge \sum_{\likeweight=2}^{\infty} (\likeweight+r)^{-a} (\likeweight-\bpdisc)^{-\bpdisc} \\
		\nonumber
		&\ge \int_{x=2}^{\infty} (x+r)^{-a} x^{-\bpdisc} \; dx \\
		\nonumber
		&= \int_{x=0}^{\infty} (x+r)^{-a} x^{-\bpdisc} \; dx - \int_{0}^{2} (x+r)^{-a} x^{-\bpdisc} \; dx \\
		\label{eq:power_lower_int}
		&\ge r^{-a - \bpdisc + 1} \frac{\Gamma(1-\bpdisc)\Gamma(a+\bpdisc-1)}{\Gamma(a)} - r^{-a} (1-\bpdisc)^{-1} 2^{1-\bpdisc}.
\end{align}

First, we consider an upper bound. To that end,
\begin{align*}
	\sum_{\likeweight=2}^{\infty} \frac{\Gamma(\likeweight+r)}{\Gamma(\likeweight+r+\bpconc)} \frac{\Gamma(\likeweight-\bpdisc)}{\Gamma(\likeweight)} 
		&\le \sum_{\likeweight=2}^{\infty} (\likeweight+r)^{-\bpconc} \left(1 - \frac{\bpconc(\bpconc+1)}{2(\likeweight+r)} + C(\likeweight+r)^{-2}\right) (\likeweight-1)^{-\bpdisc} \\
		& \textrm{for some constant $C$} \\
		&\le r^{-\bpconc - \bpdisc + 1} \frac{\Gamma(1-\bpdisc)\Gamma(\bpconc+\bpdisc-1)}{\Gamma(\bpconc)} \\
		& {} - \frac{\bpconc(\bpconc+1)}{2} r^{-\bpconc - \bpdisc} \frac{\Gamma(1-\bpdisc)\Gamma(\bpconc+1+\bpdisc-1)}{\Gamma(\bpconc+1)} - r^{-\bpconc-1} (1-\bpdisc)^{-1} 2^{1-\bpdisc} \\
		& {} + C r^{-\bpconc - \bpdisc - 1} \frac{\Gamma(1-\bpdisc)\Gamma(\bpconc+\bpdisc+1)}{\Gamma(\bpconc+2)}.
\end{align*}

For the lower bound,
\begin{align*}
	\sum_{\likeweight=2}^{\infty} \frac{\Gamma(\likeweight+r)}{\Gamma(\likeweight+r+\bpconc)} \frac{\Gamma(\likeweight-\bpdisc)}{\Gamma(\likeweight)} 
		&\ge \sum_{\likeweight=2}^{\infty} (\likeweight+r)^{-\bpconc} \left(1 - \frac{\bpconc(\bpconc+1)}{2(\likeweight+r)} - C'(\likeweight+r)^{-2}\right) (\likeweight-\bpdisc)^{-\bpdisc} \\
		&\textrm{for some constant $C'$} \\
		&\ge r^{-\bpconc - \bpdisc + 1} \frac{\Gamma(1-\bpdisc)\Gamma(\bpconc+\bpdisc-1)}{\Gamma(\bpconc)} - r^{-\bpconc} (1-\bpdisc)^{-1} 2^{1-\bpdisc} \\
		& {} - r^{-\bpconc - \bpdisc} \frac{\Gamma(1-\bpdisc)\Gamma(\bpconc+\bpdisc)}{\Gamma(\bpconc+1)} \\
		& {} - C'  r^{-\bpconc - \bpdisc -1} \frac{\Gamma(1-\bpdisc)\Gamma(\bpconc+\bpdisc+1)}{\Gamma(\bpconc+2)}.
\end{align*}

It follows from the two bounds above that
\begin{equation*}
	\sum_{\likeweight=2}^{\infty} \frac{\Gamma(\likeweight+r)}{\Gamma(\likeweight+r+\bpconc)} \frac{\Gamma(\likeweight-\bpdisc)}{\Gamma(\likeweight)}
		\sim \frac{\Gamma(1-\bpdisc)\Gamma(\bpconc+\bpdisc-1)}{\Gamma(\bpconc)} r^{-\bpconc - \bpdisc + 1}.
\end{equation*}

Since
$$
	\frac{ \Gamma(r+\bpconc+\bpdisc)}{\Gamma(r)}
		\sim r^{\bpconc + \bpdisc},
$$
it follows that
\begin{align*}
	\mbe[\sum_{k} \likeweight_{k}]
		\sim \bpmass \frac{\Gamma(1+\bpconc)}{\Gamma(1-\bpdisc) \Gamma(\bpconc + \bpdisc)}  \frac{\Gamma(1-\bpdisc)\Gamma(\bpconc+\bpdisc-1)}{\Gamma(\bpconc)} r 
		= \bpmass \frac{\bpconc}{\bpconc+\bpdisc-1} r,
\end{align*}
as was to be shown.
\end{proofof}

\begin{proofof}{\lem{Phi_r_bnbp}}
Given an atom $\bpweight_{k}$ of the beta process, the probability that the associated negative binomial count $\likeweight_{k}$ is non-zero is $1-(1-\bpweight_{k})^{r}$. It follows that
\begin{align*}
	\mbe[\sum_{k} \mbo\{\likeweight_{k} > 0\}]
		= \mbe[ \mbe[ \sum_{k} \mbo\{\likeweight_{k} > 0\} | \bpweight_{k} ] ] 
		= \mbe[ \sum_{k} 1-(1-\bpweight_{k})^{r} ] 
		= \int_{\bpweight} (1-(1-\bpweight)^{r}) \nu_{\bp}(d\bpweight),
\end{align*}
where $\nu_{\bp}$ is the intensity of beta process atoms $\{\bpweight_{k}\}$. For integer $r$, this integral was calculated by~\citet{broderick:2012:beta} to be $\sim \bpmass \bpconc \log(r)$.

Note that, in applying the result of \citet{broderick:2012:beta}, we are using the form of the negative binomial distribution to reinterpret the desired expectation as the expected number of features represented in a beta-Bernoulli process with $r$ draws from the same underlying base measure.

Now consider general $r > 1$. Let $r^{(0)} = \lfloor r \rfloor$ and $r^{(1)} = \lceil r \rceil$. Then
\begin{equation}
	\frac{ \int_{\bpweight} (1-(1-\bpweight)^{r^{(0)}}) \nu_{\bp}(d\bpweight) }{ \bpmass \bpconc \log(r^{(1)}) }
		\le \frac{ \int_{\bpweight} (1-(1-\bpweight)^{r}) \nu_{\bp}(d\bpweight) }{ \bpmass \bpconc \log(r) }
		\le \frac{ \int_{\bpweight} (1-(1-\bpweight)^{r^{(1)}}) \nu_{\bp}(d\bpweight) }{ \bpmass \bpconc \log(r^{(0)}) }
\end{equation}
by monotonicity.
Moreover,
\begin{align*}
	\frac{ \int_{\bpweight} (1-(1-\bpweight)^{r^{(0)}}) \nu_{\bp}(d\bpweight) }{ \bpmass \bpconc \log(r^{(1)}) } 
		&= \frac{ \int_{\bpweight} (1-(1-\bpweight)^{r^{(0)}}) \nu_{\bp}(d\bpweight) }{ \bpmass \bpconc \log(r^{(0)}) } \cdot
			\frac{ \bpmass \bpconc \log(r^{(0)}) }{ \bpmass \bpconc \log(r^{(1)}) } \\
		&\rightarrow 1, \quad r \rightarrow \infty.
\end{align*}
Similarly,
\begin{align*}
	\frac{ \int_{\bpweight} (1-(1-\bpweight)^{r^{(1)}}) \nu_{\bp}(d\bpweight) }{ \bpmass \bpconc \log(r^{(0)}) } 
		&\rightarrow 1, \quad r \rightarrow \infty
	\quad\quad\text{and hence}\quad\quad
	\frac{ \int_{\bpweight} (1-(1-\bpweight)^{r}) \nu_{\bp}(d\bpweight) }{ \bpmass \bpconc \log(r) }
		&\rightarrow 1, \quad r \rightarrow \infty.
\end{align*}
as was to be shown.
\end{proofof}

\begin{proofof}{\lem{Phi_r_3bnbp}}
	By the discussion in the previous proposition, this result follows from the results in~\citet{broderick:2012:beta}.
\end{proofof}

\begin{proofof}{\lem{Phi_j_r_bnbp}}
Given an atom $\bpweight_{k}$ of the beta process, the probability that the associated negative binomial count $\likeweight_{k}$ is equal to $j$ is $\negbin(j|r,\bpweight_{k})$. It follows that
\begin{align*}
	\mbe[\sum_{k} \mbo\{\likeweight_{k} = j\}]
		&= \mbe[ \mbe[ \sum_{k} \mbo\{\likeweight_{k} = j\} | \tbf{\bpweight}_{\cdot} ] ] 
		= \mbe[ \sum_{k} \negbin(j | r, \bpweight_{k})] 
		= \nu(j) 
		= \bpmass \bpconc \frac{\Gamma(j+r)}{\Gamma(j+1) \Gamma(r)} \frac{\Gamma(j) \Gamma(r+\bpconc)}{\Gamma(j+r+\bpconc)}
\end{align*}
as above.
Now we use
$
	\frac{\Gamma(r+\bpconc)}{\Gamma(r)} \sim r^{\bpconc}
$
and
$
	\frac{\Gamma(j+r)}{\Gamma(j+r+\bpconc)} \sim r^{-\bpconc}
$
to obtain
$
	\mbe[\sum_{k} \mbo\{\likeweight_{k} = j\}]
		\sim \bpmass \bpconc j^{-1}.
$
\end{proofof}

\begin{proofof}{\lem{Phi_j_r_3bnbp}}
As in the $\bnbp$ case, we have
\begin{align*}
	\mbe[\sum_{k} \mbo\{\likeweight_{k} = j\}]
		&= \nu(j)
		= \bpmass \frac{\Gamma(1+\bpconc)}{\Gamma(1-\bpdisc)\Gamma(\bpconc+\bpdisc)} \frac{\Gamma(j+r)}{\Gamma(j+1) \Gamma(r)} \frac{\Gamma(j - \bpdisc) \Gamma(r+\bpconc+\bpdisc)}{\Gamma(j+r+\bpconc)}
\end{align*}
Now we use
$
	\frac{\Gamma(r+\bpconc+\bpdisc)}{\Gamma(r)} \sim r^{\bpconc+\bpdisc}
$
and
$
	\frac{\Gamma(j+r)}{\Gamma(j+r+\bpconc)} \sim r^{-\bpconc}
$
to obtain
$
	\mbe[\sum_{k} \mbo\{\likeweight_{k} = j\}]
		\sim \bpmass \frac{\Gamma(1+\bpconc)}{\Gamma(1-\bpdisc)\Gamma(\bpconc+\bpdisc)} \frac{\Gamma(j-\bpdisc)}{\Gamma(j+1)} r^{\bpdisc}.
$
\end{proofof}

\begin{proofof}{\thm{Phi_xi_bnbp}}
Assume $\bpconc > 1$. We have from the previous discussion that
$
	\lim_{r \rightarrow \infty} \frac{\xi(r)}{\bpmass \frac{\bpconc}{\bpconc - 1} r} = 1.
$
So
$$
	\lim_{r \rightarrow \infty} \log(\xi(r)) - \log(r) = - \log\left( \bpmass \frac{\bpconc}{\bpconc - 1} \right).
$$
Hence
$
	\lim_{r \rightarrow \infty} \frac{\log(\xi(r))}{\log(r)} = 1
$
since $\log(r) \rightarrow \infty$ as $r \rightarrow \infty$.

From \lem{Phi_r_bnbp}, we also have
$
	\lim_{r \rightarrow \infty} \frac{\Phi(r)}{\bpmass \bpconc \log(r)} = 1.
$
Finally, then,
$$
	\lim_{r \rightarrow \infty} \frac{\Phi(r)}{\bpmass \bpconc \log(\xi(r))} = 1.
$$
\end{proofof}

\begin{proofof}{\thm{Phi_xi_3bnbp}}
	From above, we have
	$$
		\lim_{r \rightarrow \infty} \frac{\xi(r)}{\bpmass \frac{\bpconc}{\bpconc+\bpdisc-1} r} = 1
	\quad\quad\text{and hence}\quad\quad
		\lim_{r \rightarrow \infty} \frac{(\xi(r))^{\bpdisc}}{\left(\bpmass \frac{\bpconc}{\bpconc+\bpdisc-1} r\right)^{\bpdisc}} = 1.
	$$
	
	From \lem{Phi_r_3bnbp}, we also have
	$$
		\lim_{r \rightarrow \infty} \frac{K(r)}{\frac{\bpmass}{\bpdisc} \frac{\Gamma(\bpconc + 1)}{\Gamma(\bpconc + \bpdisc)} r^{\bpdisc}} \stackrel{a.s.}{=} 1
	\quad\quad\text{and hence}\quad\quad
		\lim_{r \rightarrow \infty} \frac{(\xi(r))^{\bpdisc} \frac{\bpmass}{\bpdisc} \frac{\Gamma(\bpconc + 1)}{\Gamma(\bpconc + \bpdisc)}}{\left(\bpmass \frac{\bpconc}{\bpconc+\bpdisc-1} \right)^{\bpdisc} K(r)} \stackrel{a.s.}{=} 1.
	$$
\end{proofof}

\begin{proofof}{\thm{Phi_j_xi_3bnbp}}
As above, we have from \lem{Phi_j_r_3bnbp} that
	$$
		\lim_{r \rightarrow \infty} \frac{\Phi_{j}(r)}{\bpmass \frac{\Gamma(1+\bpconc)}{\Gamma(1-\bpdisc)\Gamma(\bpconc+\bpdisc)} \frac{\Gamma(j-\bpdisc)}{\Gamma(j+1)} r^{\bpdisc}} = 1
	\quad\quad\text{and hence}\quad\quad
		\lim_{r \rightarrow \infty} \frac{(\xi(r))^{\bpdisc} \bpmass \frac{\Gamma(1+\bpconc)}{\Gamma(1-\bpdisc)\Gamma(\bpconc+\bpdisc)} \frac{\Gamma(j-\bpdisc)}{\Gamma(j+1)}}{\left(\bpmass \frac{\bpconc}{\bpconc+\bpdisc-1}\right)^{\bpdisc} \Phi_{j}(r)} = 1,
	$$
yielding the desired result.
\end{proofof}

\section{Conjugacy proofs} \label{app:conjugacy_proofs} 

\subsection{Reparameterized beta process and negative binomial process} \label{app:conj_params_bnbp}

\thm{rbp_nbp_conjugacy} in the main text is a corollary of \thms{finite_intens}
and \thmss{infty_intens} below.
In particular, \thms{finite_intens} and \thmss{infty_intens},
give us the form of the posterior process
when we have a general CRM prior with a Poisson process intensity 
with finite mean.  Choosing the particular Poisson process intensity 
for the $\rbp$ and choosing the distributions of the prior fixed weights 
yields the result.

\subsection{Finite Poisson process intensity}

\begin{theorem} \label{thm:finite_intens}
	Let $\bpdraw_{prior}$ be a discrete, completely random measure on $[0,1]$ with atom locations in $[0,1]$. Suppose it has the following components.
	\begin{itemize*}
		\item The ordinary component is generated from a Poisson point process with intensity $\nu(d\bpweight) \; d\atom$ such that $\nu$ is continuous and $\nu[0,1] < \infty$. In particular, the weights are in the $\bpweight$ axis, and the atom locations are in the $\atom$ axis.
		\item There are $L$ fixed atoms at locations $\fatom_{1},\ldots,\fatom_{L} \in [0,1]$. The weight of the $l$th fixed atom is a random variable with distribution $h_{l}$.
		\item There is no deterministic measure component.
	\end{itemize*}
	
	Draw a negative binomial process $\likedraw$ with shape parameter $r$ and input measure $\bpdraw_{prior}$. 	
	Let $K$ be the number of (nonzero) atoms of $\likedraw$.
	Let $\Pi = \{(\likeweight_{k}, \likeposatom_{k})\}_{k=1}^{K}$ be the pairs of observed nonzero counts and corresponding atom locations.
	
	Then the posterior process for the input measure to the negative binomial process given $\likedraw$ is a completely random measure $\bpdraw_{post}$ with the following components.
	\begin{itemize*}
		\item The ordinary component is generated from a Poisson point process with intensity
		$$
			(1-\bpweight)^{r} \nu(d\bpweight) \; d\atom.
		$$
		\item There are three sets of fixed atoms.
		\begin{enumerate*}
			\item There are the old, repeated fixed atoms. If $\fatom_{l} = \likeposatom_{k}$ for some $k$, there is a fixed atom at $\fatom_{l}$ with weight density
			$$
				c_{or}^{-1} (1-\bpweight)^{r} \bpweight^{\likeweight_{k}} h_{l}(d\bpweight) \; d\atom,
			$$
			where the $c_{or}$ is the normalizing constant:
			$$
				c_{or} = \int_{\atom=0}^{1} \int_{\bpweight=0}^{1} (1-\bpweight)^{r} \bpweight^{\likeweight_{k}} h_{l}(d\bpweight).
			$$
			\item There are the old, unrepeated fixed atoms. If $\fatom_{l} \notin \{\likeposatom_{1},\ldots,\likeposatom_{K}\}$, there is a fixed atom at $\fatom_{l}$ with weight density
			$$
				c_{ou}^{-1} (1-\bpweight)^{r} h_{l}(d\bpweight),
			$$
			where the $c_{or}$ is the normalizing constant:
			$$
				c_{ou} = \int_{\bpweight=0}^{1} (1-\bpweight)^{r} h_{l}(d\bpweight).
			$$
			\item There are the new fixed atoms. If $\likeposatom_{k} \notin \{\fatom_{1},\ldots,\fatom_{L}\}$, there is a fixed atom at $\likeposatom_{k}$ with weight density
			$$
				c_{new}^{-1} (1-\bpweight)^{r} \bpweight^{\likeweight_{k}} \nu(d\bpweight),
			$$
			where the $c_{new}$ is the normalizing constant:
			$$
				c_{new} = \int_{\bpweight=0}^{1} (1-\bpweight)^{r} \bpweight^{\likeweight_{k}} \nu(d\bpweight).
			$$
		\end{enumerate*}
		\item There is no deterministic measure component.
	\end{itemize*}
\end{theorem}

\begin{IEEEproof}
	Our proof follows the proof of beta-Bernoulli process conjugacy of \citet{Kim99}.
	Let $(\qmsp, \qmsigf)$ be the set of completely random measures on $[0,1]$ with weights in $[0,1]$ and its associated sigma algebra. Let $(\cmsp,\cmsigf)$ be the set of completely random measures on $[0,1]$ with atom weights in $\{1,2,\ldots\}$ and its associated sigma algebra. For any sets $\qmset \in \qmsigf$ and $\cmset \in \cmsigf$, let $\mathbb{P}_{prior}(\qmset \times \cmset)$ be the probability distribution induced on such sets by the 
construction of the prior measure $\bpdraw_{prior}$ and the negative binomial process $\likedraw$. Let $\mathbb{Q}(\qmset : \cmset)$ be the probability distribution induced on measures in $\qmsp$ by the proposed posterior distribution. Finally, let $\mathbb{P}_{marg}(\cmset)$ be the prior marginal distribution on counting measures in $\cmsp$. To prove the theorem, it is enough to show that, for any such sets $\qmset$ and $\cmset$, we have
\begin{equation} \label{eq:int_posterior}
	\mathbb{P}_{prior}(\qmset \times \cmset) = \int_{\likedraw \in \cmset} \mathbb{Q}(\qmset : \likedraw) \; \mathbb{P}_{marg}(\likedraw).
\end{equation}

The remainder of the proof will proceed as follows. We start by introducing some further notation. Then we will note that it is enough to prove \eq{int_posterior} for certain, restricted forms of the sets $\qmset$ and $\cmset$. Next, we will in turn find the form of each of (1) the prior distribution $\mbp_{prior}$, (2) the proposed posterior distribution $\mathbb{Q}$, and (3) the marginal count process distribution $\mbp_{marg}$ for our special sets of interest. Finally, we will show that we can integrate out the posterior with respect to the marginal in order to obtain the prior, as in \eq{int_posterior}.

Start by noting that we can write $\bpdraw_{prior}$ as
\begin{equation}
	\label{eq:mu_prior} 
	\bpdraw_{prior}(d\atom) = \sum_{j=1}^{J} \oweight_{j} \delta_{\oatom_{j}}(d\atom) + \sum_{l=1}^{L} \fweight_{l} \delta_{\fatom_{l}}(d\atom).
\end{equation}
Here, $J$ is the number of atoms in the ordinary component of $\bpdraw_{prior}$. So the total number of atoms in $\bpdraw_{prior}$ is $J + L$, and the total number of atoms in the counting measure with parameter $\bpdraw_{prior}$ is $K \le J+L$. The atom locations of the ordinary component are $\{\oatom_{j}\}$ since the fixed atom locations are at $\{\fatom_{l}\}$. Together, we have that the full set of atoms of the counting measure is some subset of the disjoint union $\{\likeposatom_{k}\}_{k=1}^{K} \subseteq \{\oatom_{j}\}_{j=1}^{J} \cup \{\fatom_{l}\}_{l=1}^{L}$. The atom weights at the fixed $\{\fatom_{l}\}$ locations are $\{\fweight_{l}\}$, and the atom weights at the ordinary component locations $\{\oatom_{j}\}$ are $\{\oweight_{j}\}$.

Let $\lambda = \nu[0,1]$, which we know to be finite by assumption. Then the number of atoms in the ordinary component is Poisson-distributed:
$$
	J \sim \pois(\lambda).
$$
And the $\{\oweight_{j}\}_{j=1}^{J}$ are independent and identically distributed random variables with values in $[0,1]$ such that each has density $\nu(d\bpweight) / \lambda$.

Next, we note that instead of general sets $\qmset$ and $\cmset$, we can restrict to sets of the form
\begin{align}
	\label{eq:finite_Mp}
	\qmset' &= \{J = \hat{J}\} \cap \bigcap_{j=1}^{\hat{J}} \{\oatom_{j} \le \hat{\oatom}_{j}, \oweight_{j} \le \hat{\oweight}_{j}\}_{j=1}^{\hat{J}} \cap \bigcap_{l=1}^{L} \{\fweight_{l} \le \hat{\fweight}_{l}\}. \\
	\label{eq:finite_Gp}
	\cmset' &= \{K = 1\} \cap \{\likeweight_{1} = \hat{\likeweight}_{1}, \likeposatom_{1} \le\hat{\likeposatom}_{1}\}. 
\end{align}
That is, in the random measure $\bpdraw_{prior}$ case, we consider a set with a fixed number $j$ of ordinary component atoms and with fixed upper bounds $\hat{\oatom}_{j}$, $\hat{\oweight}_{j}$, or $\hat{\fweight}_{l}$ on, respectively, the location of the ordinary component atoms, the weights of the ordinary component atoms, and the weights of the fixed atoms. In the counting measure $\likedraw$ case, we can restrict to a single atom with location at $\likeposatom_{1}$ and count equal to $\likeweight_{1} \in \{1,2,\ldots\}$.

With this notation and restriction in hand, we proceed to compute the prior, marginal, and posterior so that we may check whether \eq{int_posterior} holds.

\textbf{Prior.}
We first calculate the prior measure of set $\qmset'$. Recall that the number of atoms is Poisson-distributed:
\begin{equation}
	\label{eq:finite_prior_count}
	\mbp_{prior}(J = \hat{J}) = \frac{\lambda^{\hat{J}}}{\hat{J}!} e^{-\lambda}.
\end{equation}
Also, the locations of these atoms, given their number, is distributed as
\begin{equation}
	\label{eq:finite_prior_loc}
	\mathbb{P}_{prior}( \bigcap_{j=1}^{\hat{J}} \{\oatom_{j} \le \hat{\oatom}_{j}\} | J = \hat{J})
		= \hat{J}! \int_{\atom_{1}=0}^{\hat{\oatom}_{(1)}} \int_{\atom_{2}=\atom_{1}}^{\hat{\oatom}_{(2)}} \cdots \int_{\atom_{\hat{J}} = \atom_{\hat{J}-1}}^{\hat{\oatom}_{(\hat{J})}} \left( \prod_{j=1}^{\hat{J}} d\atom_{j} \right).
\end{equation}
Note that $\oatom_{(j)}$ denotes the $j$th order statistic of $\{\oatom_{j}\}_{j=1}^{\hat{J}}$, and the $\hat{J}!$ term results from enumerating the possible rearrangements of this set.
Finally, the sizes of the atoms, given their location and number, have the distribution
\begin{align}
	\mbp_{prior}\left(\bigcap_{j=1}^{J} \{\oweight_{j} \le \hat{\oweight}_{j}\} \cap \bigcap_{l=1}^{L} \{\fweight_{l} \le \hat{\fweight}_{l}\} | J =\hat{J}, \bigcap_{j=1}^{\hat{J}} \{\oatom_{j} \le \hat{\oatom}_{j}\} \right) 
		\label{eq:finite_prior_weight}
		&= \left[ \prod_{j=1}^{\hat{J}} \int_{\bpweight=0}^{\hat{\oweight}_{j}} \frac{\nu(d\bpweight)}{\lambda} \right] 
			\cdot \left[ \prod_{l=1}^{L} \int_{\bpweight=0}^{\hat{\fweight}_{l}} h_{l}(d\bpweight) \right].
\end{align}
Together, \eqs{finite_prior_count}, \eqss{finite_prior_loc}, and \eqss{finite_prior_weight} yield the prior probability of the set $\qmset'$ (\eq{finite_Mp}) describing the random measure $\bpdraw_{prior}$.

Next, we turn to the prior probability of the set $\cmset'$ describing the counting measure $\likedraw$. In this case, we condition on a particular measure $\mu \in \qmset'$. Now, in $\cmset'$, each counting measure $\likedraw$ has exactly one atom. This atom can occur either at an atom in the ordinary component of $\mu$, located at one of $\{\oatom_{j}\}_{j=1}^{J}$, or at a fixed atom of $\mu$, located at one of $\{\fatom_{l}\}_{l=1}^{L}$. We take advantage of the fact that the $\fatom_{l}$ are unique by assumption and that the $\oatom_{j}$ are a.s.\ unique and distinct from the $\fatom_{l}$ by the assumption that the distribution on locations is continuous. We also note that on the set $\{\likeposatom_{1} \le \hat{\likeposatom}_{1}\}$, we need only consider those atoms with locations at most $\hat{\likeposatom}_{1}$. Thus, we break into these two special cases as follows:
\begin{align*}
	\mbp_{prior}(K = 1, \likeweight_{1} = \hat{\likeweight}_{1}, \likeposatom_{1} \le \hat{\likeposatom}_{1} | \mu) 
		&= \sum_{j=1}^{J} \mbp_{prior}(K = 1, \likeweight_{1} = \hat{\likeweight}_{1}, \likeposatom_{1} = \oatom_{j} | \mu) \mbo\{\oatom_{j} \le \hat{\likeposatom}_{1}\} \\
		& {} + \sum_{l=1}^{L} \mbp_{prior}(K = 1, \likeweight_{1} = \hat{\likeweight}_{1}, \likeposatom_{1} = \fatom_{l} | \mu) \mbo\{\fatom_{l} \le \hat{\likeposatom}_{1}\}.
\end{align*}
The probability that the single nonzero count occurs at a particular atom is the probability that a nonzero count appears at this atom and zero counts appear at all other atoms. To express this probability, we first define a new function:
\begin{align*}
	\Phi(J, L, \tbf{\oatom}, \tbf{\oweight}, \tbf{\fweight}, \likeweight_{1}, \likeposatom)
		&= \left\{ \prod_{j=1}^{J} \left[ \negbin(0 | r, \oweight_{j}) \right]^{\mathbbm{1}\{\oatom_{j} \neq \likeposatom\}}
				\left[ \negbin(\likeweight_{1} | r, \oweight_{j}) \right]^{\mathbbm{1}\{\oatom_{j} = \likeposatom\}} \right\} \\
		& {} \cdot \left\{ \prod_{l=1}^{L} \left[ \negbin(0 | r, \fweight_{j}) \right]^{\mathbbm{1}\{\fatom_{l} \neq \likeposatom\}}
				\left[ \negbin(\likeweight_{1} | r, \fweight_{l}) \right]^{\mathbbm{1}\{\fatom_{l} = \likeposatom\}} \right\}.
\end{align*}
Here, $\negbin(x | a, b)$ is the negative binomial density. A notable special case is $\negbin(0 | a, b) = (1 - b)^{a}$.
We can write the single-atom probabilities with the $\Phi$ notation:
\begin{align*}
	\mbp_{prior}(K=1, \likeweight_{1}=\hat{\likeweight}_{1}, \likeposatom_{1}=\oatom_{j} | \mu)
		&= \Phi(J, L, \tbf{\oatom}, \tbf{\oweight}, \tbf{\fweight}, \likeweight_{1}, \oatom_{j}) \\
	\mbp_{prior}(K=1, \likeweight_{1}=\hat{\likeweight}_{1}, \likeposatom_{1}=\fatom_{l} | \mu)
		&= \Phi(J, L, \tbf{\oatom}, \tbf{\oweight}, \tbf{\fweight}, \likeweight_{1}, \fatom_{l}).
\end{align*}

We can combine the likelihood of the counting process $\likedraw$ given the random measure $\bpdraw_{prior}$ with the prior of the random measure $\bpdraw_{prior}$ to find the joint prior probability of the set $\qmset' \times \cmset'$. If we use the following notation to express the sets over which we will integrate,
\begin{align*}
	R(\tbf{\hat{\oatom}}, J) &\defeq \{\tbf{\atom} : \tbf{\atom} \in [0,1]^{J}, \atom_{1} \le \cdots \le \atom_{J}\} \cap \bigcap_{j=1}^{J} \{\atom : \atom_{j} \le \hat{\oatom}_{j}\} \\
	r(\tbf{T} = (t_{1},\ldots,t_{J}),J) &\defeq [0,t_{1}] \times \cdots \times [0,t_{J}]
\end{align*}
then we may write
\begin{align}
	\nonumber
	\mbp_{prior}(\qmset' \times \cmset')
		&= \int_{\bpdraw \in \qmset'} \mbp_{prior}(\cmset' | \bpdraw) \; d\mbp_{prior}(\bpdraw) \\
		\nonumber
		&= e^{-\lambda} \left\{ \sum_{j=1}^{\hat{J}} \left[
			\int_{\tbf{\oatom} \in R(\tbf{\hat{\oatom}}, \hat{J}), \tbf{\oweight} \in r(\tbf{\hat{\oweight}},\hat{J}), \tbf{\fweight} \in r(\tbf{\hat{\fweight}},L)}
			\mbo\{\oatom_{j} \le \hat{\likeposatom}_{1}\} \right. \right. \\
		\nonumber
		& {} \left. \cdot \Phi(\hat{J}, L, \tbf{\oatom}, \tbf{\oweight}, \tbf{\fweight}, \likeweight_{1}, \oatom_{j})
			\cdot \left( \prod_{j=1}^{\hat{J}} d\oatom_{j} \right) \cdot \left( \prod_{j=1}^{\hat{J}} \nu(d\oweight) \right)
			\cdot \left( \prod_{l=1}^{L} h_{l}(d\fweight_{l}) \right) \right] \\
		\nonumber
		& {} + \sum_{l=1}^{L} \left[
			\int_{\tbf{\oatom} \in R(\tbf{\hat{\oatom}}, \hat{J}), \tbf{\oweight} \in r(\tbf{\hat{\oweight}},\hat{J}), \tbf{\fweight} \in r(\tbf{\hat{\fweight}},L)}
			\mbo\{\fatom_{l} \le \hat{\likeposatom}_{1}\} \right. \\
		\label{eq:finite_joint_prior}
		& {} \left. \left. \cdot \Phi(\hat{J}, L, \tbf{\oatom}, \tbf{\oweight}, \tbf{\fweight}, \hat{\likeweight}_{1}, \fatom_{l})
			\cdot \left( \prod_{j=1}^{\hat{J}} d\oatom_{j} \right) \cdot \left( \prod_{j=1}^{\hat{J}} \nu(d\oweight) \right)
			\cdot \left( \prod_{l=1}^{L} h_{l}(d\fweight_{l}) \right) \right] \right\}.
\end{align}
This equation completes our prior calculation for now. We will return to it when we evaluate \eq{int_posterior} for sets $\qmset'$ and $\cmset'$.

\textbf{Proposed posterior.}
Next we consider the proposed posterior distribution $\mathbb{Q}$. Just as we calculated the probability of $\qmset' \times \cmset'$ under the measure induced by our prior generative model, we can analogously calculate the quantity $\mathbb{Q}(\qmset' : \likedraw)$ for some $\likedraw \in \cmset'$ according to the definition of $\mathbb{Q}$.

In the theorem statement, we specified a construction of completely random measure to induce the proposed posterior. In this case, the completely random measure has an ordinary component and a set of fixed atoms. Given the specific set $\cmset'$ we are considering (\eq{finite_Gp}), the set of locations of the fixed atoms is $\{\fatom_{1}, \ldots, \fatom_{L}\} \cup \{\hat{\likeposatom}_{1}\}$, where the union is not necessarily disjoint. So there are two cases we must examine: either the counting process atom is at the same location as a fixed atom of the prior random measure ($\hat{\likeposatom}_{1} = \fatom_{l}$ for some $l \in \{1,\ldots,L\}$) or it is at a different location ($\hat{\likeposatom}_{1} \notin \{\fatom_{1},\ldots,\fatom_{L}\}$).

First, we consider the case where the counting process atom location $\hat{\likeposatom}_{1}$ is the same as that of a fixed atom of the prior random measure, say $\fatom_{l^{*}}$. As before, the number of atoms in the ordinary component is Poisson-distributed with mean equal to the total Poisson point process mass
$$
	\lambda_{post} \defeq \int_{\bpweight=0}^{1} (1-\bpweight)^{r} \nu(d\bpweight).
$$
So we have (c.f.\ \eq{finite_prior_count})
\begin{equation}
	\label{eq:finite_post_count_fixed}
	\mathbb{Q}(J = \hat{J} : K = 1, \likeweight_{1} = \hat{\likeweight}_{1}, \likeposatom_{1} = \fatom_{l^{*}})
		= \frac{\lambda_{post}^{\hat{J}}}{\hat{J}!} e^{-\lambda_{post}}.
\end{equation}
Also, as for \eq{finite_prior_loc}, we can calculate the distribution of the locations of the ordinary component atoms:
\begin{align}
	\mathbb{Q}(\bigcap_{j=1}^{\hat{J}} \{\oatom_{j} \le \hat{\oatom}_{j}\}
		| J = \hat{J} : K = 1, \likeweight_{1} = \hat{\likeweight}_{1}, \likeposatom_{1} = \fatom_{l^{*}})
		\label{eq:finite_post_loc_fixed}
		&= \hat{J}! \int_{\atom_{1}=0}^{\hat{\oatom}_{(1)}} \int_{\atom_{2}=\atom_{1}}^{\hat{\oatom}_{(2)}} \cdots \int_{\atom_{\hat{J}} = \atom_{\hat{J}-1}}^{\hat{\oatom}_{(\hat{J})}} \left( \prod_{j=1}^{\hat{J}} d\atom_{j} \right).
\end{align}
And again, as in \eq{finite_prior_weight}, the sizes of the atoms, given their location and number, have the distribution
\begin{align}
	\nonumber
	\lefteqn{
	\mathbb{Q}\left(\bigcap_{j=1}^{J} \{\oweight_{j} \le \hat{\oweight}_{j}\} \cap \bigcap_{l=1}^{L} \{\fweight_{l} \le \hat{\fweight}_{l}\} | J =\hat{J}, \bigcap_{j=1}^{\hat{J}} \{\oatom_{j} \le \hat{\oatom}_{j}\}
		: K = 1, \likeweight_{1} = \hat{\likeweight}_{1}, \likeposatom_{1} = \fatom_{l^{*}} \right) } \\
		&= \left[ \prod_{j=1}^{\hat{J}} \int_{\bpweight=0}^{\hat{\oweight}_{j}}
			\frac{\negbin(0 | r, \bpweight) \nu(d\bpweight)}{\lambda_{post}} \right] 
		\label{eq:finite_post_weight_fixed}
		\left[
			\prod_{l=1}^{L} \frac{
				\int_{b=0}^{\hat{\fweight}_{l}} [\negbin(\hat{\likeweight}_{1} | r, \bpweight)]^{\mbo\{l = l^{*}\}}
				[\negbin(0 | r, \bpweight)]^{\mbo\{l \ne l^{*}\}} h_{l}(d\bpweight)
			}{
				\int_{\bpweight=0}^{1} [\negbin(\hat{\likeweight}_{1} | r, \bpweight)]^{\mbo\{l = l^{*}\}}
				[\negbin(0 | r, \bpweight)]^{\mbo\{l \ne l^{*}\}} h_{l}(d\bpweight)
			}
			\right].
\end{align}
Putting together \eqs{finite_post_count_fixed}, \eqss{finite_post_loc_fixed}, and \eqss{finite_post_weight_fixed}, we can find the proposed measure of the set $\qmset'$ given $\likedraw \in \cmset'$ for the case $\hat{\likeposatom}_{1} = \fatom_{l^{*}}$:
\begin{align}
	\nonumber
	\mathbb{Q}(\qmset' : \likedraw) 
		\nonumber
		&= \mathbb{Q}\left(J = \hat{J}, \bigcap_{j=1}^{\hat{J}} \{\oatom_{j} \le \hat{\oatom}_{j}\}, 
			\bigcap_{j=1}^{J} \{\oweight_{j} \le \hat{\oweight}_{j}\} \cap \bigcap_{l=1}^{L} \{\fweight_{l} \le \hat{\fweight}_{l}\}
			: K = 1, \likeweight_{1} = \hat{\likeweight}_{1}, \likeposatom_{1} = \fatom_{l^{*}} \right) \\
		&= C_{fixed, l^{*}}^{-1} e^{-\lambda_{post}}
			\int_{\tbf{\oatom} \in R(\tbf{\hat{\oatom}}, \hat{J}), \tbf{\oweight} \in r(\tbf{\hat{\oweight}}, \hat{J}), \tbf{\fweight} \in r(\tbf{\hat{\fweight}}, L)}
			\Phi(\hat{J}, L, \tbf{\oatom}, \tbf{\oweight}, \tbf{\fweight}, \likeweight_{1}, \fatom_{l^{*}}) \\
		\label{eq:finite_post_fixed}
		& \quad {} \cdot \left( \prod_{j=1}^{\hat{J}} d\oatom_{j} \right) \cdot \left( \prod_{j=1}^{\hat{J}} \nu(d\oweight) \right)
			\cdot \left( \prod_{l=1}^{L} h_{l}(d\fweight_{l}) \right),
\end{align}
where
$$
	C_{fixed, l^{*}}
		\defeq \prod_{l=1}^{L} \int_{\bpweight=0}^{1} [\negbin(\hat{\likeweight}_{1} | r, \bpweight)]^{\mbo\{l = l^{*}\}}
			[\negbin(0 | r, \bpweight)]^{\mbo\{l \ne l^{*}\}} h_{l}(d\bpweight).
$$

Second, we consider the case $\hat{\likeposatom}_{1} \notin \{\fatom_{1},\ldots,\fatom_{L}\}$. Then $\hat{\likeposatom}_{1} = \oatom_{j^{*}}$ for some $j^{*} \in \{1,\ldots,J\}$. Suppose that $\oatom_{j^{*}}$ is the $j_{order}$th smallest element of $\{\oatom_{1},\ldots,\oatom_{J}\}$. Note that $j_{order}$ is well-defined since the density of the $\oatom_{j}$ is continuous.
We proceed as above and start by noting that the number of atoms on either side of the location $\oatom_{j^{*}}$ is Poisson-distributed:
\begin{align}
	\nonumber
	\lefteqn{ \mathbb{Q}\left( J = \hat{J} : K = 1, \likeweight_{1} = \hat{\likeweight}_{1}, \likeposatom_{1} = \oatom_{j^{*}} \right) } \\
		\label{eq:finite_post_count_ord}
		&= \frac{\left( \lambda_{post} \oatom_{j^{*}} \right)^{j_{order}-1}}{(j_{order} - 1)!}
			e^{-\left( \lambda_{post} \oatom_{j^{*}} \right)}
			\cdot \frac{\left( \lambda_{post} (1 - \oatom_{j^{*}}) \right)^{(\hat{J} - j_{order})}}{(\hat{J} - j_{order})!}
			e^{-\left( \lambda_{post} (1 - \oatom_{j^{*}}) \right)}.
\end{align}
Further, we have the usual distribution for the atom locations on either side of $\oatom_{j^{*}}$:
\begin{align}
	\nonumber
	\lefteqn{ \mathbb{Q}\left( \bigcap_{j=1}^{\hat{J}} \{\oatom_{j} \le \hat{\oatom}_{j}\} | J = \hat{J}
		: K = 1, \likeweight_{1} = \hat{\likeweight}_{1}, \likeposatom_{1} = \oatom_{j^{*}} \right) } \\
		\nonumber
		&= (j_{order} - 1)! \int_{\atom_{1} = 0}^{\hat{\oatom}_{(1)}} \int_{\atom_{2} = \atom_{1}}^{\hat{\oatom}_{(2)}}
			\cdots \int_{\atom_{j_{order}} = \atom_{j_{order}-1}}^{\hat{\oatom}_{(j_{order})}}
			\left( \prod_{j=1}^{j_{order}-1} \frac{d\atom_{j}}{\oatom_{j^{*}}} \right) \\
		\label{eq:finite_post_loc_ord}
		& {} \cdot (\hat{J} - j_{order})! \int_{\atom_{j_{order} + 1} = \hat{\oatom}_{(j_{order})}}^{\hat{\oatom}_{(j_{order} + 1)}}
			\cdots \int_{\atom_{\hat{J}} = \atom_{\hat{J} - 1}}^{\hat{\oatom}_{(\hat{J})}}
			\left( \prod_{j=j_{order}+1}^{\hat{J}} \frac{d\atom_{j}}{1 - \oatom_{j^{*}}} \right).
\end{align}
As usual, the third step identifies the conditional weight distribution of the atom weights:
\begin{align}
	\lefteqn{ \mathbb{Q}\left( \bigcap_{j=1}^{J} \{\oweight_{j} \le \hat{\oweight}_{j}\}
	\cap \bigcap_{l=1}^{L} \{\fweight_{l} \le \hat{\fweight}_{l}\}
	| J =\hat{J}, \bigcap_{j=1}^{\hat{J}} \{\oatom_{j} \le \hat{\oatom}_{j}\} 
		: K = 1, \likeweight_{1} = \hat{\likeweight}_{1}, \likeposatom_{1} = \oatom_{j^{*}} \right) } \\
		\label{eq:finite_post_weight_ord}
		&= \left[ \prod_{j=1}^{\hat{J}} \frac{
				\int_{\bpweight=0}^{\hat{\oweight}_{j}} \left[\negbin(\hat{\likeweight}_{1} | r, \bpweight)\right]^{\mbo\{j = j^{*}\}}
				\left[\negbin(0 | r, \bpweight)\right]^{\mbo\{j \ne j^{*}\}} \nu(d\bpweight)
			}{
				\int_{\bpweight=0}^{1} \left[\negbin(\hat{\likeweight}_{1} | r, \bpweight)\right]^{\mbo\{j = j^{*}\}}
				\left[\negbin(0 | r, \bpweight)\right]^{\mbo\{j \ne j^{*}\}} \nu(d\bpweight)
			} \right] 
		\left[
			\prod_{l=1}^{L} \frac{
				\int_{\bpweight=0}^{\hat{\fweight}_{l}} \negbin(0 | r, \bpweight) h_{l}(d\bpweight)
			}{
				\int_{\bpweight=0}^{1} \negbin(0 | r, \bpweight) h_{l}(d\bpweight) 
			} \right].
\end{align}
So, combining \eqs{finite_post_count_ord}, \eqss{finite_post_loc_ord}, and \eqss{finite_post_weight_ord}, we find that the proposed posterior distribution in the case $\hat{\likeposatom}_{1} = \oatom_{j^{*}}$ is
\begin{align}
	\nonumber
	\mathbb{Q}(\qmset' : \likedraw) 
		\nonumber
		&= \mathbb{Q}\left(J = \hat{J}, \bigcap_{j=1}^{\hat{J}} \{\oatom_{j} \le \hat{\oatom}_{j}\}, 
			\bigcap_{j=1}^{J} \{\oweight_{j} \le \hat{\oweight}_{j}\} \cap \bigcap_{l=1}^{L} \{\fweight_{l} \le \hat{\fweight}_{l}\}
			: K = 1, N_{1} = n_{1}, S_{1} = \oweight_{j^{*}} \right) \\
		\nonumber
		&= C_{ord}^{-1} e^{-\lambda_{post}}
			\int_{\tbf{\oatom} \in R(\tbf{\hat{\oatom}}, \hat{J}), \oweight \in r(\tbf{\hat{\oweight}}, \hat{J}), \fweight \in r(\tbf{\hat{\fweight}}, L)}
			\Phi(\hat{J}, L, \tbf{\oatom}, \tbf{\oweight}, \tbf{\fweight}, \hat{\likeweight}_{1}, \oatom_{j^{*}}) \\
		\label{eq:finite_post_ord}
		& \quad {} \cdot \left( \prod_{j=1}^{\hat{J}} d\oatom_{j} \right) \cdot \left( \prod_{j=1}^{\hat{J}} \nu(d\oweight) \right)
			\cdot \left( \prod_{l=1}^{L} h_{l}(d\fweight_{l}) \right),
\end{align}
where
$$
	C_{ord}
		\defeq
			\left(
				\int_{b=0}^{1} \negbin(\hat{\likeweight}_{1} | r, \bpweight) \nu(d\bpweight)
			\right)
			\cdot \left(
				\prod_{l=1}^{L} \int_{\bpweight=0}^{1} \negbin(0 | r, \bpweight) h_{l}(d\bpweight)
			\right).
$$

Putting together the cases $\hat{\likeposatom}_{1} = \fatom_{l^{*}}$ for some $l^{*}$ (\eq{finite_post_fixed}) and $\hat{\likeposatom}_{1} \notin \{\fatom_{1},\ldots,\fatom_{L}\}$ (\eq{finite_post_ord}), we obtain the full proposed posterior distribution:
\begin{align}
	\nonumber
	\mathbb{Q}(\qmset' : \likedraw) 
		\nonumber
		&= \mathbb{Q}\left(J = \hat{J}, \bigcap_{j=1}^{\hat{J}} \{\oatom_{j} \le \hat{\oatom}_{j}\}, 
			\bigcap_{j=1}^{J} \{\oweight_{j} \le \hat{\oweight}_{j}\} \cap \bigcap_{l=1}^{L} \{\fweight_{l} \le \hat{\fweight}_{l}\}
			: K = 1, \likeweight_{1} = \hat{\likeweight}_{1}, \likeposatom_{1} = \hat{\likeposatom}_{1} \right) \\
		\nonumber
		&= \sum_{l^{*}=1}^{L} \mbo\{\hat{\likeposatom}_{1} = \fatom_{l^{*}}\} \; C_{fixed, l^{*}}^{-1} e^{-\lambda_{post}}
			\int_{\tbf{\oatom} \in R(\tbf{\hat{\oatom}}, \hat{J}), \tbf{\oweight} \in r(\tbf{\hat{\oweight}}, \hat{J}), \tbf{\fweight} \in r(\tbf{\hat{\fweight}}, L)}
			\Phi(\hat{J}, L, \tbf{\oatom}, \tbf{\oweight}, \tbf{\fweight}, \hat{\likeweight}_{1}, \fatom_{l^{*}}) \\
		\nonumber
		& \quad {} \cdot \left( \prod_{j=1}^{\hat{J}} d\oatom_{j} \right) \cdot \left( \prod_{j=1}^{\hat{J}} \nu(d\oweight) \right)
			\cdot \left( \prod_{l=1}^{L} h_{l}(d\fweight_{l}) \right) \\
		\nonumber
		& \quad {} + \mbo\left\{ \hat{\likeposatom}_{1} \notin \{\fatom_{1},\ldots,\fatom_{L}\} \right\} \; C_{ord}^{-1} e^{-\lambda_{post}}
			\int_{\tbf{\oatom} \in R(\tbf{\hat{\oatom}}, \hat{J}), \tbf{\oweight} \in r(\tbf{\hat{\oweight}}, \hat{J}), \fweight \in r(\tbf{\hat{\fweight}}, L)}
			\Phi(\hat{J}, L, \tbf{\oatom}, \tbf{\oweight}, \tbf{\fweight}, \hat{\likeweight}_{1}, \oatom_{j^{*}}) \\
		\label{eq:finite_proposed_post}
		& \quad {} \cdot \left( \prod_{j=1}^{\hat{J}} d\oatom_{j} \right) \cdot \left( \prod_{j=1}^{\hat{J}} \nu(d\oweight) \right)
			\cdot \left( \prod_{l=1}^{L} h_{l}(d\fweight_{l}) \right).
\end{align}

\textbf{Counting process marginal.}
With the prior and proposed posterior in hand, it remains to calculate the marginal distribution of the counting process. Then we may integrate out the proposed posterior with respect to the counting process marginal in order to obtain the prior (\eq{int_posterior}). Since we are focusing on counting process sets $G'$ of the form in \eq{finite_Gp}, we aim to calculate
$$
	\mbp_{marg}(K = 1, \likeweight_{1} = \hat{\likeweight}_{1}, \likeposatom_{1} \le \hat{\likeposatom}_{1}).
$$

In our calculations above, we also worked with a set of prior measure $\mu \in \qmset'$ and therefore worked with a set of locations for the ordinary component atoms. In this case, we will need to calculate the probability of zero counts in an interval where the number and location of the ordinary component atoms is integrated out. Let $\likedraw'\{\atom\}$ be the counting process that includes exactly those counts at ordinary component atoms and not the counts at fixed atoms; we can see, e.g., that $\likedraw'\{\atom\} \le \likedraw\{\atom\}$ at all $\atom$. Further, similar to \eq{mu_prior}, let $\bpdraw_{ord}$ be the random measure composed only of those atoms in the ordinary component of $\bpdraw_{prior}$:
$$
	\bpdraw_{ord} = \sum_{j=1}^{J} \oweight_{j} \delta_{\oatom_{j}}.
$$
Then we are interested in the quantity:
\begin{align*}
	\mbe\left[ \mbo\{\forall t \in (\atom_{1}, \atom_{2}), \likedraw'\{t\} =  0\} \right] 
		&= \mbe\left[ \prod_{t \in (\atom_{1},\atom_{2})} (1 - \bpdraw_{ord}\{t\})^{r} \right] 
		= \prod_{t \in (\atom_{1},\atom_{2})} \left(1 - \mbe\left[ 1 - (1 - \bpdraw_{ord}\{t\})^{r} \right] \right),
\end{align*}
where the last equality follows from the independence of $\bpdraw_{prior}$ across increments.

Now define a new process $\bpdraw' \defeq 1 - (1 - \bpdraw_{ord})^{r}$. This process has intensity $\nu'$, which can be obtained by a change of variables from the Poisson process intensity $\nu$ of $\bpdraw_{ord}$. We will find it notationally useful to refer to $\nu'$ though we do not calculate it here. Also, let $\bar{\bpdraw'}$ be the mean process of $\bpdraw'$: $\bar{\bpdraw'}(d\atom) \defeq \mbe\left[ \bpdraw'(d\atom) \right]$. With this notation in hand, we can write
\begin{align*}
	\lefteqn{ \mbe\left[ \mbo\{\forall t \in (\atom_{1},\atom_{2}), \bpdraw'\{t\} =  0\} \right] } \\
		&= \prod_{t \in (\atom_{1},\atom_{2})} \left(1 - \bar{\bpdraw'}\{t\} \right)
			= \exp\left\{ - \int_{t=\atom_{1}}^{\atom_{2}} \bar{\bpdraw'}\{t\} \right\}
			= \exp\left\{ - \int_{t=\atom_{1}}^{\atom_{2}} \int_{\bpweight=0}^{1} \bpweight' \; \nu'(d\bpweight') \right\} \\
		&= \exp\left\{ - (\atom_{2} - \atom_{1}) \int_{\bpweight=0}^{1} (1 - (1 - \bpweight)^{r}) \; \nu(d\bpweight) \right\}.
\end{align*}

As usual, we consider two separate cases. First, suppose $\likeposatom_{1} = \fatom_{l^{*}}$ for some $l^{*} \in \{1,\ldots,L\}$. Then, using independence of increments of the prior random measure and counting process, we find
\begin{align}
	\nonumber
	\mbp_{marg}(K = 1, \likeweight_{1} = \hat{\likeweight}_{1}, \likeposatom_{1} = \fatom_{l^{*}})
		\nonumber
		&= \mbp_{marg}(\likedraw\{\fatom_{l^{*}}\} = \hat{\likeweight}_{1})\ \mbp_{marg}(\forall l \ne l^{*}, \likedraw\{\fatom_{l}\} = 0) 
		\nonumber
		\ \mbp_{marg}(\forall t \in (0, 1), \likedraw'\{t\} =  0) \\
		\nonumber
		&= \left( \prod_{l=1}^{L} \int_{\bpweight=0}^{1} \left[ \negbin(0 | r, \bpweight) \right]^{\mbo\{l \ne l^{*}\}} \left[ \negbin(\hat{\likeweight}_{1} | r, \bpweight) \right]^{\mbo\{l = l^{*}\}} h_{l}(d\bpweight) \right) \\
		\nonumber
		& \quad {} \cdot \exp\left\{ - (1 - 0) \int_{\bpweight=0}^{1} (1 - (1 - \bpweight)^{r}) \; \nu(d\bpweight) \right\} \\
		\label{eq:finite_marg_fixed}
		&= e^{-\lambda + \lambda_{post}} C_{fixed, l^{*}}.
\end{align}

Next, suppose $\likeposatom_{1} \notin \{\fatom_{1}, \ldots, \fatom_{L}\}$. Then
\begin{align}
	\nonumber
	 \mbp_{marg}(K = 1, \likeweight_{1} = \hat{\likeweight}_{1}, \likeposatom_{1} \le \hat{\likeposatom}_{1}) 
		\nonumber
		&= \int_{\atom = 0}^{\hat{\likeposatom}_{1}} \mbp_{marg}(K = 1, \likeweight_{1} = \hat{\likeweight}_{1} | \likeposatom_{1} = \hat{\likeposatom}) d\mbp_{marg}(\likeposatom_{1} \le \atom) \\
		\nonumber
		&= \int_{\atom = 0}^{\hat{\likeposatom}_{1}} \mbp_{marg}(\likedraw\{\atom\} = \hat{\likeweight}_{1}) \mbp_{marg}(\forall l, \likedraw(\fatom_{l}) = 0) \\
		\nonumber
		& \quad {} \cdot \mbp_{marg}(\forall t \in (0,1) \backslash \{\atom\}, \likedraw'\{t\} =  0) d\mbp_{marg}(\likeposatom_{1} \le \atom) \\
		&= \int_{\atom = 0}^{\hat{\likeposatom}_{1}} \left( \int_{\bpweight=0}^{1} \negbin(\hat{\likeweight}_{1} | r, \bpweight) \nu(d\bpweight) \right) 
		\label{eq:finite_marg_ord}
  	   \left[ \prod_{l=1}^{L} \int_{\bpweight=0}^{1} \negbin(0 | r, \bpweight) h_{l}(d\bpweight) \right] \; d\atom.
\end{align}

\textbf{Checking integration.}
The final step is to note that we may integrate out the proposed posterior in \eq{finite_proposed_post} with respect to the marginal described by \eqs{finite_marg_fixed} and \eqss{finite_marg_ord} to obtain the joint prior in \eq{finite_joint_prior}. This integration is exactly the one we desired from \eq{int_posterior} in the special case of sets of the form $\qmset'$ in \eq{finite_Mp} and $\cmset'$ in \eq{finite_Gp}, as was to be shown.
\end{IEEEproof}

\subsection{Infinite Poisson process intensity}

\begin{theorem}
	\label{thm:infty_intens}
	\thm{finite_intens} still applies when the intensity measure $\nu$ does not necessarily have a finite integral $\nu[0,1]$ but satisfies the (weaker) condition
	\begin{equation}
		\label{eq:infty_finite_mean}
		\int_{\bpweight=0}^{1} \bpweight \; \nu(d\bpweight) < \infty.
	\end{equation}
\end{theorem}

\begin{IEEEproof}
The main idea behind the proof of \thm{infty_intens} is to take advantage of the finiteness condition in \eq{infty_eps_finite} to construct a sequence of finite intensity measures tending to the true intensity measure of the process. We will use the known form of the posterior in the finite case from \thm{finite_intens} to deduce the form of the posterior in the case where $\nu$ merely satisfies the weaker condition in \eq{infty_finite_mean}, which we note implies
\begin{equation}
	\label{eq:infty_eps_finite}
	\forall \epsilon > 0, \nu[\epsilon, \infty) < \infty.
\end{equation}
We therefore start by defining the sequence of (finite) measures $\nu_{n}$ by
\begin{equation}
	\label{eq:nu_trunc}
	\nu_{n}(A) \defeq \int_{\bpweight \in A} \mbo\{ \bpweight > 1/n \} \nu(d\bpweight), 
		\quad \textrm{ for all measurable $A \subset [0,1]$}.
\end{equation}
Further, we may generate a random measure $\bpdraw_{prior, n}$ as described by the prior in \thm{finite_intens} with Poisson point process intensity $\nu_{n}$. And we may generate a counting process $\likedraw_{n}$ with parameters $r$ and $\bpdraw_{prior,n}$ as described in \thm{finite_intens}.

As before, let $\mbp_{prior}$ be the prior distribution on the prior random measure $\bpdraw_{prior}$ and the counting process $\likedraw$. Let $\mbe_{prior}$ denote the expectation with respect to this distribution. Further, let $\mbp_{marg}$ represent the marginal distribution on the counting process from $\mbp_{prior}$. And let $\mathbb{Q}( \qmset : \cmset )$ represent the proposed posterior distribution on sets $\qmset \in \qmsp$ given any set $\cmset \in \cmsigf$. We use the same notation, but with $n$ subscripts, to denote the case with finite intensity $\nu_{n}$.

Our proof will take advantage of Laplacian-style characterizations of distributions. In particular, we note that in order to prove \thm{infty_intens}, it is enough to show that, for arbitrary continuous and nonnegative functions $f$ and $g$ (i.e., $f, g \in C^{+}[0,1]$), we have
\begin{align}
	\nonumber
	\lefteqn{ \int_{\bpdraw \in \qmsp} \int_{\likedraw \in \cmsp} 
		\exp\left\{ -\int_{\atom=0}^{1} ( g(\atom) \bpdraw\{\atom\} + f(\atom) \likedraw\{\atom\} ) \right\}
		\; d\mathbb{Q}( \bpdraw : \likedraw ) \; d\mbp_{marg}(\likedraw) } \\
		\label{eq:goal_laplace}
		&= \mbe_{prior} \left[ \exp\left\{ - \int_{\atom=0}^{1} ( g(\atom) \bpdraw\{\atom\} + f(\atom) \likedraw\{\atom\} ) \right\} \right].
\end{align}

By \lem{infty_conv_dist}, we have the following limit for all $f, g \in C^{+}[0,1]$ as $n \rightarrow \infty$:
\begin{align}
	\nonumber
	\lefteqn{ \mbe_{prior,n} \left[ \exp\left\{ - \int_{\atom=0}^{1} ( g(\atom) \bpdraw\{\atom\} + f(\atom) \likedraw\{\atom\}) ) \right\} \right] } \\
		\nonumber
		&\rightarrow \mbe_{prior} \left[ \exp\left\{ - \int_{\atom=0}^{1} ( g(\atom) \bpdraw\{\atom\} + f(\atom) \likedraw\{\atom\} ) \right\} \right].
\end{align}
Therefore, by \eq{goal_laplace} and the observation that \thm{finite_intens} holds under the finite intensity $\nu_{n}$, we see that it is enough to show that
\begin{align}
	\nonumber
	\lefteqn{ \int_{\bpdraw \in \qmsp_{n}} \int_{\likedraw \in \cmsp} 
		\exp\left\{ -\int_{\atom=0}^{1} ( g(\atom) \bpdraw\{\atom\} + f(\atom) \likedraw\{\atom\} ) \right\}
		\; d\mathbb{Q}_{n}( \bpdraw : \likedraw ) \; d\mbp_{marg,n}(\likedraw) } \\
		\label{eq:infty_laplace_lim}
		&\rightarrow \int_{\bpdraw \in \qmsp} \int_{\likedraw \in \cmsp} 
			\exp\left\{ -\int_{\atom=0}^{1} ( g(\atom) \bpdraw\{\atom\} + f(\atom) \likedraw\{\atom\} ) \right\}
			\; d\mathbb{Q}( \bpdraw : \likedraw ) \; d\mbp_{marg}(\likedraw), \quad n \rightarrow \infty.
\end{align}

Define
\begin{align}
	\label{eq:infty_def_psi_n}
	\Psi_{n}(\likedraw)
		&\defeq \int_{\bpdraw \in \qmsp_{n}} \exp\left\{ - \int_{\atom=0}^{1} g(\atom) \bpdraw\{\atom\} \right\}
			d\mathbb{Q}_{n}( \bpdraw : \likedraw ) \\
	\label{eq:infty_def_psi}
	\Psi(\likedraw)
		&\defeq \int_{\bpdraw \in \qmsp} \exp\left\{ - \int_{\atom=0}^{1} g(\atom) \bpdraw\{\atom\} \right\}
			d\mathbb{Q}( \bpdraw : \likedraw ).
\end{align}

By \lem{infty_psi_n_psi}, we have
\begin{equation}
	\label{eq:infty_psi_diff_lim}
	\int_{\likedraw \in \cmsp} \exp\left\{ - \int_{\atom=0}^{1} f(\atom) \likedraw\{\atom\} \right\}
		(\Psi_{n}(\likedraw) - \Psi(\likedraw)) d\mbp_{marg,n}(\likedraw)
		\rightarrow 0.
\end{equation}
And \lem{infty_conv_dist} together with the fact that $\exp\left\{ - \int_{\atom=0}^{1} f(\atom) \likedraw\{\atom\} \right\} \Psi(\likedraw)$ is a bounded function of $\likedraw$ yields
\begin{equation}
	\label{eq:infty_pmarg_diff_lim}
	\int_{\likedraw \in \cmsp} \exp\left\{ - \int_{\atom=0}^{1} f(\atom) \likedraw\{\atom\} \right\}
		\Psi(\likedraw) (d\mbp_{marg,n}(\likedraw) - d\mbp_{marg}(\likedraw))
		\rightarrow 0.
\end{equation}

Combining \eqs{infty_psi_diff_lim} and \eqss{infty_pmarg_diff_lim} yields the desired limit in
\eq{infty_laplace_lim}.

\end{IEEEproof}

\begin{lemma} \label{lem:infty_conv_dist}
	Let $\bpdraw_{prior,n}$ be a completely random measure with a finite set of fixed atoms in $[0,1]$
	and with the Poisson process
	intensity $\nu_{n}$ in \eq{nu_trunc}, where $\nu$ satisfies \eq{infty_finite_mean}. Let $\likedraw_{n}$ be drawn as a negative binomial process
	with parameters $r$ and $\bpdraw_{prior,n}$. Similarly, let $\bpdraw_{prior}$ be a completely 
	random measure with Poisson process intensity $\nu$, and let $\likedraw$ be drawn as a negative binomial
	process with parameters $r$ and $\bpdraw_{prior}$. Then
	$$
		(\bpdraw_{prior,n}, \likedraw_{n}) \stackrel{d}{\rightarrow} (\bpdraw_{prior}, \likedraw)
	$$
\end{lemma}

\begin{IEEEproof}
It is enough to show that, for all $f,g \in C^{+}[0,1]$, we have
\begin{align}
	\nonumber
	\lefteqn{ \mbe_{prior,n} \left[ \exp\left\{ - \int_{\atom=0}^{1} ( g(\atom) \bpdraw_{prior,n} \{\atom\} + f(\atom) \likedraw_{n} \{\atom\} ) \right\} \right] } \\
		\nonumber
		&\rightarrow \mbe_{prior} \left[ \exp\left\{ - \int_{\atom=0}^{1} ( g(\atom) \bpdraw_{prior} \{\atom\} + f(\atom) \likedraw\{\atom\} ) \right\} \right],
			\quad n \rightarrow 0.
\end{align}

We can construct a new completely random measure, $\hat{\bpdraw}_{n}$, by keeping only those jumps from $\bpdraw_{prior}$ (generated with intensity $\nu$) that are either at the fixed atom locations or have height at least $1/n$. Then $\hat{\bpdraw}_{n} \eqd \bpdraw_{prior,n}$ for $\bpdraw_{prior,n}$ generated with intensity $\nu_{n}$. Let $\hat{\likedraw}_{n}$ be the counting process generated with parameters $r$ and $\hat{\bpdraw}_{n}$. Then it is enough to show
\begin{align}
	\nonumber
	\lefteqn{ \mbe_{prior} \left[ \exp\left\{ - \int_{\atom=0}^{1} ( g(\atom) \hat{\bpdraw}_{n}\{\atom\} + f(\atom) \hat{\likedraw}_{n}\{\atom\} ) \right\} \right] } \\
		\nonumber
		&\rightarrow \mbe_{prior} \left[ \exp\left\{ - \int_{\atom=0}^{1} ( g(\atom) \bpdraw_{prior} \{\atom\} + f(\atom) \likedraw\{\atom\} ) \right\} \right],
			\quad n \rightarrow 0.
\end{align}

Let $\hat{\bpdraw}_{n}^{-} = \bpdraw_{prior} - \hat{\bpdraw}_{n}$ be the completely random measure consisting only of an ordinary component with jumps of size less than $1/n$. Let $\hat{\likedraw}^{-}_{n}$ be a counting process with parameters $r$ and $\hat{\bpdraw}_{n}^{-}$. Then, using the independence of $\hat{\bpdraw}_{n}$ and $\hat{\bpdraw}^{-}_{n}$, we have
\begin{align*}
	\lefteqn{ \mbe_{prior} \left[ \exp\left\{ - \int_{\atom=0}^{1} ( g(\atom) \hat{\bpdraw}_{n}\{\atom\} + f(\atom) \hat{\likedraw}_{n}\{\atom\} ) \right\} \right] } \\
		&= \mbe_{prior} \left[ \exp\left\{ - \int_{\atom=0}^{1} ( g(\atom) \bpdraw_{prior}\{\atom\} + f(\atom) \likedraw\{\atom\} ) \right\} \right] 
		\mbe_{prior} \left[ \exp\left\{ - \int_{\atom=0}^{1} ( g(\atom) \hat{\bpdraw}_{n}^{-}\{\atom\} + f(\atom) \hat{\likedraw}_{n}^{-}\{\atom\} ) \right\} \right].
\end{align*}
So it is enough to show that
\begin{align}
	\label{eq:infty_distr_limit_zero}
	\mbe_{prior} \left[ \exp\left\{ - \int_{\atom=0}^{1} ( g(\atom) \hat{\bpdraw}_{n}^{-}\{\atom\} + f(\atom) \hat{\likedraw}_{n}^{-}\{\atom\} ) \right\} \right]
		&\rightarrow 1, \quad n \rightarrow \infty.
\end{align}

In order to show \eq{infty_distr_limit_zero} holds, we establish the following upper bounds:
\begin{equation}
	\label{eq:inf_lem_upper_bound_one}
	 \mbe_{prior} \left[ \exp\left\{ - \int_{\atom=0}^{1} ( g(\atom) \hat{\bpdraw}_{n}^{-}\{\atom\} + f(\atom) \hat{\likedraw}_{n}^{-}\{\atom\} ) \right\} \right]
	 	\le 1,
\end{equation}
and
\begin{align}
	\nonumber
	\int_{\atom=0}^{1} ( g(\atom) \hat{\bpdraw}_{n}^{-}\{\atom\} + f(\atom) \hat{\likedraw}_{n}^{-}\{\atom\} )
		\nonumber
		&\le  (\max_{\atom} g(\atom)) \hat{\bpdraw}_{n}^{-}[0,1] + (\max_{\atom} f(\atom) )\hat{\likedraw}_{n}^{-}[0,1]. 
\end{align}

Henceforth we use the shorthand $c \defeq (\max_{\atom} g(\atom))$ and $c' \defeq (\max_{\atom} f(\atom))$. These quantities are finite by the assumptions on $g$ and $f$.
Choose $\epsilon > 0$.
Further define the events
$$
	A_{B} \defeq \{\hat{\bpdraw}_{n}^{-}[0,1] > \epsilon\}
	\quad \textrm{and} \quad
	A_{I} \defeq \{\hat{\likedraw}_{n}^{-}[0,1] > \epsilon\}.
$$
By Chebyshev's inequality, 
$$
	\mbp(A_{B,n}) < \mbe\left[ \hat{\bpdraw}_{n}^{-}[0,1] \right] / \epsilon
	\quad \textrm{and} \quad
	\mbp(A_{I,n}) < \mbe\left[ \hat{\likedraw}_{n}^{-}[0,1] \right] / \epsilon.
$$

Using these definitions, we can write
\begin{align}
	\nonumber
	\lefteqn{ \mbe_{prior} \left[ \exp\left\{ - \int_{\atom=0}^{1} ( g(\atom) \hat{\bpdraw}_{n}^{-}\{\atom\} + f(\atom) \hat{\likedraw}_{n}^{-}\{\atom\} ) \right\} \right] } \\
		\nonumber
		&\ge \mbe_{prior} \left[ \exp\left\{ -c \hat{\bpdraw}_{n}^{-}[0,1] - c' \hat{\likedraw}_{n}^{-}[0,1] \right\} \right] \\
		\nonumber
		&\ge 
			\mbe_{prior} \left[ \mbo(A^{C}_{B,n} \cap A^{C}_{I,n}) \exp\left\{ -c \hat{\bpdraw}_{n}^{-}[0,1] - c' \hat{\likedraw}_{n}^{-}[0,1] \right\} \right] \\
		\label{eq:inf_lem_event_bound}
		&\ge \mbp_{prior} (A^{C}_{B,n} \cap A^{C}_{I,n}) \cdot \exp\left\{ - c \epsilon - c' \epsilon \right\}.
\end{align}

Now $\mbp_{prior} (A^{C}_{B,n} \cap A^{C}_{I,n}) = 1 - \mbp_{prior}(A_{B,n} \cup A_{I,n})$. And
\begin{align*}
	\mbp_{prior} (A_{B,n} \cup A_{I,n})
		&\le \mbp_{prior}(A_{B,n}) + \mbp_{prior}(A_{I,n}) \\
		&\le \epsilon^{-1}\left\{ \mbe\left[ \hat{\bpdraw}_{n}^{-}[0,1] \right]
			+ \mbe\left[ \hat{\likedraw}_{n}^{-}[0,1] \right] \right\} 
		\rightarrow 0, \quad n \rightarrow \infty,
\end{align*}
where the last line follows by noting
\begin{align*}
	\mbe\left[ \hat{\bpdraw}_{n}^{-}[0,1] \right]
		&= \int_{\bpweight=0}^{1/n} \bpweight \nu(d\bpweight)
		\rightarrow 0, \quad n \rightarrow 0,
\end{align*}
since $\nu$ is continuous and $\int_{\bpweight=0}^{1} \bpweight \nu(d\bpweight) < \infty$ by assumption,
and
\begin{align*}
	\mbe\left[ \hat{\likedraw}_{n}^{-}[0,1] \right]
		&= \sum_{m=1}^{\infty} \int_{\bpweight=0}^{1/n} C \bpweight^{-1} (1-\bpweight)^{\bpconc - 1} \binom{m+r-1}{m} (1-\bpweight)^{r} \bpweight^{m} \; d\bpweight \\
		&\textrm{where $C$ is a constant in $n$ (c.f.\ \eq{dens_pois_proc_and_neg_bin})} \\
		&= C \sum_{m} (2/n)^{m} \int_{0}^{1/2}  (\tilde{\bpweight})^{m-1} (1-(2/n) \tilde{\bpweight})^{r+\bpconc-1} \binom{m+r-1}{m} \; d\tilde{\bpweight} \\
		&\le C 2^{r+\bpconc-1} \sum_{m} (2/n)^{m} \int_{0}^{1/2} \tilde{\bpweight}^{m-1} (1-\tilde{\bpweight})^{r+\bpconc-1} \binom{m+r-1}{m} \; d\tilde{\bpweight} \\
		&\le 2^{r+\bpconc-1} (1/n) \sum_{m=1}^{\infty} C \int_{0}^{1} \tilde{\bpweight}^{m-1} (1-\tilde{\bpweight})^{r+\bpconc-1} \binom{m+r-1}{m} \; d\tilde{\bpweight} \\
		&= 2^{r+\bpconc-1} (1/n) C',
\end{align*}
where $C'$ is a constant in $n$ (by \lem{xi_r_bnbp}). The final line goes to zero as $n \rightarrow 0$.

So $\mbp_{prior} (A^{C}_{B,n} \cap A^{C}_{I,n}) \rightarrow 1$ as $n \rightarrow \infty$, and the
bound in \eq{inf_lem_event_bound} yields:
$$
	\lim_{n \rightarrow \infty} \mbe_{prior} \left[ \exp\left\{ - \int_{\atom=0}^{1} ( g(\atom) \hat{\bpdraw}_{n}^{-}\{\atom\} + f(\atom) \hat{\likedraw}_{n}^{-}\{\atom\} ) \right\} \right]
		\ge \exp\left\{ - c \epsilon - c' \epsilon \right\}.
$$
Since this result is true for every $\epsilon > 0$, we must have
$$
	\lim_{n \rightarrow \infty} \mbe_{prior} \left[ \exp\left\{ - \int_{\atom=0}^{1} ( g(\atom) \hat{\bpdraw}_{n}^{-}\{\atom\} + f(\atom) \hat{\likedraw}_{n}^{-}\{\atom\} ) \right\} \right]
		\ge 1.
$$
Together with \eq{inf_lem_upper_bound_one}, this equation gives the desired result.
\end{IEEEproof}

\begin{lemma}
	\label{lem:infty_psi_n_psi}
	For $\Phi_{n}$ and $\Phi$ defined in, respectively, \eqs{infty_def_psi_n} and \eqss{infty_def_psi}, we have the limit in \eq{infty_psi_diff_lim}:
	\begin{equation}
	\label{eq:infty_psi_diff_lim_state}
	\int_{\likedraw \in \cmsp} \exp\left\{ - \int_{\atom=0}^{1} f(\atom) \likedraw\{\atom\} \right\}
		(\Psi_{n}(\likedraw) - \Psi(\likedraw)) d\mbp_{marg,n}(\likedraw)
		\rightarrow 0.
	\end{equation}
\end{lemma}

\begin{IEEEproof}
We start by choosing $n$ large enough so that (1) the difference between the ordinary components in the truncated case and the non-truncated case are, in some sense, small enough and (2) the number of atoms in the truncated case is bounded with high probability. Under these two conditions, we will then show that $\Psi_{n}(\likedraw)$ and $\Psi(\likedraw)$ are sufficiently close in value by examining in turn each of the various types of atoms in the proposed posterior.

Therefore, choose $\epsilon > 0$. First note that by the assumption of finite integration of $\nu$ (\eq{infty_finite_mean}) we can choose $n_{0}$ such that for all $n > n_{0}$ we have
\begin{equation}
	\label{eq:infty_mean_less_eps}
	\int_{\bpweight = 0}^{1/n} \bpweight \nu(d\bpweight) < \epsilon.
\end{equation}
This choice implies the existence of $n_{1}$ such that for all $n > n_{1}$ and all $\likeweight \ge 1$ we have \eq{infty_mean_less_eps} as well as
\begin{equation}
	\label{eq:infty_weight_less_eps}
	\int_{\bpweight = 0}^{1/n} b^{\likeweight} (1-\bpweight)^{r} \nu(d\bpweight) < \epsilon.
\end{equation}

Second, since $\likedraw \sim \mbp_{marg,n}$ approaches $\likedraw \sim \mbp_{marg}$ in distribution by \lem{infty_conv_dist}, there exist constants $K'$ and $n_{2}$ such that the number of atoms $K_{n}$ of $\likedraw_{n}$ satisfies
$$
	\mbp_{marg}(K_{n} > K') < \epsilon \quad \textrm{for all} \quad n > n_{2}.
$$

It remains to use these conditions to bound
$$
	\int_{\likedraw \in \cmsp} \exp\left\{ - \int_{\atom=0}^{1} f(\atom) \likedraw\{\atom\} \right\}
		(\Psi_{n}(\likedraw) - \Psi(\likedraw)) d\mbp_{marg,n}(\likedraw).
$$
For instance, since $\Psi_{n}(\likedraw)$ and $\Psi(\likedraw)$ are both bounded between zero and one, we have that
\begin{align}
	\nonumber
	\lefteqn{ \left| \int_{\likedraw \in \cmsp} \exp\left\{ - \int_{\atom = 0}^{1} f(\atom) \likedraw\{\atom\} \right\}
		(\Psi_{n}(\likedraw) - \Psi(\likedraw)) d\mbp_{marg,n}(\likedraw) \right| } \\
		\label{eq:infty_2eps_plus_bound}
		&\le 2\epsilon + \int_{\substack{\likedraw \in \cmsp \\ K_{n} \le K'}}
			\left| \Psi_{n}(\likedraw) - \Psi(\likedraw) \right|
			d\mbp_{marg,n}(\likedraw).
\end{align}

Next, we need to bound the second term on the righthand side of \eq{infty_2eps_plus_bound}. To that end, we break $\Psi_{n}$ and $\Psi$ into their three constituent parts: the fixed atoms from the prior, the new fixed atoms in the proposed posterior, and the ordinary component in the proposed posterior. For $\Psi_{n}$, we have
\begin{align*}
	\Psi_{n}(\likedraw)
		&= \int_{\bpdraw \in \qmsp_{n}} \exp\left\{ -\int_{\atom = 0}^{1} g(\atom) \bpdraw\{\atom\} \right\}
			\; d\mathbb{Q}(\bpdraw : \likedraw) \\
		&= \int_{\bpdraw \in \qmsp_{n}} \exp\left\{
				-\sum_{\atom: \likedraw\{\atom\} \ge 1, \atom \notin \{\fatom_{1},\ldots,\fatom_{L}\}}
					g(\atom) \bpdraw\{\atom\}
				-\sum_{l=1}^{L} g(\fatom_{l}) \bpdraw\{\fatom_{l}\}
				-\int_{\atom=0}^{1} g(\atom) \bpdraw_{ord}\{\atom\}
			\right\}
			\; d\mathbb{Q}(\bpdraw : \likedraw) \\
		&= \left[
				\prod_{\atom: \likedraw\{\atom\} \ge 1, \atom \notin \{\fatom_{1},\ldots,\fatom_{L}\}} 
					\int_{\bpdraw \in \qmsp_{n}}
						\exp\left\{ -g(\atom) \bpdraw\{\atom\} \right\}
					\; d\mathbb{Q}(\bpdraw : \likedraw)
			\right] \\
		& \quad {} \cdot \left[
				\prod_{l=1}^{L}
					\int_{\bpdraw \in \qmsp_{n}}
						\exp\left\{ -g(\fatom_{l}) \bpdraw\{\fatom_{l}\} \right\}
					\; d\mathbb{Q}(\bpdraw : \likedraw)
			\right] \left[
					\int_{\bpdraw \in \qmsp_{n}}
						\exp\left\{ - \int_{\atom = 0}^{1} g(\atom) \bpdraw_{ord}\{ \atom \} \right\}
					\; d\mathbb{Q}(\bpdraw : \likedraw)
				\right] \\
		&\textrm{by the independence of these components under $\mathbb{Q}(\bpdraw : \likedraw)$} \\
		&= \left[
				\prod_{\atom: \likedraw\{\atom\} \ge 1, \atom \notin \{\fatom_{1},\ldots,\fatom_{L}\}}
				c_{new,n}^{-1} \int_{\bpweight = 0}^{1} \exp\{-g(\atom) \bpweight\} \bpweight^{\likedraw\{\atom\}} (1-\bpweight)^{r} \nu_{n}(d\bpweight)
			\right] \\
		& \quad {} \cdot \left[ 
				\prod_{l=1}^{L}
					\int_{\bpdraw \in \qmsp_{n}}
						\exp\left\{ -g(\fatom_{l}) \bpdraw\{\fatom_{l}\} \right\}
					\; d\mathbb{Q}(\bpdraw : \likedraw)
			\right] \left[
				\exp\left\{ 
					-\int_{\bpweight = 0}^{1} \int_{\atom = 0}^{1}
						\left( 1 - e^{-g(\atom) \bpweight} \right) (1-\bpweight)^{r}
					\; d\atom \; \nu_{n}(d\bpweight)
				\right\}
			\right].
\end{align*}
The final factor results from Campbell's theorem.
The analogous formula holds for $\Psi$ by removing the $n$ subscripts.

With the formulas for $\Psi_{n}$ and $\Psi$ in hand, we turn again to our desired bound.
We follow Lemma 3 of \citet{Kim99} in using the following fact:
for $x_{1},\ldots,x_{M}, y_{1},\ldots,y_{M} \in \mathbb{R}$ and $|x_{m}|, |y_{m}| \le 1$ for all m, 
we have 
$$
	\left| \prod_{m=1}^{M} x_{m} - \prod_{m=1}^{M} y_{m} \right|
		\le \sum_{m=1}^{M} |x_{m} - y_{m}|.
$$

In particular, we apply this inequality to transform the difference in $\Psi_{n}$ and $\Psi$ into separate differences in each component, where we note that the prior fixed atom component is shared and therefore disappears. First, for notational convenience, define
\begin{align*}
	C_{n}(I,\atom) &\defeq
		\int_{\bpweight = 0}^{1} \exp\{-g(\atom) \bpweight\} \bpweight^{\likedraw\{\atom\}} (1-\bpweight)^{r} \nu_{n}(d\bpweight) \\
	C(I,\atom) &\defeq \int_{\bpweight = 0}^{1} \exp\{-g(\atom) \bpweight\} \bpweight^{\likedraw\{\atom\}} (1-\bpweight)^{r} \nu(d\bpweight)
\end{align*}
Then
\begin{align*}
	\lefteqn{ \int_{\substack{\likedraw \in \cmsp \\ K_{n} \le K'}} \left| \Psi_{n}(\likedraw) - \Psi(\likedraw) \right|
		\; d\mbp_{marg}(\likedraw) } \\
		&\le \int_{\substack{\likedraw \in \cmsp \\ K_{n} \le K'}} \left\{
			\left[ \sum_{\atom: \likedraw\{\atom\} \ge 1, \atom \notin \{\fatom_{1},\ldots,\fatom_{L}\}}
				\left|
					\left[c_{new,n}(I\{\psi\}) \right]^{-1} C_{n}(I,\atom)
				 - \left[c_{new}(I\{\psi\})\right]^{-1} C(I,\atom)
				\right| \right] \right. \\
		& \quad \left.
				{} + \left| 
					\exp\left\{ -\int_{\bpweight = 0}^{1} \int_{\atom = 0}^{1} \left( 1 - e^{-g(\atom) \bpweight} \right) (1-\bpweight)^{r} \; d\atom \; \nu_{n}(d\bpweight) \right\}
				- \exp\left\{ -\int_{\bpweight = 0}^{1} \int_{\atom = 0}^{1} \left( 1 - e^{-g(\atom) \bpweight} \right) (1-\bpweight)^{r} \; d\atom \; \nu(d\bpweight) \right\}
				\right|
			\right\}.
\end{align*}
We note that $|c_{new,n}(I\{\psi\}) - c_{new}(I\{\psi\})| \le \epsilon$ and $|C_{n}(I,\atom) - C(I,\atom)| \le \epsilon$. And finally the difference in the two exponential terms is at most $\epsilon$. So for large enough $n$ and hence small enough $\epsilon$ we have
\begin{align*}
	\lefteqn{ \int_{\substack{\likedraw \in \cmsp \\ K_{n} \le K'}} \left| \Psi_{n}(\likedraw) - \Psi(\likedraw) \right|
		\; d\mbp_{marg}(\likedraw) } \\
		&\le \epsilon K' \max_{\atom: \likedraw\{\atom\} \ge 1, \atom \notin \{\fatom_{1},\ldots,\fatom_{L}\}}
			\left[
				c_{new}(I\{\psi\})^{-1} + C(I,\atom)
			\right] + \epsilon
\end{align*}
Together with \eq{infty_2eps_plus_bound}, this bound completes the proof.
\end{IEEEproof}

\section{Posterior Inference Details}

\subsection{Exact Gibbs slice sampler} 
\label{app:infinite-sampler-details}
We sample $\bpweight_{d,k}$ and  $\atom_k$ from their Gibbs conditionals as follows:

\textbf{Sample $\atom_k$.}  The conditional posterior of $\atom_k$ given $\tbf{\topic}_{\doubcdot}$ and $\tbf{\obs}_\doubcdot$ is proportional to
\begin{align*}
H(d\atom_k) \prod_{d=1}^D\prod_{n=1}^{\clustsize_d} F(d\obs_{d,n}\mid \atom_k)^{\indic(z_{d,n}=k)}.
\end{align*}
This has a closed form when $H$ is conjugate to $F(\atom_k)$ and may otherwise be sampled using a generic univariate sampling procedure (e.g., random-walk Metropolis-Hastings or slice sampling).

\textbf{Sample $\bpweight_{d,k}$.} By beta-negative binomial conjugacy, the conditional posterior of $\bpweight_{d,k}$ given $\topic_{d,.}$ and $\bpweight_{0,k}$ is a beta distribution,
\begin{align*}
 \bpweight_{d,k} \sim \tb(\bpmass_d\bpconc_d \bpweight_{0,k} + \clustsize_{d,k},\bpconc_d(1-\bpmass_d\bpweight_{0,k}) + r_d),
\end{align*}
where $\clustsize_{d,k} \defeq \sum_n \mathbb{I}(\topic_{d,n} = k)$. 

\textbf{Sample $\bpweight_{0,k}$.}
To sample the shared beta process weights $\bpweight_{0,k}$, we turn to the size-biased construction of the beta process introduced by~\cite{ThibauxJo07}
\begin{align*}
	\bpdraw_0 = \sum_{\round=0}^{\infty}\sum_{i=1}^{C_{\round}} \bpweight_{0,\round,i} \delta_{\atomloc{\round,i}},
\end{align*}
where 
\begin{equation*}
	C_{\round} \indep \pois\left(\frac{\bpconc_0\bpmass_0}{\bpconc_0+\round}\right),\quad  \bpweight_{0,\round,i} \indep \tb(1,\bpconc_0+\round),\quad  \text{ and }\quad \atomloc{\round,i} \iid \base.
\end{equation*}

If we order the atoms by the rounds in which they were drawn, then  
the $k$-th atom overall was drawn in round $\round_k$, where
$$
	\round_k \defeq \min\left\{\round : \sum_{j=0}^\round C_j \geq k\right\}.
$$
Conditional on the round indices $(\round_k)_{k=1}^\infty$, we have
\begin{align*}
	\bpdraw_0  = \sum_{k=1}^{\infty}\bpweight_{0,k} \delta_{\atom_k}\quad
\end{align*}
for
$$
	\bpweight_{0,k} \indep \tb(1,\bpconc_0+\round_k)\quad \text{and}\quad \atom_k \iid \base.
$$
The conditional density of $\bpweight_{0,k}$ given the remaining variables is therefore
proportional to
\begin{align}\label{eq:b0-uncollapsed}
 (1-\bpweight_{0,k})^{\bpconc_0+\round_k-1}\prod_{d=1}^D \frac{1}{\Gamma(\bpmass_d\bpconc_d\bpweight_{0,k})\Gamma(\bpconc_d(1-\bpmass_d\bpweight_{0,k}))}\left(\frac{\bpweight_{d,k}}{1-\bpweight_{d,k}}\right)^{\bpmass_d\bpconc_d \bpweight_{0,k}} 
\end{align}
and may be sampled using random-walk Metropolis-Hastings.

It remains then to sample the latent round indices $\round_k$ or, equivalently, their
differences  $\rounddiff_k \defeq \round_k - \round_{k-1}$, where $m_0 \defeq 0$ for notational convenience.
Let $f_m$ and $F_m$ denote the pmf and cdf of the $\pois(\frac{\bpconc_0\bpmass_0}{\bpconc_0+m})$ distribution respectively,
and define $C_{\round,j} \defeq \sum_{k=1}^j\indic(\round_k = m)$.
Since $C_{\round} = \sum_{k=1}^\infty\indic(\round_k = m) \sim \pois(\frac{\bpconc_0\bpmass_0}{\bpconc_0+m})$, it follows that
\begin{align*}
	\mbp(\rounddiff_k < 0 \mid (\rounddiff_j)_{j=1}^{k-1}) &= 0,\\
	\mbp(\rounddiff_k = 0 \mid (\rounddiff_j)_{j=1}^{k-1}) &= \frac{1- F_{\round_{k-1}}(C_{\round_{k-1},k-1})}{1- F_{\round_{k-1}}(C_{\round_{k-1},k-1}-1)} 
\end{align*}
for $\round_{k-1} = \sum_{j=1}^{k-1} \rounddiff_j$, and 
\begin{align*}	
	\mbp(\rounddiff_k = \rounddiff \mid (\rounddiff_j)_{j=1}^{k-1}) =
		&\frac{f_{\round_{k-1}}(C_{\round_{k-1},k-1})}{1- F_{\round_{k-1}}(C_{\round_{k-1},k-1}-1)}
		(1 - f_{\round_{k-1}+\rounddiff}(0))\prod_{g=1}^{\rounddiff-1} f_{\round_{k-1}+g}(0) 
\end{align*}
for all $\rounddiff\in\N$.
The conditional distribution of $\rounddiff_k$ given $(\rounddiff_j)_{j=1}^{k-1}$ and $\bpweight_{0,k}$ is then
$$
	p(\rounddiff_k \mid (\rounddiff_j)_{j=1}^{k-1}, \bpweight_{0,k}) \propto (1-\bpweight_{0,k})^{\rounddiff_k}(\bpconc_0+\rounddiff_k+\round_{k-1})p(\rounddiff_k \mid (\rounddiff_j)_{j=1}^{k-1}),
$$
which cannot be normalized in closed form due to the infinite summation.
To permit posterior sampling of $\rounddiff_k$, we introduce an auxiliary variable $v_k$ with conditional
distribution
$$
	v_k \sim \unif(0, \zeta_{0,\rounddiff_k}(1-\bpweight_{0,k})^{\rounddiff_k})
$$ 
where $(\zeta_{0,\rounddiff})_{\rounddiff=1}^\infty$ is a fixed positive sequence with $\lim_{\rounddiff\to\infty} \zeta_{0,\rounddiff} = 0$.
Given $v_k$, we may slice sample $\rounddiff_k$ from the finite distribution
$$
	p(\rounddiff_k \mid (\rounddiff_j)_{j=1}^{k-1}, \bpweight_{0,k}) \propto \frac{\indic(v_k \leq \zeta_{0,\rounddiff_k}(1-\bpweight_{0,k})^{\rounddiff_k})}{\zeta_{0,\rounddiff_k}}(\bpconc_0+\rounddiff_k+\round_{k-1}) p(\rounddiff_k \mid(\rounddiff_j)_{j=1}^{k-1}).
$$

\subsection{Collapsed sampling} 
\label{app:collapsed-sampling}
In \eq{b0-uncollapsed}, we sampled $\bpweight_{0,k}$ conditional on  $\tbf{\bpweight}_{\cdot,k}$.
A more efficient alternative is to integrate  $\tbf{\bpweight}_{\cdot,k}$ out of this conditional.
We exploit the conjugacy of the beta and negative binomial distributions to derive the
conditional distribution of $\clustsize_{d,k}$ given $\bpweight_{0,k}$, $\bpmass_d$, $\bpconc_d$, and $r_d$:
\begin{align*}
	p(\clustsize_{d,k} \mid \bpweight_{0,k},\bpmass_d,\bpconc_d,r_d) 
	&= \int p(\clustsize_{d,k}\mid \bpweight_{d,k},r_d)p(\bpweight_{d,k}\mid \bpweight_{0,k},\bpmass_d,\bpconc_d)d{\bpweight_{d,k}} \\
	&= \int \frac{\Gamma(\clustsize_{d,k}+r_d)}{\clustsize_{d,k}!\ \Gamma(r_d)} \frac{\Gamma(\bpconc_d)\bpweight_{d,k}^{\clustsize_{d,k}+\bpmass_d\bpconc_d\bpweight_{0,k}-1} (1-\bpweight_{d,k})^{r_d+\bpconc_d(1-\bpmass_d\bpweight_{0,k})-1}}{\Gamma(\bpmass_d\bpconc_d\bpweight_{0,k})\ \Gamma(\bpconc_d(1-\bpmass_d\bpweight_{0,k}))} d{\bpweight_{d,k}} \\
	&= \frac{\Gamma(\clustsize_{d,k}+r_d)}{\clustsize_{d,k}!\ \Gamma(r_d)} \frac{\Gamma(\bpconc_d)\ \Gamma(\clustsize_{d,k}+\bpmass_d\bpconc_d\bpweight_{0,k})\ \Gamma(r_d+\bpconc_d(1-\bpmass_d\bpweight_{0,k}))}{\Gamma(\clustsize_{d,k}+r_d+\bpconc_d)\ \Gamma(\bpmass_d\bpconc_d\bpweight_{0,k})\ \Gamma(\bpconc_d(1-\bpmass_d\bpweight_{0,k}))} .
\end{align*}
The conditional density of $\bpweight_{0,k}$ with $\tbf{\bpweight}_{\cdot,k}$ integrated out now takes the form 
\begin{align*}
 (1-\bpweight_{0,k})^{\bpconc_0+\round_k-1}\prod_{d=1}^D 
\frac{\Gamma(\clustsize_{d,k}+\bpmass_d\bpconc_d\bpweight_{0,k})\ \Gamma(r_d+\bpconc_d(1-\bpmass_d\bpweight_{0,k}))}{\Gamma(\bpmass_d\bpconc_d\bpweight_{0,k})\ \Gamma(\bpconc_d(1-\bpmass_d\bpweight_{0,k}))}
\end{align*}
and may be sampled using random-walk Metropolis-Hastings.

\subsection{Finite approximation Gibbs sampler}
\label{app:finite-sampler-details}
The full conditional distribution of $\bpweight_{0,k}$ under the finite approximation of \eq{bp-finite-approximation} is proportional to
\begin{align*}
 \bpweight_{0,k}^{\bpconc_0\bpmass_0/K-1}(1-\bpweight_{0,k})^{\bpconc_0(1-\bpmass_0/K)-1}\prod_{d=1}^D \frac{1}{\Gamma(\bpmass_d\bpconc_d\bpweight_{0,k})\Gamma(\bpconc_d(1-\bpmass_d\bpweight_{0,k}))}\left(\frac{\bpweight_{d,k}}{1-\bpweight_{d,k}}\right)^{\bpmass_d\bpconc_d \bpweight_{0,k}},
\end{align*}
while the conditional density with $\tbf{\bpweight}_{\cdot,k}$ integrated out is proportional to
\begin{align*}
 \bpweight_{0,k}^{\theta_0\bpmass_0/K-1}(1-\bpweight_{0,k})^{\theta_0(1-\bpmass_0/K)-1}\prod_{d=1}^D 
\frac{\Gamma(\clustsize_{d,k}+\bpmass_d\bpconc_d\bpweight_{0,k})\ \Gamma(r_d+\bpconc_d(1-\bpmass_d\bpweight_{0,k}))}{\Gamma(\bpmass_d\bpconc_d\bpweight_{0,k})\ \Gamma(\bpconc_d(1-\bpmass_d\bpweight_{0,k}))}.
\end{align*}
Random-walk Metropolis-Hastings may be used to sample $\bpweight_{0,k}$ from either distribution.

With this approximation in hand, we sample $\lambda_{d,k}$, $\bpweight_{d,k}$, and $\atom_k$ precisely as described in \mysec{infinite-sampler}.
Since the number of components is finite, no auxiliary slice variables are needed to sample the component indices.
Hence, we may sample $\topic_{d,n}$ from its discrete conditional distribution
\begin{align*}
 \mbp(\topic_{d,n} = k) \propto F(d\obs_{d,n}\mid \atom_{k}) \lambda_{d,k}
\end{align*}
given the remaining variables.

\end{document}